\documentclass[sigconf]{acmart}

\usepackage[linesnumbered,ruled,vlined,algo2e]{algorithm2e}
\usepackage{amsmath}
\usepackage{enumitem}
\usepackage{hyperref}
\usepackage[utf8]{inputenc}
\usepackage{multirow}

\newtheorem{problem}{Problem}

\AtBeginDocument{%
  \providecommand\BibTeX{{%
    \normalfont B\kern-0.5em{\scshape i\kern-0.25em b}\kern-0.8em\TeX}}}

\begin{document}

\title{A Critical Re-evaluation of Benchmark Datasets for
      (Deep) Learning-Based Matching Algorithms}

\author{George Papadakis}
\orcid{0000-0002-7298-9431}
\affiliation{%
  \institution{National and Kapodistrian University of Athens}
  \city{Athens}
  \country{Greece}
}
\email{gpapadis@uoa.gr}

\author{Nishadi Kirielle}
\orcid{0000-0002-6503-0302}
\affiliation{%
  \institution{The Australian National University}
  \city{Canberra}
  \postcode{2600}
  \country{Australia}
}
\email{nishadi.kirielle@anu.edu.au}

\author{Peter Christen}
\orcid{0000-0003-3435-2015}
\affiliation{%
  \institution{The Australian National University}
  \city{Canberra}
  \postcode{2600}
  \country{Australia}
}
\email{peter.christen@anu.edu.au}

\author{Themis Palpanas}
\orcid{0000-0002-8031-0265}
\affiliation{%
  \institution{Universit{\'e} Paris Cit{\'e} \& French University Institute (IUF)}
  \city{Paris}
  \country{France}
}
\email{themis@mi.parisdescartes.fr}

\begin{abstract}
Entity resolution (ER) is the process of identifying records that refer to the same entities within one or across multiple databases. Numerous techniques have been developed to tackle ER challenges over the years, with recent emphasis placed on machine and  deep learning methods for the matching phase. 
However, the quality of the benchmark datasets typically used in the experimental evaluations of learning-based matching algorithms has not been examined in the literature. To cover this gap, we propose four different approaches to assessing the difficulty and appropriateness of 13 established datasets: two theoretical approaches, which involve new measures of linearity and existing measures of complexity, and two practical approaches: the difference between the best non-linear and linear matchers, as well as the difference between the best learning-based matcher and the perfect oracle. Our analysis demonstrates that most of the popular datasets pose rather easy classification tasks. As a result, they are not suitable for properly evaluating learning-based matching algorithms. To address this issue, we propose a new methodology for yielding benchmark datasets. We put it into practice by creating four new matching tasks, and we verify that these new benchmarks are more challenging and therefore more suitable for further advancements in the field.
\end{abstract}

\maketitle

\section{Introduction}

Entity Resolution (ER) aims to 
identify and link records that refer to the same entity across databases,
called \textit{duplicates}~\cite{Naumann2010morgan}. 
ER has been an
active topic of research 
since the 1950s~\cite{Newcombe1959science},  while various learning-based ER techniques, both supervised and
unsupervised, have been developed in the past two decades. 
For overviews of ER, we refer the reader to recent books and
surveys~\cite{Binette2022scadv,Christen2020springer,Dong2015morgan,Papadakis2021morgan}.

ER faces several major challenges. First, databases typically contain no unique global entity identifiers that would allow an exact join to identify those records that refer
to the same entities. As a result, \textit{matching} methods compare
quasi-identifiers (QIDs)~\cite{Christen2020springer}, such as names
and addresses of people, or titles and authors of publications.
The assumption here is that the more similar their QIDs are, the
more likely the corresponding records are to be matching. Second,
as databases are getting larger, comparing all possible pairs of
records is infeasible, due to the quadratic cost. Instead,
\textit{blocking}, \textit{indexing}, or \textit{filtering}
techniques~\cite{Papadakis2020csur} typically identify the
candidate pairs or groups of records that are forwarded to
matching.

In recent years, a diverse range of methods based on
machine learning (ML)~\cite{DBLP:journals/pvldb/KondaDCDABLPZNP16} and especially deep
learning (DL) has been developed to address the first challenge, namely
matching~\cite{Barlaug2021acm,Mudgal2018sigmod}. Due to the
similarity of ER to natural language processing tasks, such as
machine translation or entity extraction and recognition, many
DL-based matching techniques leverage relevant technologies like
pre-trained language models. The experimental results reported
have been outstanding, as these methods maximize matching
effectiveness in many benchmark
datasets~\cite{DBLP:journals/jdiq/LiLSWHT21,Mudgal2018sigmod,DBLP:conf/edbt/BrunnerS20}. 

However, the quality of these benchmark datasets has been overlooked in the literature -- the sole exception is the analysis of the large portion of entities shared by training and testing sets, which results in low performance in the case of unseen test entities \cite{Wang2022arxiv}. Existing ER benchmark datasets typically treat matching as a binary classification task that applies to a set of candidate pairs generated after blocking, which typically has a significant impact on the resulting performance \cite{Dong2015morgan,DBLP:series/synthesis/2015Christophides}. In general, a loose blocking approach achieves high recall, ensuring that all positive instances (i.e., matching pairs) are included, at the cost of many negative ones (i.e., non-matching pairs) with low similarity, which can thus be easily discarded by a learning-based matching algorithm, even a linear one. In contrast, a strict blocking approach might sacrifice a small part of the positive instances, but mostly includes highly similar negative ones, which involve nearest neighbors and are harder to be classified, thus requiring more effective, complex and non-linear learning-based matching algorithms. Nevertheless, most existing datasets lack any documentation about the blocking process that generated their candidate pairs, i.e., no information is provided about which blocking method was used, how it was configured and which attributes provided the textual evidence for creating blocks.
As a result, there is a large deviation in core characteristics like the imbalance ratio between the existing benchmarks and those created through a principled approach that employs a fine-tuned state-of-the-art blocking method, as documented in Section \ref{sec:methodology}.

In this paper, we aim to cover the above gap in the literature by
proposing a principled framework for assessing the quality of benchmark datasets for learning-based matching algorithms. It consists of two types of measures. First,
\textit{a-priori} measures theoretically estimate the appropriateness of a benchmark dataset, based exclusively on the characteristics of its classes. We propose novel measures that estimate the degree of linearity in a benchmark dataset as well as existing complexity measures that are applied to ER benchmarks for the first time. Second,
\textit{a-posteriori} measures rely on the performance of matching algorithms.

To put these measures into practice, we consider seven open-source, non-linear ML- and DL-based matching algorithms, which include the state-of-the-art techniques in the field. We complement them with novel matching algorithms, which perform linear classification, thus estimating the baseline performance of learning-based methods. These two types of algorithms allow for estimating the real advantage of non-linear learning-based matching algorithms over simple linear ones, as well as their distance from the ideal matcher, i.e., the perfect oracle.

When applying our a-priori and a-posteriori measures on widely used
benchmark ER datasets, our experimental evaluation
shows that most of these datasets are inappropriate for evaluating the full potential of complex matchers, such as DL-based ones. To address this issue, we propose a novel way of constructing benchmarks from the same original data based on blocking and the knowledge of the complete ground truth, i.e., the real set of matching entities. We apply all our a-priori and a-posteriori measures to the new benchmark datasets, demonstrating that they form harder classification tasks
that highlight the advantages of DL-based matching algorithms. To the best of our knowledge, these topics have not been examined in the literature before. 

Overall, we make the following contributions:
\begin{itemize}[leftmargin=*]
    \item In Section~\ref{sec:theoreticalMeasures}, we coin novel theoretical measures for \textit{a-priori} assessing the difficulty of ER benchmark datasets. We also introduce two novel aggregate measures that leverage a series of matching algorithms to \textit{a-posteriori} assess the difficulty of ER benchmarks.
    \item In Section \ref{sec:practicalMeasures}, we introduce a taxonomy of DL-based matching methods that facilitates the understanding of their functionality, showing that we consider a representative sample of the recent developments in the field. We also define a new family of linear learning-based matching algorithms, whose performance depends heavily on the difficulty of ER benchmarks. These algorithms lay the ground for estimating the two a-posteriori measures.
    \item In Section \ref{sec:datasetAnalysis}, we perform the first systematic evaluation of 13 popular ER benchmarks, demonstrating experimentally that most of them are too easy to classify to properly assess the expected improvements of novel matching algorithms in real ER scenarios.
    \item In Section \ref{sec:methodology}, we propose a novel methodology for creating new ER benchmarks and experimentally demonstrate that they are more suitable for assessing the benefits of DL-based matchers.
\end{itemize}

All our experiments can be reproduced through a Docker image\footnote{\url{https://github.com/gpapadis/DLMatchers/tree/main/dockers/mostmatchers}}.

\section{Problem Definition}
\label{sec:pb}

The goal of ER is to identify duplicates, i.e., different
records that describe the same real-world entities. To
this end, an ER matching algorithm receives as input a set
of candidate record pairs ${C}$. These are likely matches
that are produced by a blocking or filtering
technique~\cite{Papadakis2020csur}, which is used to 
reduce the inherently quadratic computational cost
(instead of considering all possible pairs, it
restricts the search space to highly similar ones).
For each record pair $(r_i,r_j) \in C$, a matching
algorithm decides whether $r_i \equiv r_j$ or not,
where $\equiv$ indicates that they are duplicates,
referring to the same entity. The resulting set of
matching pairs is denoted by ${M}$, and the non-matching
pairs by ${N}$ (where $C = M \cup N$ and $M \cap N =
\emptyset$).

This task naturally lends itself to a binary classification setting. In this case, $C$ constitutes the testing set, which is accompanied by a training and a validation set, ${T}$ and ${V}$, respectively, with record pairs of known label, such that $C$, $T$ and $V$ are mutually exclusive. As a result, the performance of matching is typically assessed through the
\textit{F-Measure} (${F1}$)~\cite{Christen2023csur}, which is the harmonic mean of recall (${Re}$) and precision (${Pr}$), i.e., $F1=2\cdot Re \cdot Pr / (Re + Pr)$, where $Re$ expresses the portion of existing duplicates that are classified as such, i.e., $Re = |G \cap M| / |G|$ with ${G}$ denoting the ground truth (the set of true duplicates), and $Pr$ determines the portion of detected matches that correspond to duplicates, i.e., $Pr = |G \cap M| / |M|$~\cite{Christen2012springer,Han18}. 
All these measures are defined in $[0,1]$, with
higher values indicating higher effectiveness. Note that
we do not consider evaluation measures used for blocking,
such as pairs completeness and pairs
quality~\cite{Christen2012springer}, because the blocking
step lies outside of our analysis.

In this given context, we formally define matching as
follows:

\begin{problem}[Matching] Given a testing set of candidate pairs
$C$ along with a training and a validation set, $T$ and $V$, respectively, such
that $C \cap T = \emptyset$, $C \cap V = \emptyset$, and
$T \cap V = \emptyset$, train a binary classification
model that splits the elements of $C$ into the set of
matching and non-matching pairs, $M$ and $N$ respectively,
such that $F1$ is maximized.
\end{problem}

Note that by considering the candidate pairs generated through
blocking or filtering techniques, this definition is
generic enough to cover any type of ER. There are three main types: (1)
\textit{deduplication}~\cite{Naumann2010morgan}, also known as \textit{Dirty ER} \cite{DBLP:journals/csur/ChristophidesEP21}, where the input comprises a single database with duplicates in itself; (2) \textit{record linkage} (RL)~\cite{Christen2020springer}, also known as \textit{Clean-Clean} ER \cite{DBLP:journals/csur/ChristophidesEP21}, where the input involves two individually duplicate-free, but overlapping databases; (3) \textit{multi-source ER} \cite{DBLP:conf/sigmod/SagiGBBA16}, where the goal is to identify matching records across multiple duplicate-free data sources. In this work, we follow the literature on DL-based matching algorithms, considering exclusively matching algorithms for record linkage \cite{Mudgal2018sigmod,DBLP:conf/edbt/BrunnerS20,DBLP:conf/www/ChenSZ20,DBLP:journals/pvldb/0001LSDT20,DBLP:conf/ijcai/FuHHS20}.

\section{Measures of Difficulty}
\label{sec:theoreticalMeasures}

We now describe two types of theoretical measures and two practical ones for a-priori and a-posteriori assessing the difficulty of ER benchmark datasets. All operate in a \textit{schema-agnostic} manner that considers all attribute values in every record, disregarding the attribute structures in the given data sources. In preliminary experiments, we also explored schema-aware settings, applying the same measures to specific attribute values. These settings, though, showed no significant difference in performance in comparison to the schema-agnostic settings for both types of theoretical measures. Thus, we omit them for brevity, but report them in the Appendix.

\subsection{Degree of Linearity}
\label{sec:linearityDegrees}

\begin{algorithm2e}[t]
\DontPrintSemicolon
\SetKw{KwBy}{by}
\small
\KwIn{The training, validation and testing sets $T, V, C$, respectively, and the similarity measure $sim$}
\KwOut{The linearity degree $F1^{max}_{sim}$, and the best threshold $t_{best}$}
$S \gets \{\}$, $D \gets T \cup V \cup C$\\
\ForEach{$(r_i,r_j) \in D$}{
    $T_i \gets tokens(r_i)$, $T_j \gets tokens(r_j)$\\
    $S \gets S \cup (sim(T_i, T_j), (r_i,r_j))$
}
$F1^{max}_{sim} \gets 0$, $t_{best} \gets 0$ \\
\For{$t\gets 0.01$ \KwTo $0.99$ \KwBy $0.01$}{
    $M \gets \{\}$, $N \gets \{\}$\\
    \ForEach{$(sim_{i,j}, (r_i,r_j)) \in S$}{
        \lIf(\tcp*[h]{A match}){$t \leq sim_{i,j}$}{
            $M \gets M \cup \{(r_i,r_j)\}$
        } \lElse(\tcp*[h]{A non-match}) {
            $N \gets N \cup \{(r_i,r_j)\}$
        } 
    }
    \If{$F1^{max}_{sim} < F1(M, N, D)$}{
        $F1^{max}_{sim} \gets F1(M, N, D)$, $t_{best} \gets t$
    }
}
\textbf{return} $F1^{max}_{sim}$, $t_{best}$
\caption{Estimating the degree of linearity}
\label{algo:linearityDegree}
\end{algorithm2e}

To assess the difficulty of matching benchmarks, we introduce two new measures for estimating the success of a \textit{linear classifier}. The higher their scores are, the easier it is for \textit{any} supervised matching algorithm to achieve high effectiveness on the corresponding dataset. This means that benchmarks with a high degree of linearity are not suitable for highlighting the differences between complex, non-linear classifiers like those leveraging deep learning in Section \ref{sec:dlMatchers}. Instead, datasets with a low degree of linearity are more likely to stress the pros and cons of each matcher.

In this context, we propose Algorithm \ref{algo:linearityDegree}, which relies on all labels in a benchmark dataset. First, it merges the training with the validation and the testing sets into a single dataset $D$ in line 1. Then, for every candidate pair $c_{i,j}=(r_i, r_j) \in D$, it creates two token sequences,
$T_i$ and $T_j$, where $T_x$ comprises the set of 
tokens in all attribute values in record $r_x$, after converting all tokens to lower-case (lines 2 and 3). A similarity score per pair, $sim(c_{i,j})=sim(T_i,T_j) \in [0, 1]$, is then calculated based on $T_i$ and $T_j$ and added to the set of similarities $S$ along with the candidate pair in line 4.
Finally, it classifies all labelled pairs using a threshold
$t$ with the following rule: if $t \leq sim(c_{i,j})$, we
have a matching pair (line 9)
otherwise a non-matching pair (line 10).
In line 6, the algorithm loops over all thresholds in $[0.01, 0.99]$ with an increment
of $0.01$, and identifies the threshold that results in the
highest F-measure value (lines 11 and 12). We denote this maximum F1 as the
degree of linearity, $F1_{sim}^{max}$, which is returned as output along with the corresponding threshold, $t_{best}$ in line 13. 

In this work, we consider two similarity measures between the token sequences $T_i$ and $T_j$ of the candidate pair $c_{i,j}=(r_i, r_j)$:
\begin{enumerate}
    \item The Cosine similarity, which is defined as:
    \begin{equation}
        CS(c_{i,j})= |T_i \cap T_j|/\sqrt{|T_i| \times |T_j|}.
    \label{eq:cosine}
    \end{equation}
    \item The Jaccard similarity, which is defined as:
    \begin{equation}
        JS(c_{i,j})= |T_i \cap T_j|/(|T_i| + |T_j| - |T_i \cap T_j|).
    \label{eq:jaccard}
    \end{equation}
\end{enumerate}
They yield two degrees of linearity: $F1_{CS}^{max}$ and $F1_{JS}^{max}$, respectively. By considering their maximum possible value, these measures indicate the optimal performance of a linear matching algorithm, with F1=1.0
indicating perfect separation (no false matches and no
false non-matches). Note that other measures such as the Dice or Overlap similarities~\cite{Christen2012springer} could be employed, however these are linearly dependent on the Cosine and Jaccard similarities and therefore do not provide additional useful information.

\subsection{Complexity Measures} 
\label{sec:complexityMeasures}
Measures for estimating the complexity of imbalanced classification tasks have been summarized and extended in \cite{DBLP:conf/ijcnn/BarellaGSLC18,lorena2018complex}. They serve the same purpose as the degree of linearity, determining whether a benchmark is suitable for comparing learning-based matching algorithms. In this case, the lower the average score of a dataset is, the easier is the corresponding classification task. Collectively, these measures consider versatile and comprehensive evidence that is complementary to the degree of linearity. As a result, there are datasets where the degree of linearity is low, but the average complexity score suggests otherwise, indicating that simple patterns suffice for a high effectiveness, and vice versa.

Essentially, there are five
types of such measures, as shown in Table \ref{tb:features}, which summarizes them:

\begin{enumerate}[leftmargin=*]
 \item The \textit{feature overlapping measures} assess how
    discriminative the numeric features are. $f_1$ denotes the
    maximum Fisher’s discriminant ratio, $f_{1v}$ alters $f_1$
    by taking projections into account, $f_2$ expresses the
    volume of the overlapping region, $f_3$ captures the
    maximum individual feature efficiency (in separating the
    two classes), and $f_4$ is the collective feature efficiency
    measure, which summarizes the overall discriminatory power
    of all features. A low value in at least one of these
    measures indicates an easy classification~task.

\item The \textit{linearity measures} check how effective the
    hyperplane defined by a linear SVM classifier is in
    separating the two classes. $l_1$ sums the error distance
    of misclassified instances, $l_2$ is the error rate of the
    linear classifier, and $l_3$ stands for the non-linearity
    of the linear classifier (measuring the error rate on randomly generated
    synthetic instances of
    interpolated same-class training pairs).

\item The \textit{neighborhood measures} characterize the
    decision boundary between the two classes, taking into
    account the class overlap in local neighborhoods according
    to the Gower distance~\cite{gower1971general}. $n_1$
    estimates the fraction of borderline instances after
    constructing a minimum spanning tree, $n_2$ is the ratio
    formed by the sum of distances of each instance to its
    nearest neighbor from the same class in the numerator and
    its nearest neighbor from other class in the denominator,
    $n_3$ is the error rate of a kNN classifier with $k=$1 that
    is trained through leave-one-out cross validation, $n_4$
    differs from $n_3$ in that it uses a neural network as a
    classifier, $t_1$ is the number of hyperspheres centered
    at an instance required to cover the entire dataset
    divided by the total number of instances, and $lsc$ is
    the average cardinality of the local set per instance, 
    which includes the instances from the
    other class that are closer than its nearest neighbor
    from the same class.

\item The \textit{network measures} model a dataset as a
    graph, whose nodes correspond to instances and the
    edges connect pairs of instances with a Gower
    distance lower than a threshold. Edges between instances
    of a different class are pruned after the construction of
    the graph. \textit{Density} measures the portion of
    retained edges over all possible pairs of instances,
    \textit{clsCoef} is the average number of retained edges
    per node divided by the neighborhood size before the
    pruning, and \textit{hub} assesses the average influence
    of the nodes (for each node, it sums the number of its
    links, weighting each neighbor by the number of its own
    links).

\item The \textit{dimensionality measures} evaluate data
    sparsity in three ways: $t_2$ calculates the average number
    of instances per dimension, $t_3$  
    estimates the PCA components required for representing
    95\% of data variability, while $t_4$ assesses the
    portion of relevant dimensions by dividing those considered
    by $t_3$ with the original ones of $t_2$.

\end{enumerate}

\begin{table}[t]\centering
\caption{{\small Definition of Complexity Measures}}
\vspace{-7pt}
{\small
\begin{tabular}{ | l | l |}
\hline
$f_1$ & maximum Fisher’s discriminant ratio\\
$f_{1v}$ & directional-vector maximum Fisher’s discriminant ratio\\
$f_2$ & volume of the overlapping region \\
$f_3$ & max individual feature efficiency in separating the classes \\
\hline
\multicolumn{2}{c}{(a) Feature-based measures}\\
\hline
$l_1$ & sum of the error distance by linear programming \\
$l_2$ & error rate of linear SVM classifier \\
\hline
\multicolumn{2}{c}{(b) Linearity measures}\\
\hline
$n_1$ & fraction of borderline points \\
$n_2$ & ratio of intra/extra class nearest neighbor distance\\
$n_3$ & error rate of the nearest neighbor classifier\\
$n_4$ & non-linearity of the nearest neighbor classifier \\
$t_1$ & fraction of hyperspheres covering data \\
$lsc$ & local set average cardinality \\
    \hline
\multicolumn{2}{c}{(c) Neighborhood measures}\\
\hline
$den$ & average density of the network \\
$cls$ & custering coefficient\\
$hub$ & hub score \\
\hline
\multicolumn{2}{c}{(d) Network measures}\\
\hline
$c_1$ & entropy of class proportions \\
$c_2$ & imbalance ratio \\
\hline
\multicolumn{2}{c}{(e) Class balance measures}\\
	\end{tabular}
	}
	\vspace{-14pt}
	\label{tb:features}
\end{table}

All these measures yield values in $[0,1]$, with higher values
indicating more complex classification tasks. To put them into practice,
we transform each dataset into a set of features using the same methodology as in Section \ref{sec:linearityDegrees}: we represent every pair of candidates $c_{i,j} \in D$ by the two-dimensional feature vector $f_{i,j}=[CS(c_{i,j}), JS(c_{i,j})]$, where $CS(c_{i,j})$ and $JS(c_{i,j})$ are the Cosine and Jaccard similarities defined in Equations \ref{eq:cosine} 
and \ref{eq:jaccard}, respectively.

\subsection{Practical Measures}
\label{ref:aggmeas}

The above a-priori measures provide no evidence about the actual performance of learning-based matching algorithms on a particular benchmark. To cover this aspect, we complement them with two a-posteriori measures that encapsulate the performance of the matching algorithms in Section \ref{sec:quantitativeAnalysis}. These measures help to identify benchmarks that contain a considerable portion of non-linearly separable candidate pairs, thus yielding low scores for the a-priori measures, but are still not suitable for benchmarking matching algorithms. There are two conditions for these cases: (i) a linear matching algorithm achieves a performance comparable to the top-performing non-linear ones, and (ii) the maximum F1 score among all learning-based matching algorithms is very close to the maximum possible score of F1=1. Only datasets satisfying none of these conditions are suitable for benchmarking supervised matching algorithms, despite their low a-priori scores.

Our practical measures include two ML-based and five DL-based matching algorithms, each combined with different configurations, as we describe in more detail in Section \ref{sec:quantitativeAnalysis}. Overall, we consider seven state-of-the-art algorithms, which together provide a representative performance of non-linear, learning-based techniques. Especially the DL-based algorithms cover all subcategories in our taxonomy, as shown in Table \ref{tb:taxonomy}. Along with the above six linear classifiers, they yield two novel, aggregate measures for assessing the advantage of the non-linear and the potential of all learning-based matchers: 
\begin{enumerate}[leftmargin=*]
    \item \textit{Non-linear boost} (\textsf{NLB}) is defined as the difference between the maximum F1 of all considered ML- and DL-based matching algorithms and the maximum F1 of all linear ones. The larger its value is, the greater is the advantage of non-linear classifiers, due to the high difficulty of an ER benchmark. In contrast, values close to zero
    indicate trivial ER benchmarks with linearly separable classes.
    \item \textit{Learning-based margin} (\textsf{LBM}) is defined as the distance between 1 and the maximum F1 of all considered learning-based matching algorithms. The higher its value is for a 
    benchmark, the more room for improvements there is. 
    Low values, close to zero, indicate datasets where learning-based matchers already exhibit practically perfect performance.
\end{enumerate}

\section{Matching Algorithms}
\label{sec:practicalMeasures}

To quantify the practical measures, we use
three types of matchers. Each one is presented in a different subsection.

\subsection{DL-based Matching Algorithms}
\label{sec:dlMatchers}

\textbf{Selection Criteria.}
In our analysis, we consider as many DL-based matching algorithms as possible in order to get a reliable estimation on this type of algorithms on each dataset. To this end, we consider algorithms that satisfy the following four selection criteria:

\begin{enumerate}
    \item \textit{Publicly available implementation:} All DL-based
  algorithms involve hyperparameters that affect
  their performance to a large extent, but for brevity or due to limited space, their description and
  fine-tuning  are typically omitted in the context
  of a scientific publication. Reproducing experiments can therefore
  be a challenging task that might bias the results of our experimental
  analysis. In fact, as our experimental results in Section~\ref{sec:quantitativeAnalysis} demonstrate, it is also challenging to reproduce the performance of publicly available matching algorithms. To avoid such issues, we exclusively consider methods with
  a publicly released implementation.
  \item \textit{No auxiliary data sources:} Practically, all DL-based
  matching algorithms leverage deep neural networks in combination with
  embedding techniques, which transform every input record into a
  (dense) numerical vector. To boost time efficiency, these embeddings
  typically rely on pre-trained corpora, such as fastText~\cite{DBLP:journals/tacl/BojanowskiGJM17} or
  BERT-based models~\cite{devlin2018bert,lan2019albert,liu2019roberta,DBLP:journals/air/AcheampongNC21}. Despite the different sources of
  embedding vectors, this approach is common to all methods we analyze,
  ensuring a fair comparison. However, any additional source of
  background knowledge is excluded from our analysis, such as an
  external dataset, or a knowledge-base that could be used for
  transfer learning~\cite{Zhao2019autoem,Kasai2019acl}.
  \item \textit{Scope:} We exclusively consider RL, 
   excluding methods for multi-source ER~\cite{DBLP:journals/pvldb/PeetersB21}
  and for entity alignment \cite{zhang2022benchmark,DBLP:journals/pvldb/LeoneHAGW22}.
  \item \textit{Guidelines:} We exclude open-source algorithms that have publicly released their implementation, but provide neither instructions nor examples on using it (despite contacting their authors).
\end{enumerate}

Due to the first criterion, we could not include well-known techniques like 
\textit{Seq2SeqMatcher}~\cite{Nie2019cikm},
\textit{GraphER}~\cite{Li2020aaai},
\textit{CorDEL}~\cite{DBLP:conf/icdm/WangSWDJ20}, \textit{EmbDI}~\cite{DBLP:conf/sigmod/CappuzzoPT20} (despite contacting its authors) and \textit{Leva} \cite{DBLP:conf/sigmod/ZhaoF22}. The second criterion
excludes DL-based methods that aim to reduce the size of the training
set through transfer and active learning approaches, such as
\textit{Auto-EM}~\cite{Zhao2019autoem},
\textit{DeepMatcher+}~\cite{Kasai2019acl},
\textit{DIAL}~\cite{Jain2021vldb} and
\textit{DADER}~\cite{Tu2022sigmod}. 
The third criterion leaves out methods on tasks other than
matching, like \textit{Name2Vec}~\cite{Foxcroft2019name2vec} 
and \textit{Auto-ML}~\cite{Paganelli2021edbt}, methods crafted for multi-source ER like \textit{JointBERT}~\cite{DBLP:journals/pvldb/PeetersB21}, as well as all DL-based methods targeting the entity alignment problem~\cite{DBLP:journals/aiopen/ZengLHLF21,zhang2022benchmark} (these require non-trivial adaptations for matching). The fourth criterion prevented us from including \textit{MCAN} \cite{DBLP:conf/www/ZhangNWST20} and \textit{HIF-KAT} \cite{DBLP:conf/acl/YaoLDLY0LZD20}, as we could not run their code without guidelines. Finally, we exclude \textit{DeepER}~\cite{DBLP:journals/pvldb/EbraheemTJOT18}, since it is subsumed by DeepMatcher~\cite{Mudgal2018sigmod}, as explained below.

\textbf{Taxonomy.} To facilitate a better
understanding of the DL-based matching algorithms, we propose a new taxonomy that is formed by the following three dimensions:

\begin{enumerate}[leftmargin=*]
\item \textit{Language model type:} We distinguish methods as
  being \textit{static} or \textit{dynamic}. The former leverage
  pre-trained embedding techniques that associate every token with
  the same embedding vector, regardless of its context. 
  Methods such as word2vec \cite{mikolov2013efficient, mikolov2013distributed}, Glove \cite{pennington2014glove} and
  fastText \cite{DBLP:journals/tacl/BojanowskiGJM17} fall into this category. The opposite is true for dynamic methods, which leverage BERT-based language models~\cite{devlin2018bert,lan2019albert,liu2019roberta,DBLP:journals/air/AcheampongNC21}
  that generate context-aware embedding vectors. 
  Based on the context of every token,  they support polysemy, where the same word has different
  meanings (e.g., `bank' as an institution and `bank' as the edge
  of a river) as well as synonymy, where different words have
  identical or similar meanings (e.g., `job' and `profession').
\item \textit{Schema awareness:} We distinguish methods as being
  \textit{homogeneous} and \textit{heterogeneous}. In the RL settings we are considering,
  the former require that both input databases have the same 
  or at least aligned schemata, unlike methods in the latter category.
\item \textit{Entity similarity context.} We distinguish methods as
  being \textit{local} and \textit{global}. The former receive as
  input the textual description of two entities, and based on their
  encoding and the ensuing similarity, they decide whether they are
  matching or not, using a binary classifier. Global methods, on
  the other hand, leverage contextual information, which goes beyond
  the textual representation of a pair of records and their respective
  embedding vectors. For example, contextual information can leverage
  knowledge from the entire input datasets (e.g., overall term
  salience), or from the relation between candidate pairs.
\end{enumerate}

\begin{table}[t]\centering
\renewcommand{\tabcolsep}{2.8pt}
{\small
\caption{Taxonomy of the selected DL-based ER methods.}
 \begin{tabular}{ | l | c | c | c |} \cline{1-4}
 \multicolumn{1}{|c|}{DL-based} & Token embedding & Schema    & Entity similarity \\
 \multicolumn{1}{|c|}{algorithm} & context         & awareness & context \\
 \hline \hline
 DeepMatcher & Static & Homogeneous & Local \\
 EMTransformer & Dynamic & Heterogeneous & Local \\
 GNEM & Static, Dynamic & Homogeneous & Global \\
 HierMatcher & Dynamic & Heterogeneous & Local \\
 DITTO & Dynamic & Heterogeneous & Local \\
 \hline
 \end{tabular}
 \label{tb:taxonomy}
 }
\end{table}

Regarding the first dimension, we should stress that the dynamic
approaches that leverage transformer language models cast matching as a
\textit{sequence-pair classification problem}. All QID attribute
values in a record are concatenated into a single string
representation called \textit{sequence}. Then, every candidate
pair is converted into the following string representation that
forms the input to the neural classifier:
\texttt{``$[CLS]$ Sequence 1 $[SEP]$ Sequence 2 $[SEP]$"}, where 
$[CLS]$ and $[SEP]$ are special tokens that designate the beginning
of a new candidate pair and the end of each entity description,
respectively. In practice, every input should involve up to 512
tokens, which is the maximal attention span of transformer
models~\cite{DBLP:conf/edbt/BrunnerS20}.
These methods also require fine-tuning on a task-specific training set, while
their generated vectors are much larger than those generated by static embeddings (768 versus 300 dimensions \cite{DBLP:journals/pvldb/PaganelliBBG22,DBLP:journals/pvldb/EbraheemTJOT18}).

Based on these three dimensions, Table~\ref{tb:taxonomy} shows
that the considered DL-based matching algorithms cover \textit{all} types defined by our taxonomy, providing a representative sample of the field.

\textbf{Methods Overview.}
We now describe the five DL-based methods satisfying our selection criteria in chronological order.

\textit{DeepMatcher}~\cite{Mudgal2018sigmod}\footnote{\url{https://github.com/anhaidgroup/deepmatcher}}
(Jun. 2018) proposes a framework for DL-based matching algorithms that generalizes \textit{DeepER} \cite{DBLP:journals/pvldb/EbraheemTJOT18}, the first such
algorithm in the literature (which does not conform to the first selection criterion). DeepMatcher's architectural template contains three main modules: (1) The
attribute embedding module converts every word of an attribute value into a
\underline{static} embedding vector using an existing pre-trained model, such as fastText~\cite{DBLP:journals/tacl/BojanowskiGJM17}. (2) The attribute similarity vector module operates in a \underline{homogeneous} way that summarizes the sequence of token embeddings in each attribute and then obtains a similarity vector between every pair of candidate records (\underline{local} 
functionality). DeepMatcher provides four different solutions for summarizing
attributes, including a smooth inverse frequency method, a sequence aware RNN
method, a sequence alignment attention model, and a hybrid model. (3)
The classification module employs a two-layer fully connected ReLU HighwayNet \cite{DBLP:conf/icml/ZillySKS17}, followed by a softmax layer for classification. 

\textit{EMTransformer} \cite{DBLP:conf/edbt/BrunnerS20}\footnote{\url{https://github.com/brunnurs/entity-matching-transformer}} (Mar. 2020) employs \underline{dynamic} token embeddings using
attention-based transformer models like BERT~\cite{DBLP:conf/naacl/DevlinCLT19},
XLNet~\cite{DBLP:conf/nips/YangDYCSL19},
RoBERTa~\cite{DBLP:journals/corr/abs-1907-11692}, or
DistilBERT~\cite{DBLP:journals/corr/abs-1910-01108}. These models are applied in
an out-of-the-box manner, because no task-specific architecture is developed. In
each case, the smallest pre-trained model is used in order to ensure low run-times
even on commodity hardware. 
To handle noise, especially in the form of misplaced attribute values (e.g., name associated with profession),
it leverages a schema-agnostic setting that concatenates all attribute values per entity (\underline{heterogeneous} approach). 
EMTransformer processes every pair of records independently of all others through
a \underline{local} operation. 

\textit{GNEM} \cite{DBLP:conf/www/ChenSZ20}\footnote{\url{https://github.com/ChenRunjin/GNEM}} (Apr. 2020) is a \underline{global} approach that considers the
relations between all candidate pairs that are formed after blocking.
At its core lies a graph, where every node corresponds to a candidate record pair
and nodes with at least one common record are connected with weighted edges that
have weights proportional to their similarity. Due to blocking, the order of this
graph is significantly lower than the Cartesian product of the input datasets.
Pre-trained embeddings, either \underline{static} like fastText \cite{DBLP:journals/tacl/BojanowskiGJM17} or
\underline{dynamic} like the BERT-based ones~\cite{DBLP:journals/air/AcheampongNC21}, are used to represent the record
pair in every node. The edge weights consider the semantic difference between the
representation of the different records. To simultaneously calculate the matching
likelihood between all candidate pairs, the interaction module, which relies on a
gated graph convolutional network~\cite{DBLP:conf/aaai/ChenLLLZS20}, propagates (dis)similarity 
signals
between all nodes of the graph. GNEM can be considered as a generalization of 
DeepMatcher and EMTransformer, as it can leverage the pair representations of these local methods. GNEM assumes that all input records are described by the same schema,
therefore involving a \underline{homogeneous} operation.

\textit{DITTO} \cite{DBLP:journals/pvldb/0001LSDT20,DBLP:journals/jdiq/LiLSWHT21}\footnote{\url{https://github.com/megagonlabs/ditto}} (Sep. 2020) extends
EMTransformer's straightforward application of \underline{dynamic} BERT-based
models in three ways: (1) It incorporates domain knowledge while serializing
records. It uses a named entity recognition model to identify entity types
(such as persons or dates), as well as regular expressions for identities (like
product ids). Both are surrounded with special tokens: $[LAST]$...$[/LAST]$.
DITTO also allows users to normalize values such as numbers and to expand
abbreviations using a dictionary. (2) Its \underline{heterogeneous}
functionality summarizes long attribute values that exceed the 512-tokens
limit of BERT by keeping only the tokens that do not correspond to
stop-words and have a high TF-IDF weight. (3) DITTO uses data augmentation
to artificially produce additional training instances. It does this by
randomly applying five different operators that delete or shuffle parts
of the tokens or attribute values in a serialized pair of records, or
swap the two records. The original and the distorted representation are then
interpolated to ensure that the same label applies to both of them. 
The resulting classifier processes
every candidate record pair independently of the others in a
\underline{local} manner.

\textit{HierMatcher} \cite{DBLP:conf/ijcai/FuHHS20}\footnote{\url{https://github.com/casnlu/EntityMatcher}} (Jan. 2021) constructs a hierarchical neural network
with four layers that inherently addresses schema heterogeneity and noisy,
misplaced attribute values. (1) The \textit{representation layer} tokenizes
each attribute value, converts every token into its pre-trained
fastText~\cite{DBLP:journals/tacl/BojanowskiGJM17} embedding, and associates
it with its contextual vector through a bi-directional gated recurrent unit (GRU) \cite{Goodfellow-et-al-2016}. This means that HierMatcher is a
\underline{dynamic} approach, even though it does not leverage transformer
language models. (2) The \textit{token matching layer} maps every token of
one record to the most similar token from the other record in a pair, regardless of the respective attributes. The cross-attribute token alignment performed by
this layer turns HierMatcher into a \underline{heterogeneous} approach. (3)
The \textit{attribute matching layer} applies an attribute-aware mechanism to
adjust the contribution of every token in an attribute value according to its
importance, which minimizes the impact of redundant or noisy words. (4) The
\textit{entity matching layer} builds a comparison vector by concatenating the
outcomes of the previous layer and produces the matching likelihood for the given
pair of records. This means that HierMatcher disregards information from other
candidate pairs, operating in a~\underline{local} entity similarity context.

\subsection{Non-neural, Non-linear ML-based Methods}
\label{sec:mlMethods}

Two ML-based methods are typically used in the literature, both of which
satisfy the selection criteria in Section \ref{sec:dlMatchers}.

\textit{Magellan}~\cite{DBLP:journals/pvldb/KondaDCDABLPZNP16}\footnote{\url{https://github.com/anhaidgroup/py_entitymatching}} (Sep. 2016)
combines traditional ML classifiers with a set of automatically extracted features that are
based on similarity functions. This means that every candidate pair is
represented as a numerical feature vector, where every dimension
corresponds to the score of a particular similarity function for a
specific attribute. Magellan 
implements the following established functions: affine, Hamming and edit distances, as well as Jaro, Jaro-Winkler, Needleman-Wunsch, Smith-Waterman, Jaccard, Monge Elkan, Dice, Cosine and overlap similarity~\cite{Christen2012springer}. The resulting
feature vectors can be used as training data (assuming ground
truth data is available) on several state-of-the-art classifiers,
namely random forest, logistic regression, support vector machines,
and decision trees~\cite{DBLP:journals/pvldb/KondaDCDABLPZNP16}.

\textit{ZeroER}~\cite{wu2020sigmod}\footnote{\url{https://github.com/chu-data-lab/zeroer}} (Jun. 2020) is an unsupervised method that does not require any training data. It extends the expectation-maximisation approach~\cite{Herzog2007springer}
by employing Gaussian mixture models to capture the distributions
of matches and non-matches and by taking into consideration
dependencies between different features. ZeroER 
applies feature regularisation by considering the overlap between
the match and non-match distributions for each feature, to prevent
certain features from dominating the classification outcomes. 
It also incorporates the transitive
closure~\cite{Christen2012springer} into the generative model as a
constraint through a probabilistic inequality. 
For feature generation, ZeroER uses
Magellan~\cite{DBLP:journals/pvldb/KondaDCDABLPZNP16}. 

\subsection{Non-neural, Linear Supervised Methods}
\label{sec:newSupervisedMethods}

We now present our new linear matching algorithms for efficiently assessing the difficulty of a benchmark dataset. They are inspired from the degree of linearity, but unlike Algorithm \ref{algo:linearityDegree}, they do not report a characteristic of the dataset based on the tokens in the entity profiles of all labeled instances. They use only the training and validation sets for their learning, while being more flexible, using character n-grams and language models for the similarity estimation between each pair of profiles (not just tokens). Their performance can be measured in terms of effectiveness, time and space complexity, thus being suitable for a holistic comparison with the matching algorithms proposed in the literature. 

Algorithm~\ref{algo:supervisedMethods} details our methods. For each pair of records in the training set $T$, $c_{i,j}=(r_i, r_j)$, it extracts the features from their QID attributes, e.g., by tokenizing all the  values on whitespaces (lines 2 and 3). Using the individual features, it calculates and stores the feature vector $f_{i,j}$ for each candidate pair $(r_i,r_j)$ in lines 4 and 5. For example, $f_{i,j}$ could be formed by comparing the tokens of the individual features with similarity measures like Cosine, Jaccard and Dice. Note that all dimensions in $f_{i,j}$ are defined in $[0,1]$. 

Next, the algorithm identifies the threshold that achieves the highest F-measure per feature, when applied to the instances of the training set. Line 6 loops over all thresholds in $[0.01, 0.99]$ with a step increment of 0.01,  and lines 7 to 11 generate a set of matching and non-matching pairs for each individual feature $f$. Using these sets, it estimates the maximum F-measure per feature $f$, $F1_{max}^T[f]$, along with the corresponding threshold (lines 12 to 14).

The resulting configurations are then applied in lines 15 to 24 to the validation set in order to identify the overall best feature and the respective threshold. For each candidate pair in $V$ (line 16), the algorithm extracts the individual features and the corresponding feature vector (lines 17 and 18), using the same functions that were applied to the training instances in lines 3 and 4. Rather than storing the feature vectors, it directly applies the best threshold per feature to distinguish the matching from the non-matching pairs of each dimension (lines 19 to 21). This lays the ground for estimating $F1$ per feature and the maximum $F1$ over all features in lines 22 to 24. 

Finally, the algorithm applies the identified best feature and threshold to the testing instances in order to estimate the overall $F1$ in lines 25 to 29. For each candidate pair in $C$, it estimates only the value of the feature with the highest performance over the validation set (lines 26 to 27). Using the respective threshold, it distinguishes the matching from the non-matching pairs (lines 29 and 30) and returns the corresponding $F1$ in line 31.

\begin{table*}[t]\centering
\renewcommand{\tabcolsep}{4.2pt}
\caption{{
The established datasets for DL-based matching algorithms. $|D_x|$ is the number of records, $|A|$ the number of attributes, and $|I_x|$, $|P_x|$, $|N_x|$ are the numbers of labelled, positive and negative instances in the training ($x$=tr) or the testing~($x$=te)~set. IR is the class imbalance~ratio.
}}
	\begin{tabular}{ | l | c |  c | r | r | c || r | r | r | r | r | r | r || l |}
		\cline{1-14}
		 \multicolumn{1}{|c|}{} & 
		 $D_1$ & 
		 $D_2$ & 
         \multicolumn{1}{c|}{$|D_1|$} & 
         \multicolumn{1}{c|}{$|D_2|$} & 
         $|A|$ &
         \multicolumn{1}{c|}{$|I_{tr}|$} & 
         \multicolumn{1}{c|}{$|I_{te}|$} & 
         \multicolumn{1}{c|}{$|P_{tr}|$} &
         \multicolumn{1}{c|}{$|P_{te}|$} & 
         \multicolumn{1}{c|}{$|N_{tr}|$} & 
         \multicolumn{1}{c|}{$|N_{te}|$} &
         \multicolumn{1}{c||}{IR} & 
         References \\
        \hline \hline
        \multicolumn{14}{|c|}{(a) Structured datasets}\\
        \hline
        $D_{s1}$ & DBLP & ACM & 2,616 & 2,294 & 4 & 7,417 & 2,473 & 1,332 & 444 & 6,085 & 2,029 & 18.0\% & \cite{DBLP:conf/ijcai/FuHHS20,DBLP:journals/pvldb/0001LSDT20,Mudgal2018sigmod}\\
        $D_{s2}$ & DBLP & Google Scholar & 2,616 & 64,263 & 4 & 17,223 & 5,742 & 3,207 & 1,070 & 14,016 & 4,672 & 18.6\% & \cite{DBLP:conf/ijcai/FuHHS20,DBLP:journals/pvldb/0001LSDT20,Mudgal2018sigmod}\\
        $D_{s3}$ & iTunes & Amazon & 6,907 & 55,923 & 8 & 321 & 109 & 78 & 27 & 243 & 82 & 24.3\% & \cite{DBLP:journals/pvldb/0001LSDT20,Mudgal2018sigmod}\\
        $D_{s4}$ & Walmart & Amazon & 2,554 & 22,074 & 5 & 6,144 & 2,049 & 576 & 193 & 5,568 & 1,856 & 9.4\% & \cite{DBLP:conf/ijcai/FuHHS20,DBLP:conf/www/ChenSZ20,DBLP:journals/pvldb/0001LSDT20,Mudgal2018sigmod}\\
        $D_{s5}$ & BeerAdvo & RateBeer & 4,345 & 3,000 & 4 & 268 & 91 & 40 & 14 & 228 & 77 & 14.9\% & \cite{DBLP:journals/pvldb/0001LSDT20,Mudgal2018sigmod}\\
        $D_{s6}$ & Amazon & Google Products & 1,363 & 3,226 & 3 & 6,874 & 2,293 & 699 & 234 & 6,175 & 2,059 & 10.2\% & \cite{DBLP:conf/ijcai/FuHHS20,DBLP:conf/www/ChenSZ20,DBLP:journals/pvldb/0001LSDT20,Mudgal2018sigmod}\\
        $D_{s7}$ & Fodors & Zagat & 533 & 331 & 6 & 567 & 189 & 66 & 22 & 501 & 167 & 11.6\% & \cite{DBLP:journals/pvldb/0001LSDT20,Mudgal2018sigmod}\\
        \hline
        \multicolumn{14}{|c|}{(b) Dirty datasets}\\
        \hline
        $D_{d1}$ & DBLP & ACM & 2,616 & 2,294 & 4 & 7,417 & 2,473 & 1,332 & 444 & 6,085 & 2,029 & 18.0\% & \cite{DBLP:conf/ijcai/FuHHS20,DBLP:conf/edbt/BrunnerS20,DBLP:journals/pvldb/0001LSDT20,Mudgal2018sigmod}\\
        $D_{d2}$ & DBLP & Google Scholar & 2,616 & 64,263 & 4 & 17,223 & 5,742 & 3,207 & 1,070 & 14,016 & 4,672 & 18.6\% & \cite{DBLP:conf/ijcai/FuHHS20,DBLP:conf/edbt/BrunnerS20,DBLP:journals/pvldb/0001LSDT20,Mudgal2018sigmod}\\
        $D_{d3}$ & iTunes & Amazon & 6,907 & 55,923 & 8 & 321 & 109 & 78 & 27 & 243 & 82 & 24.3\% & \cite{DBLP:conf/edbt/BrunnerS20,DBLP:journals/pvldb/0001LSDT20,Mudgal2018sigmod}\\
        $D_{d4}$ & Walmart & Amazon & 2,554 & 22,074 & 5 & 6,144 & 2,049 & 576 & 193 & 5,568 & 1,856 & 9.4\% & \cite{DBLP:conf/ijcai/FuHHS20,DBLP:conf/edbt/BrunnerS20,DBLP:journals/pvldb/0001LSDT20,Mudgal2018sigmod}\\
	    \hline
	    \multicolumn{14}{|c|}{(c) Textual datasets}\\
	    \hline
        $D_{t1}$ & Abt & Buy & 1,081 & 1,092 & 3 & 5,743 & 1,916 & 616 & 206 & 5,127 & 1,710 & 10.7\% & \cite{DBLP:conf/www/ChenSZ20,DBLP:conf/edbt/BrunnerS20,DBLP:journals/pvldb/0001LSDT20,Mudgal2018sigmod} \\
        $D_{t2}$ & CompanyA & CompanyB & 28,200 & 28,200 & 1 & 67,596 & 22,503 & 16,859 & 5,640 & 50,737 & 16,863 & 24.9\% & \cite{DBLP:journals/pvldb/0001LSDT20,Mudgal2018sigmod}\\
        \hline
	\end{tabular}
	\label{tb:commonDatasets}
\end{table*}

We call this algorithm \textit{Efficient Supervised Difficulty Estimation} (\textbf{ESDE}). Its advantages are its linear space complexity, given that it stores one feature vector per training instance, as well as its linear time complexity: it goes through the training set 100 times, $|F|$ times through the validation set (where $|F|$ denotes the dimensionality of the feature vector and is practically constant in each dataset, as explained below), and once through the testing set. It is also versatile, accommodating diverse feature vectors like the following:

\begin{algorithm2e}[t]
\DontPrintSemicolon
\SetKw{KwBy}{by}
\footnotesize
\KwIn{Training $T$, validation $V$ and testing $C$ sets \& the set of features~$F$}
\KwOut{The F-Measure on the testing set $F1$}
$F_T \gets \{\}$, $F1^T_{max}[] \gets \{\}$, $t^T_{best}[] \gets \{\}$ \\
\ForEach(\tcp*[h]{Training phase}){$(r_i,r_j) \in T$}{
    $F_i \gets getFeatures(r_i)$, $F_j \gets getFeatures(r_j)$\\
    $f_{i,j} \gets getFeatureVector(F_i, F_j)$\\
    $F_T \gets F_T \cup (f_{i,j}, (r_i,r_j))$
}
\For{$t\gets 0.01$ \KwTo $0.99$ \KwBy $0.01$}{
    $M_F[] \gets \{\}$, $N_F[] \gets \{\}$ \tcp*{arrays with the list of matching ($M$) and non-matching ($N$) pairs per feature}
    \ForEach{$(f_{i,j}, (r_i,r_j)) \in F_T$}{
        \ForEach(\tcp*[h]{get category of pairs per feature}){$f \in f_{i,j}$}{
            \lIf(\tcp*[h]{A match}){$t \leq f$}{
                $M[f] \gets M[f] \cup \{(r_i,r_j)\}$
            } \lElse(\tcp*[h]{A non-match}) {
                $N[f] \gets N[f] \cup \{(r_i,r_j)\}$
            }
        }
    }
    \ForEach{$f \in F$}{
        \If{$F1^T_{max}[f] < F1(M[f], N[f], T)$}{
            $F1^T_{max}[f] \gets F1(M[f], N[f], T)$, $t^T_{best}[f] \gets t$
        }
    }
}
$F1^V_{max} \gets 0$, $f_{best} \gets null$, $t_{best} \gets 0$, $M_F[] \gets \{\}$, $N_F[] \gets \{\}$\\
\ForEach(\tcp*[h]{Validation phase}){$(r_i,r_j) \in V$}{
    $F_i \gets getFeatures(r_i)$, $F_j \gets getFeatures(r_j)$\\
    $f_{i,j} \gets getFeatureVector(F_i, F_j)$\\
    \ForEach{$f \in f_{i,j}$}{
        \lIf(\tcp*[h]{A match}){$t^T_{best}[f] \leq f$}{
            $M[f] \gets M[f] \cup \{(r_i,r_j) \}$
        } \lElse(\tcp*[h]{A non-match}) {
            $N[f] \gets N[f] \cup \{(r_i,r_j) \}$
        }
    }
}
\ForEach{$f \in F$}{
    \If{$F1^V_{max} < F1(M[f], N[f], V)$}{
        $F1^V_{max} \gets F1(M[f], N[f], V)$, $f_{best} \gets f$, $t_{best} \gets t^T_{best}[f]$
    }
}
$M \gets \{\}$, $N \gets \{\}$\\
\ForEach(\tcp*[h]{Testing phase}){$(r_i,r_j)\in C$}{
    $s_{best} \gets getFeature(r_i, r_j, f_{best})$\\
    \lIf(\tcp*[h]{A match}){$t_{best} \leq s_{best}$}{
        $M \gets M \cup \{ (r_i,r_j) \}$
    } \lElse(\tcp*[h]{A non-match}) {
        $N \gets N \cup \{ (r_i,r_j) \}$
    }
}
\textbf{return} $F1(M, N, C)$
\caption{Efficient Supervised Difficulty Estimation}
\label{algo:supervisedMethods}
\end{algorithm2e}

\begin{enumerate}[leftmargin=*]
    \item \textit{Schema-agnostic ESDE} (\textbf{SA-ESDE}). The function $getFeatures()$ represents every record $r_i$ by the set of its distinct tokens, i.e., $F_i = T_i$. Every candidate pair $c_{i, j}$ is represented by the vector $f_{i,j} = [CS_{i,j}, DS_{i,j}, JS_{i,j}]$, where $CS_{i,j}=\frac{|F_i \cap F_j|}{\sqrt{|F_i|\times |F_j|}}$ stands for the Cosine, $DS_{i,j}=\frac{2 \times|F_i \cap F_j|}{|F_i| + |F_j|}$ for the Dice, and $JS_{i,j}=\frac{|F_i \cap F_j|}{|F_i| + |F_j| - |F_i \cap F_j|}$ for the Jaccard similarity. Hence, $|F|=3$.
    \item \textit{Schema-based ESDE} (\textbf{SB-ESDE}) applies the functions of SA-ESDE to every attribute of the given records. In other words, it considers every attribute independently of the others, yielding a feature vector with the Cosine, Dice and Jaccard similarity of the tokens per attribute. As a result, $|F| = 3\times|A|$, where $|A|$ denotes the number of attributes in the input dataset.
    \item \textit{Schema-agnostic Q-gram-based ESDE} (\textbf{SAQ-ESDE}) alters SA-ESDE to work with character q-grams ($q \in [2,10]$) instead of tokens. This means that $F_i = [2$$-$$grams(r_i), 3$$-$$grams(r_i)$, 
    $...$, $10$$-$$grams(r_i)]$, where $x$$-$$grams(r_i)$ returns the set of $x$-grams from all attribute values of record $r_i$. The feature vector includes the Cosine, Dice and Jaccard similarity per $q$, such that~$|F|=30$.
    \item \textit{Schema-based Q-gram-based ESDE} (\textbf{SBQ-ESDE}). In a similar vein, this algorithm alters SB-ESDE so that it works with character q-grams ($2 \le q \le 10$) instead of tokens. Every attribute value is converted into its set of $q$-grams, which are then used to calculate the Cosine, Dice and Jaccard similarities of this attribute. $F_i = [2$$-$$grams(r_i, a_1),  ..., 10$$-$$grams(r_i, a_{|A|})]$, where $|A|$ is the number of attributes in the input dataset. Hence, the dimensionality of the resulting feature vector is $|F|=30\times|A|$.
    \item \textit{Schema-agnostic FastText ESDE} (\textbf{SAF-ESDE}). In this algorithm, the function $getFeatures()$ represents every record $r_i$ by the 300-dimensional pre-trained fastText embedding vector of the concatenation of all its attribute values, $v_i$. Every candidate pair $c_{i, j}$ is then represented by the feature vector $f_{i,j} = [CS_{i,j}, ES_{i,j},$ $WS_{i,j}]$, where $CS_{i,j}$ is the Cosine similarity of the vectors $v_i$ and $v_j$, $ES_{i,j}$ is their Euclidean similarity, which is defined as $ES_{i,j}$=1/(1+$ED(v_i, v_j)$, where $ED(v_i, v_j)$ stands for the Euclidean distance of the two vectors, and $WS_{i,j}$ is the Wasserstein similarity between $v_i$ and $v_j$, which is derived from the Wasserstein distance \cite{villani2009wasserstein} (a.k.a., earth mover’s distance) with the same equation combining $ES_{i,j}$ and $ED$. Hence, $|F|=3$.
    \item \textit{Schema-based FastText ESDE} (\textbf{SBF-ESDE}) applies the previous feature generation function to each attribute, hence the dimensionality of its feature vector is $|F| = 3\times|A|$, with $|A|$ standing for the number of attributes in the given dataset.
    \item \textit{Schema-agnostic S-GTR-T5 ESDE} (\textbf{SAS-ESDE}) has the same functionality as \textsf{SAF-ESDE}, but replaces the fastText embeddings with the 768-dimensional pre-trained S-GTR-T5 ones, one of the state-of-the-art SentenceBERT models~\cite{ni2021large}.
    \item \textit{Schema-based S-GTR-T5 ESDE} (\textbf{SBS-ESDE}) alters \textsf{SBF-ESDE} so that it uses S-GTR-T5 rather than fastText embeddings.
\end{enumerate}

\section{Analysis of Existing Benchmarks}
\label{sec:datasetAnalysis}

We now assess how challenging the datasets commonly used in the evaluation of DL-based matching
algorithms are. These are the datasets used in the
experimental evaluation of DeepMatcher and are 
available on its repository~\cite{DMdataRepository} (which
also provides details on their content and generation).
The characteristics of these datasets are shown
in Table~\ref{tb:commonDatasets}. Note that every dataset
is split into specific training, validation, and testing
sets, with the same ratio of~3:1:1.

The rightmost column in Table~\ref{tb:commonDatasets}
cites the methods from Section~\ref{sec:dlMatchers}
that use each dataset in their experimental study. In
short, DeepMatcher and DITTO have used all these datasets
in their experiments, followed by HierMatcher,
EMTransformer and GNEM, which use 7, 5 and 3 datasets,
respectively. No other dataset is shared by
at least two of these methods. Finally, it is worth
noting that the dirty datasets $D_{d1}$ to $D_{d4}$
were generated from the structured datasets $D_{s1}$ to
$D_{s4}$ by injecting artificial noise in the following
way: for each record, the value of every attribute except
``title'' was randomly assigned to its ``title'' with
50\% probability~\cite{DMdataRepository,Mudgal2018sigmod}.

\begin{figure*}[t]
\centering
\includegraphics[width=0.47\textwidth]{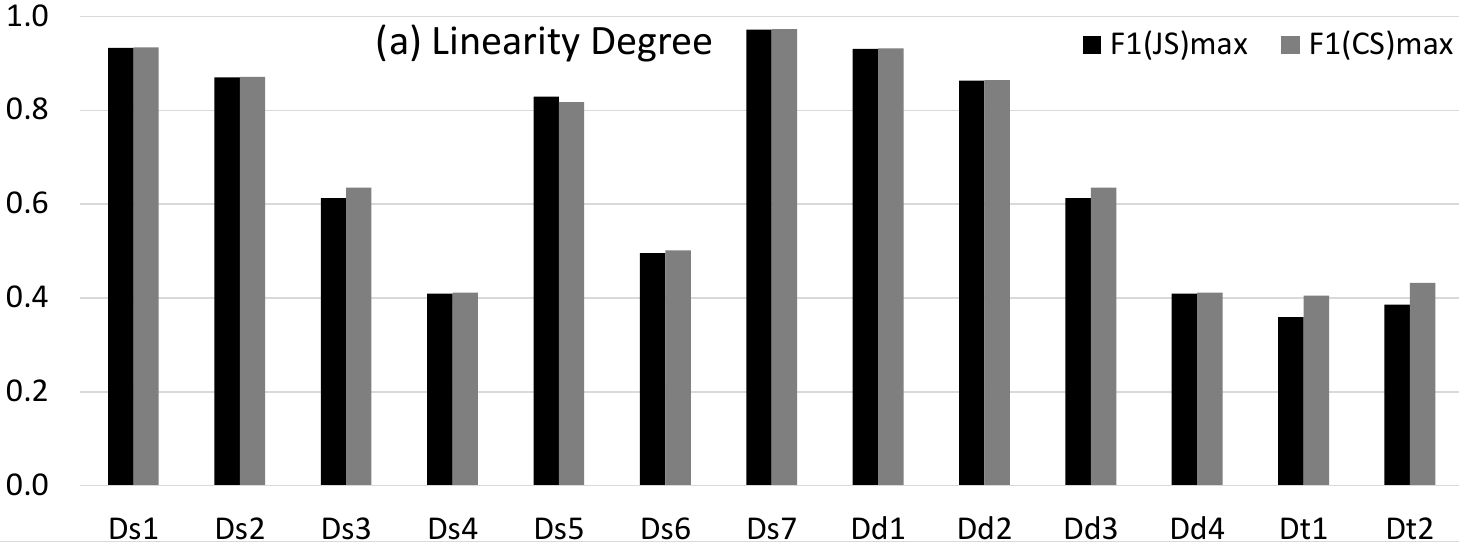}
\hspace{5pt}
\includegraphics[width=0.47\textwidth]{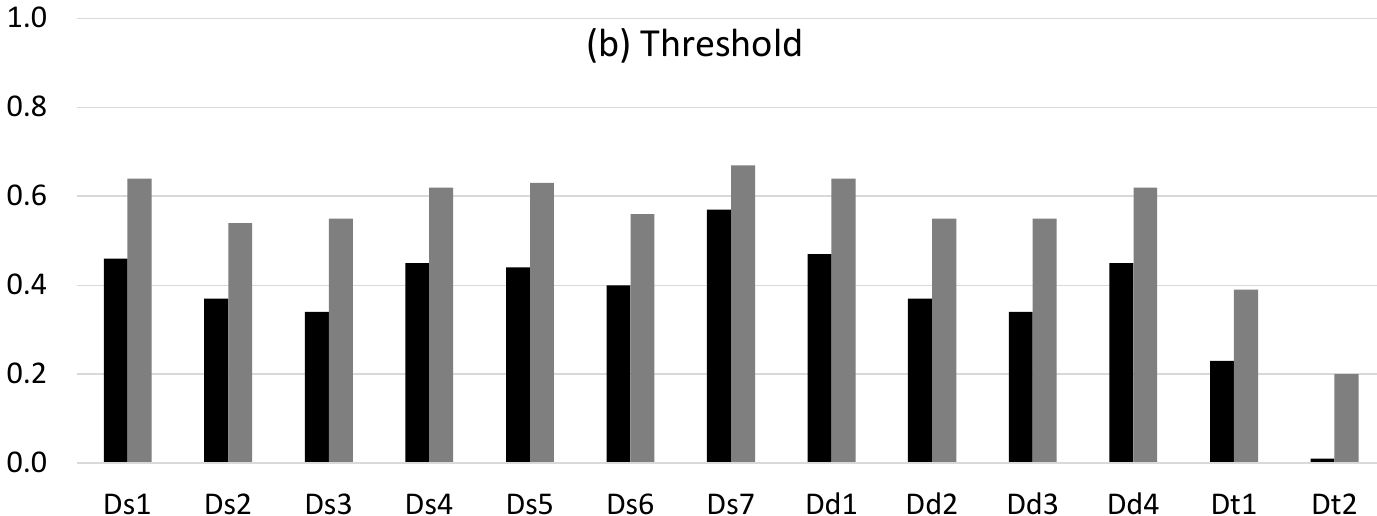}
\caption{ Degree of linearity per dataset in Table \ref{tb:commonDatasets} (left) and the respective threshold (right) with respect to Equations \ref{eq:cosine} and \ref{eq:jaccard}.}
\label{fig:theoreticalLnearity}
\end{figure*}

\subsection{Theoretical Measures}

\textbf{Degree of Linearity.}
The results for each benchmark dataset are shown in
Figure \ref{fig:theoreticalLnearity}.
We observe that the linearity of six datasets exceeds 0.8
(with three exceeding 0.9) for both similarity measures, which indicates
rather easy classification tasks, as the two classes can be
separated by a linear classifier with high accuracy. The
maximum values are obtained with $D_{s7}$, where indeed 
practically all DL-based algorithms achieve a perfect F-measure of F1=1.0 \cite{DBLP:conf/www/ZhangNWST20}. 

\begin{figure*}[t]
\centering
\includegraphics[width=0.24\textwidth]{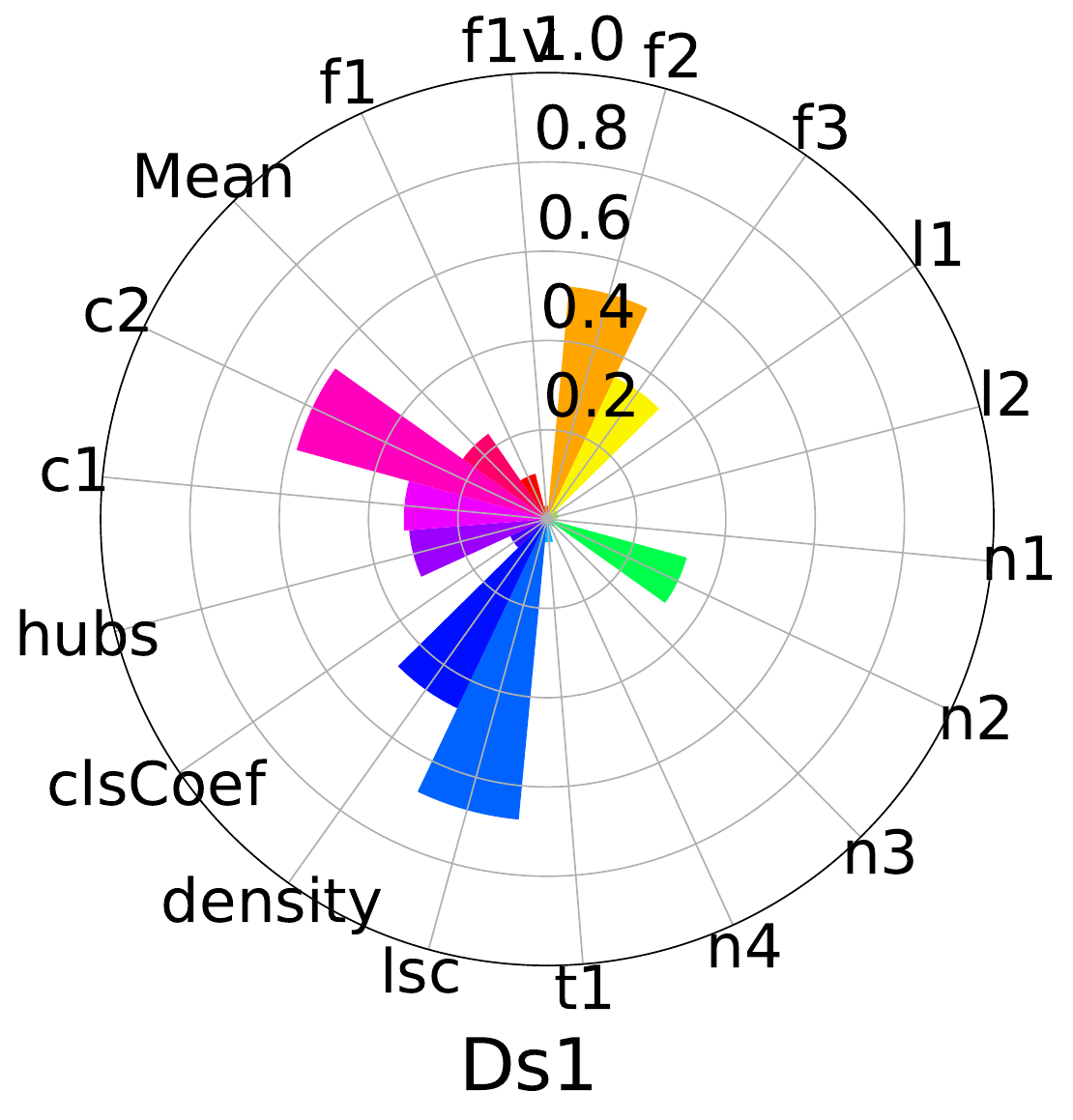}
\includegraphics[width=0.24\textwidth]{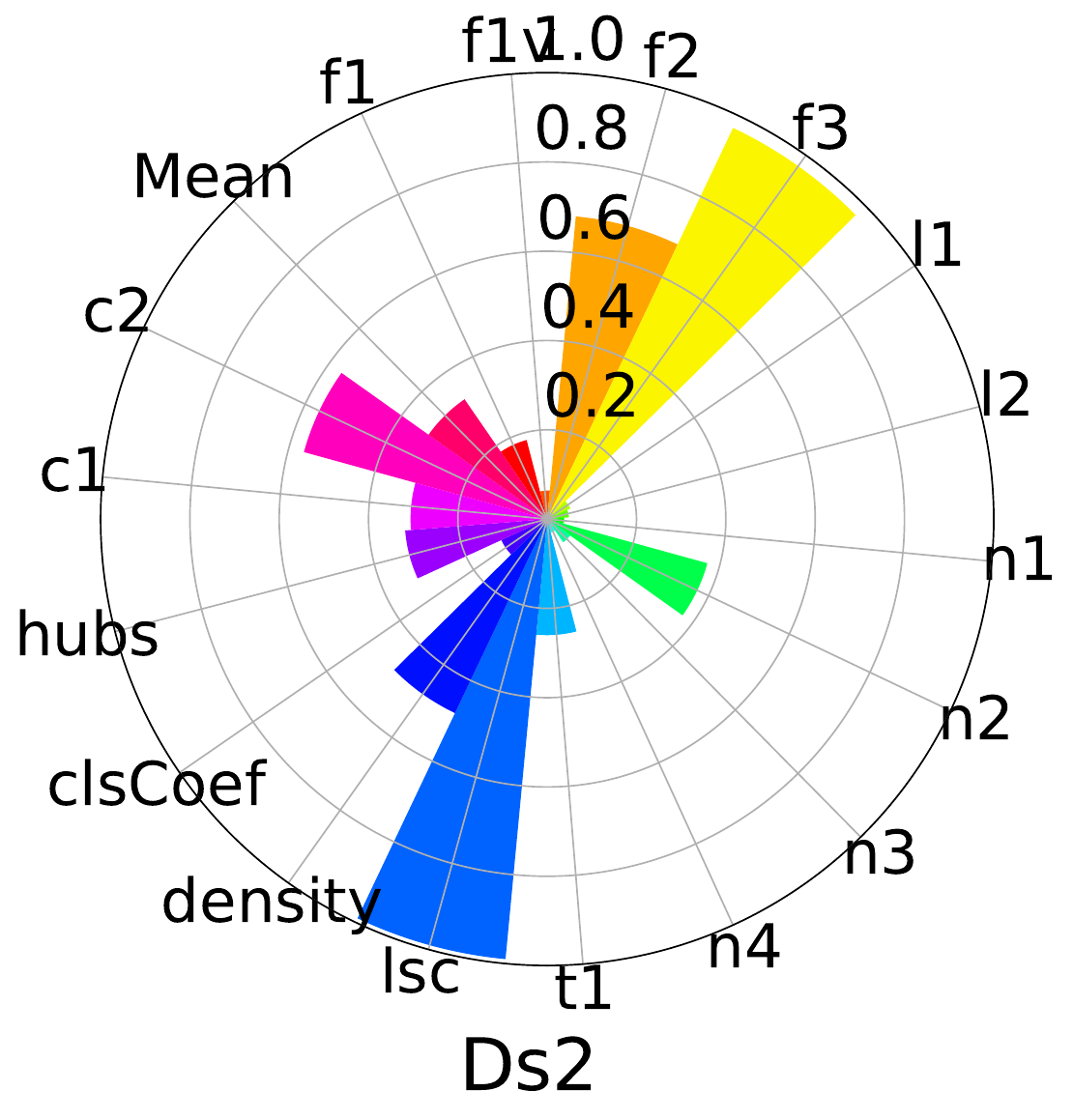}
\includegraphics[width=0.24\textwidth]{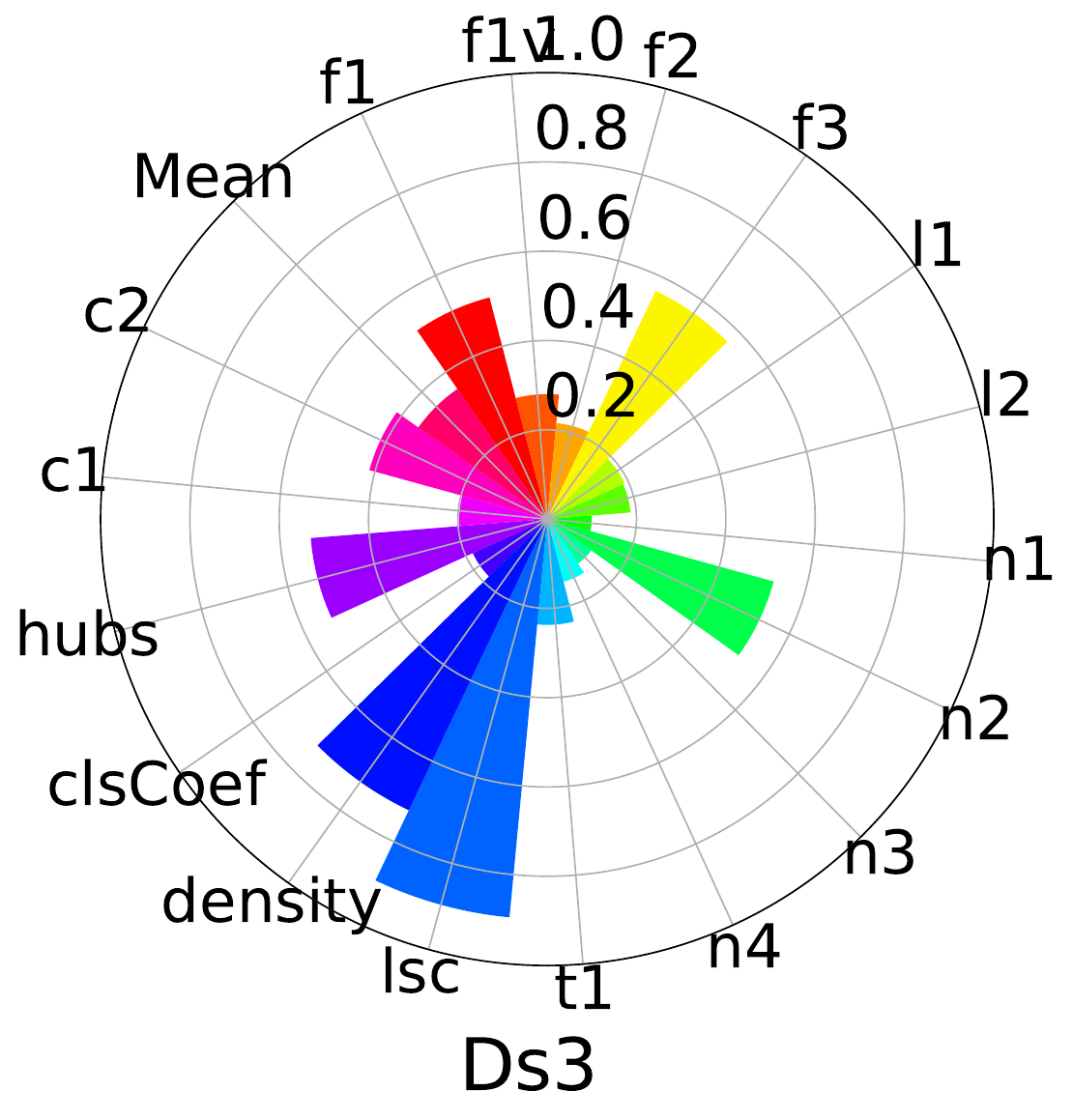}
\includegraphics[width=0.24\textwidth]{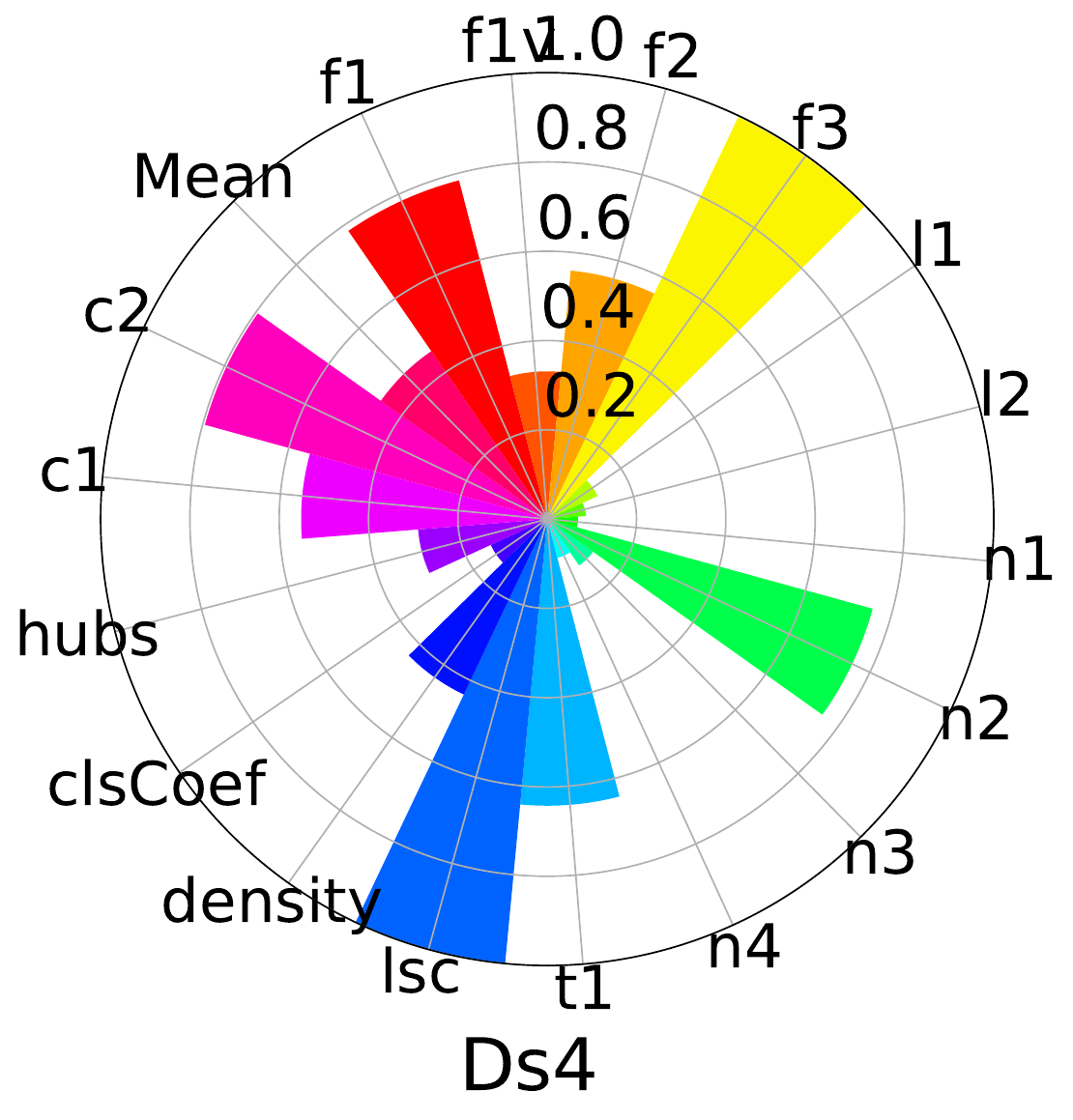}
\includegraphics[width=0.24\textwidth]{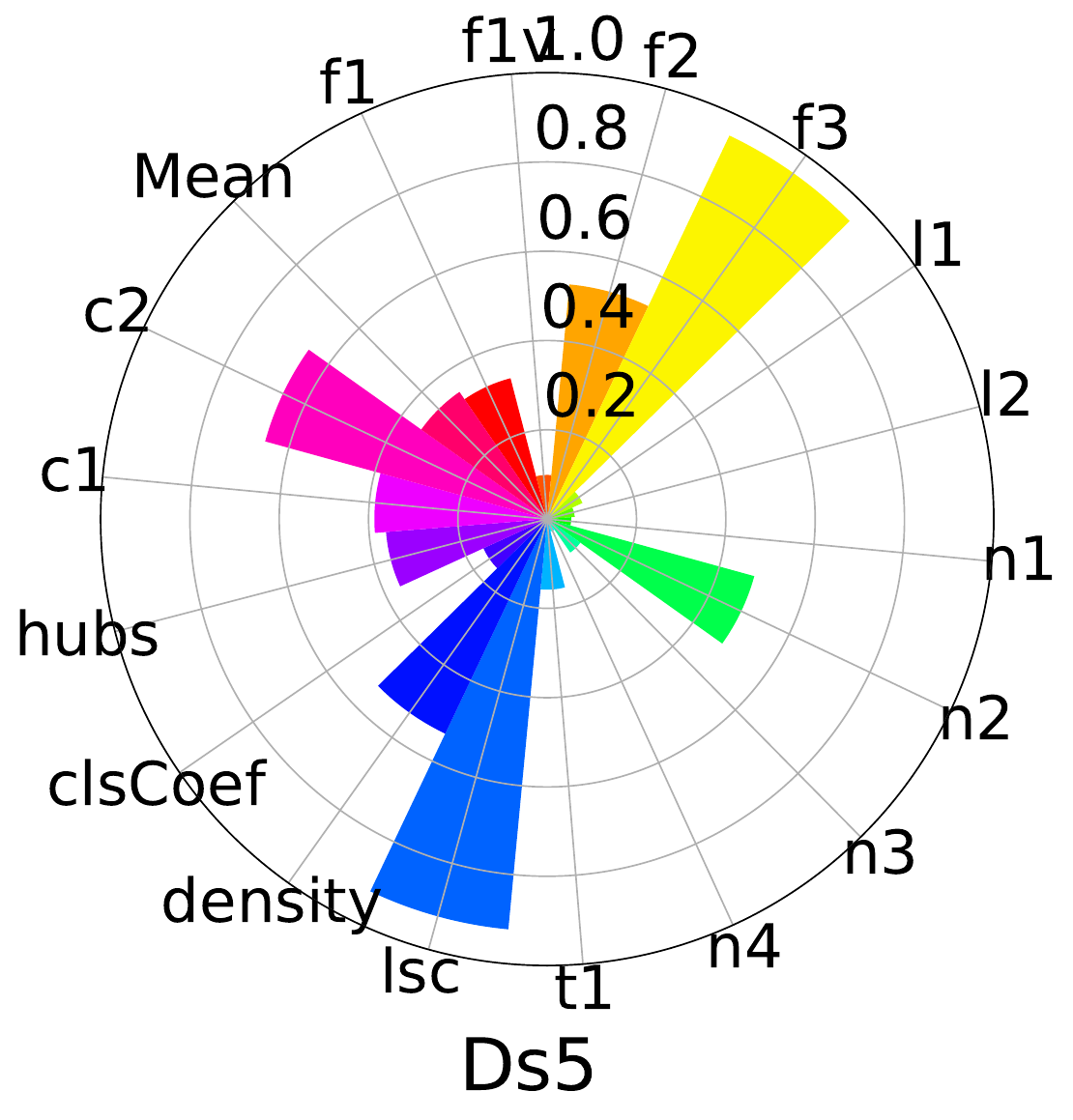}
\includegraphics[width=0.24\textwidth]{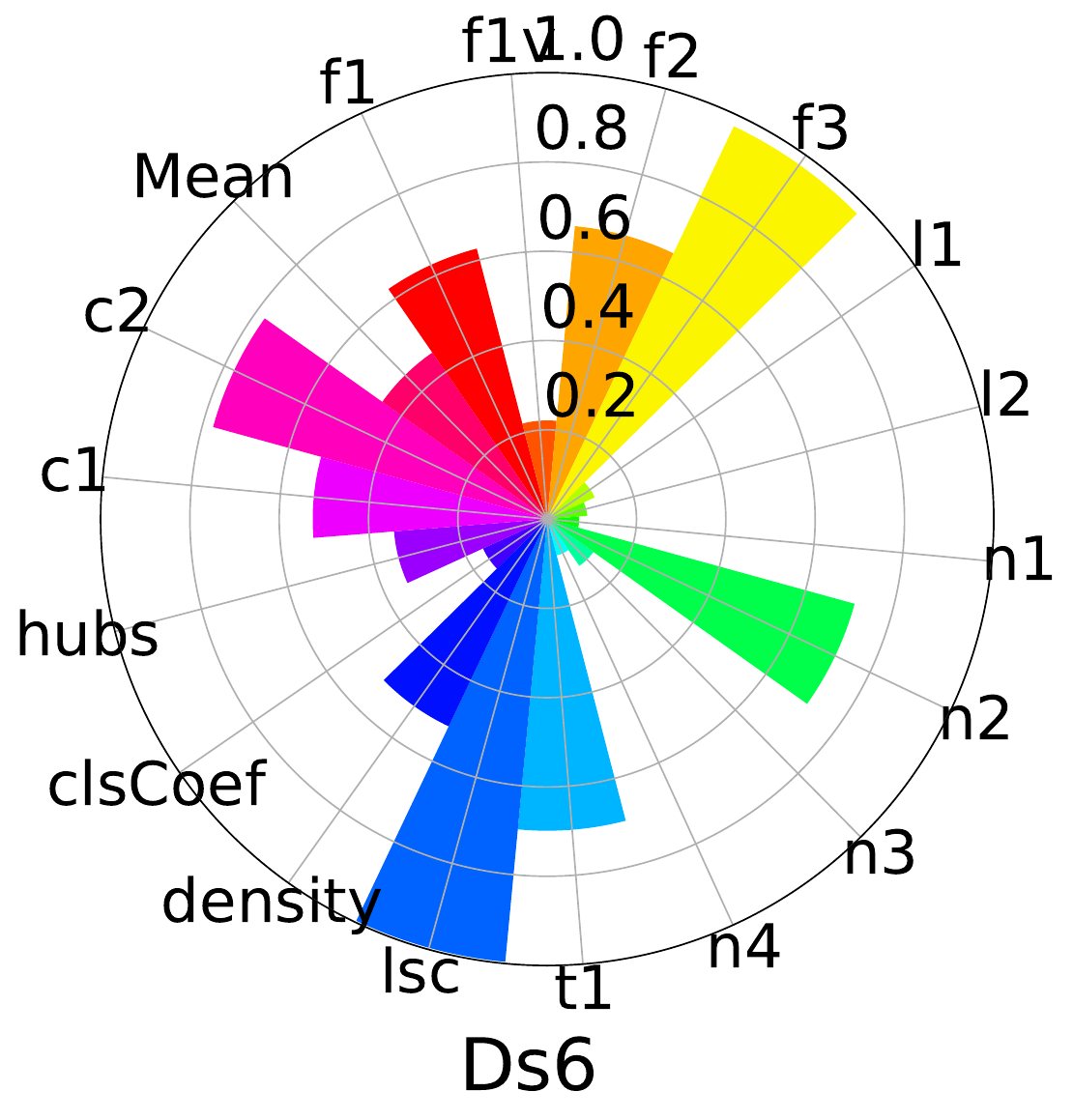}
\includegraphics[width=0.24\textwidth]{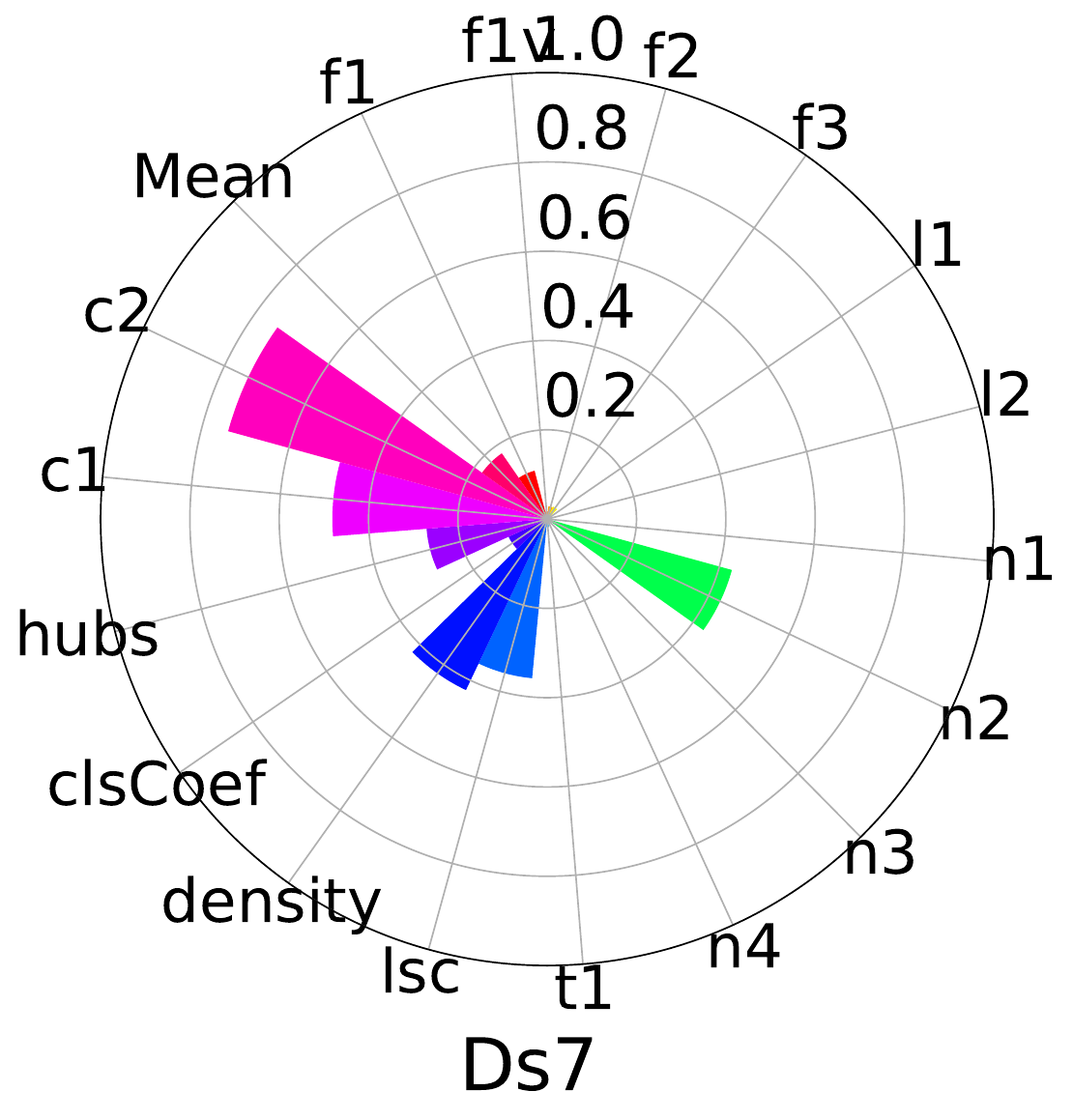}
\includegraphics[width=0.24\textwidth]{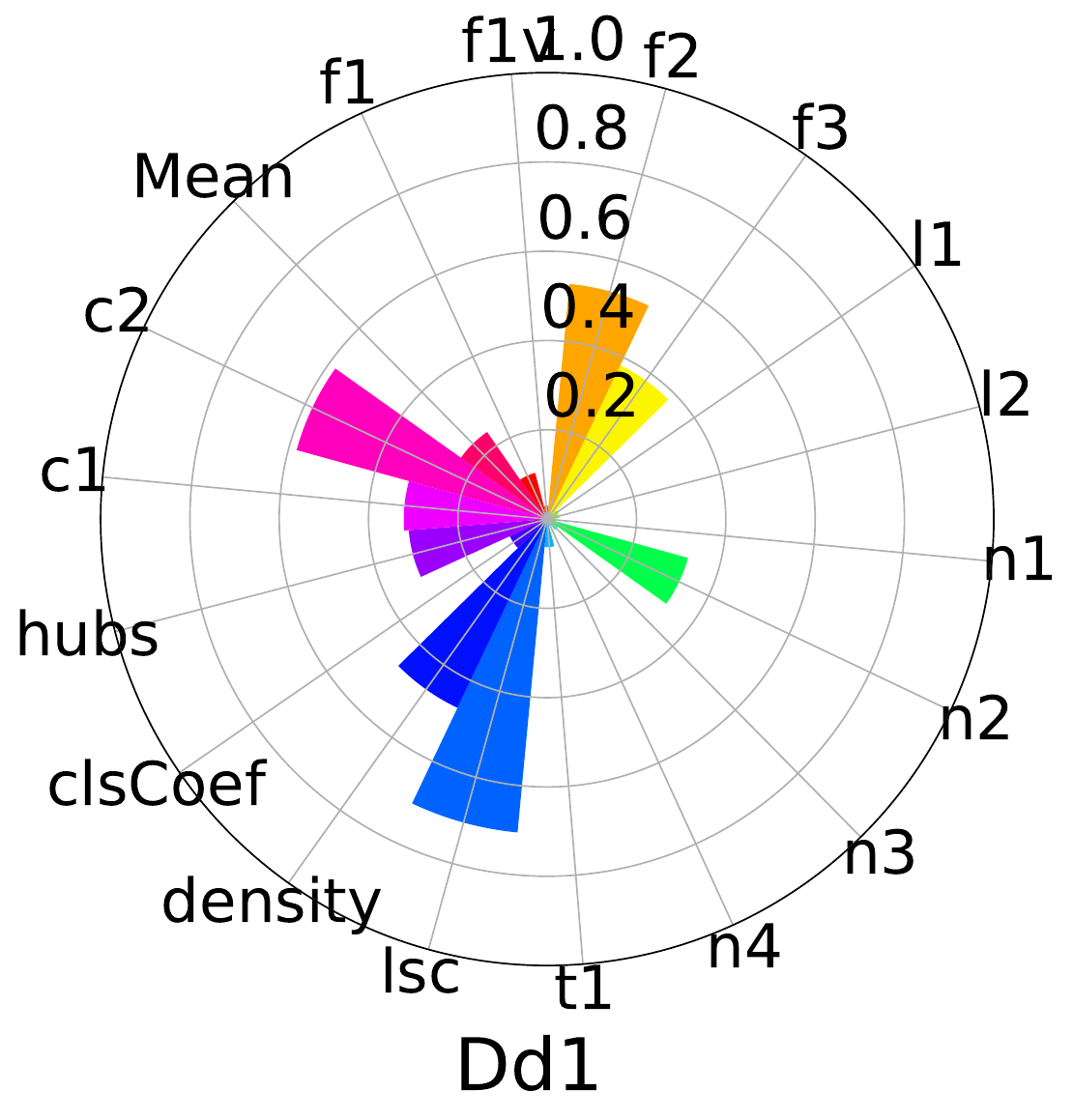}
\includegraphics[width=0.195\textwidth]{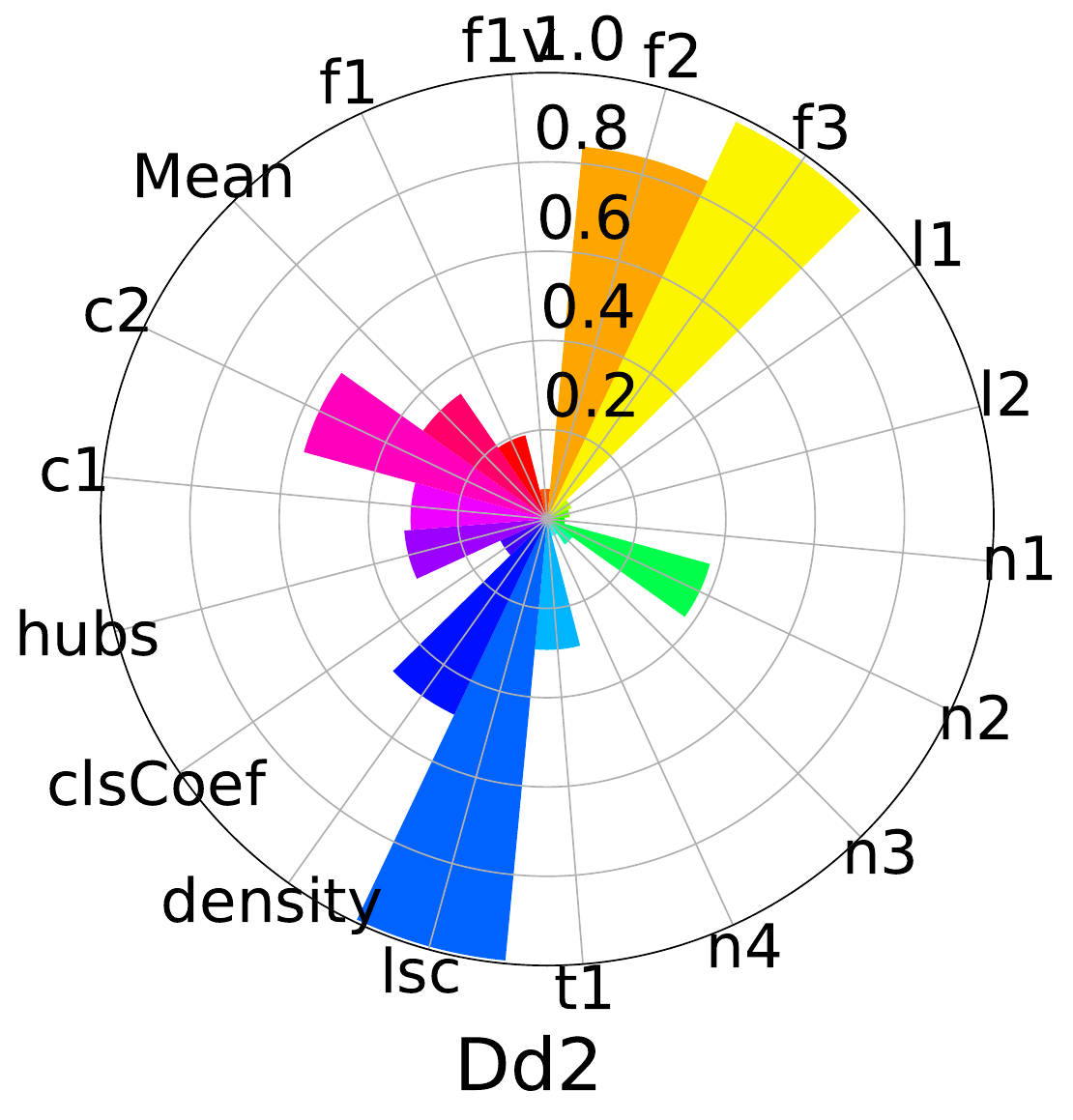}
\includegraphics[width=0.195\textwidth]{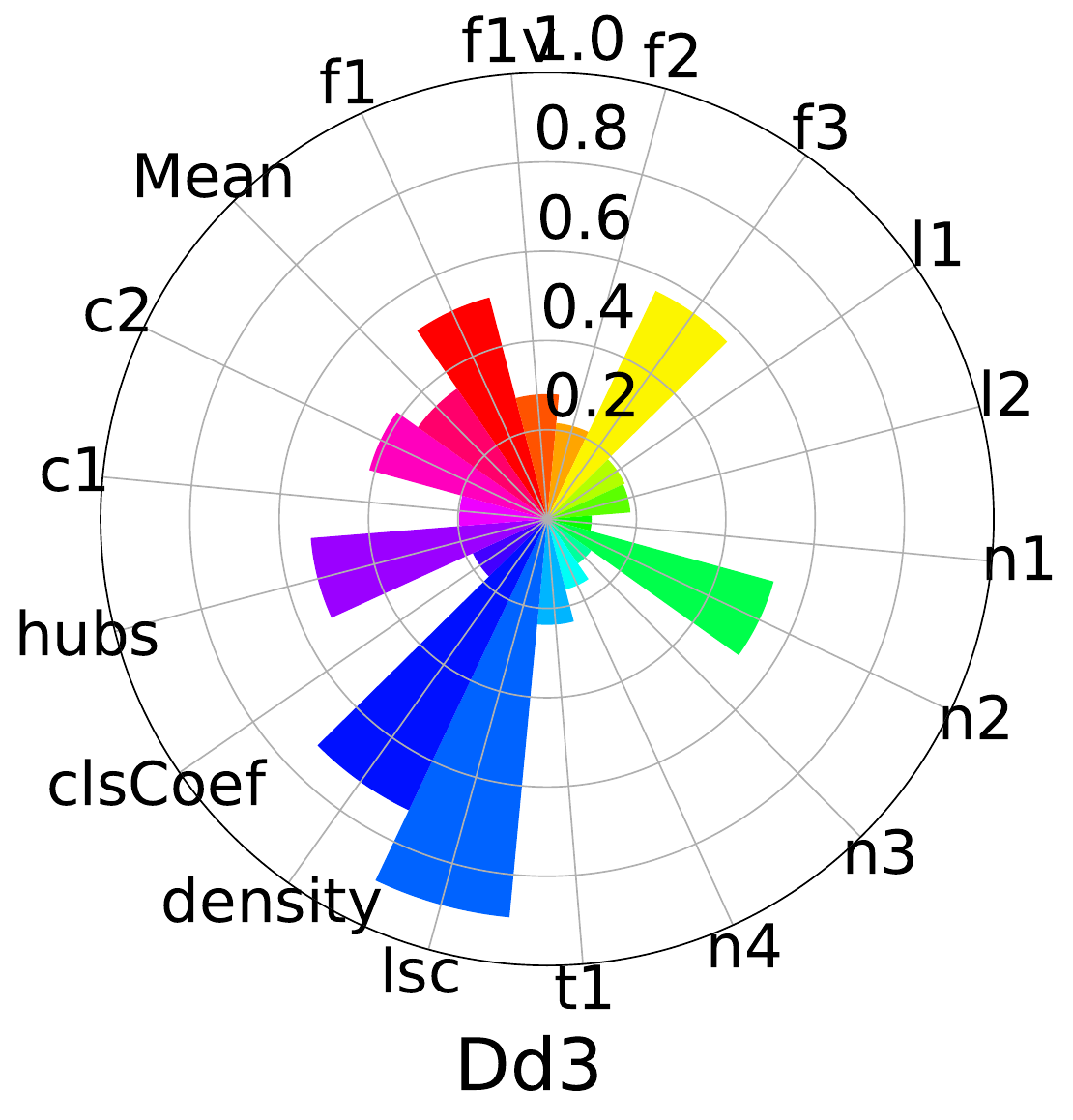}
\includegraphics[width=0.195\textwidth]{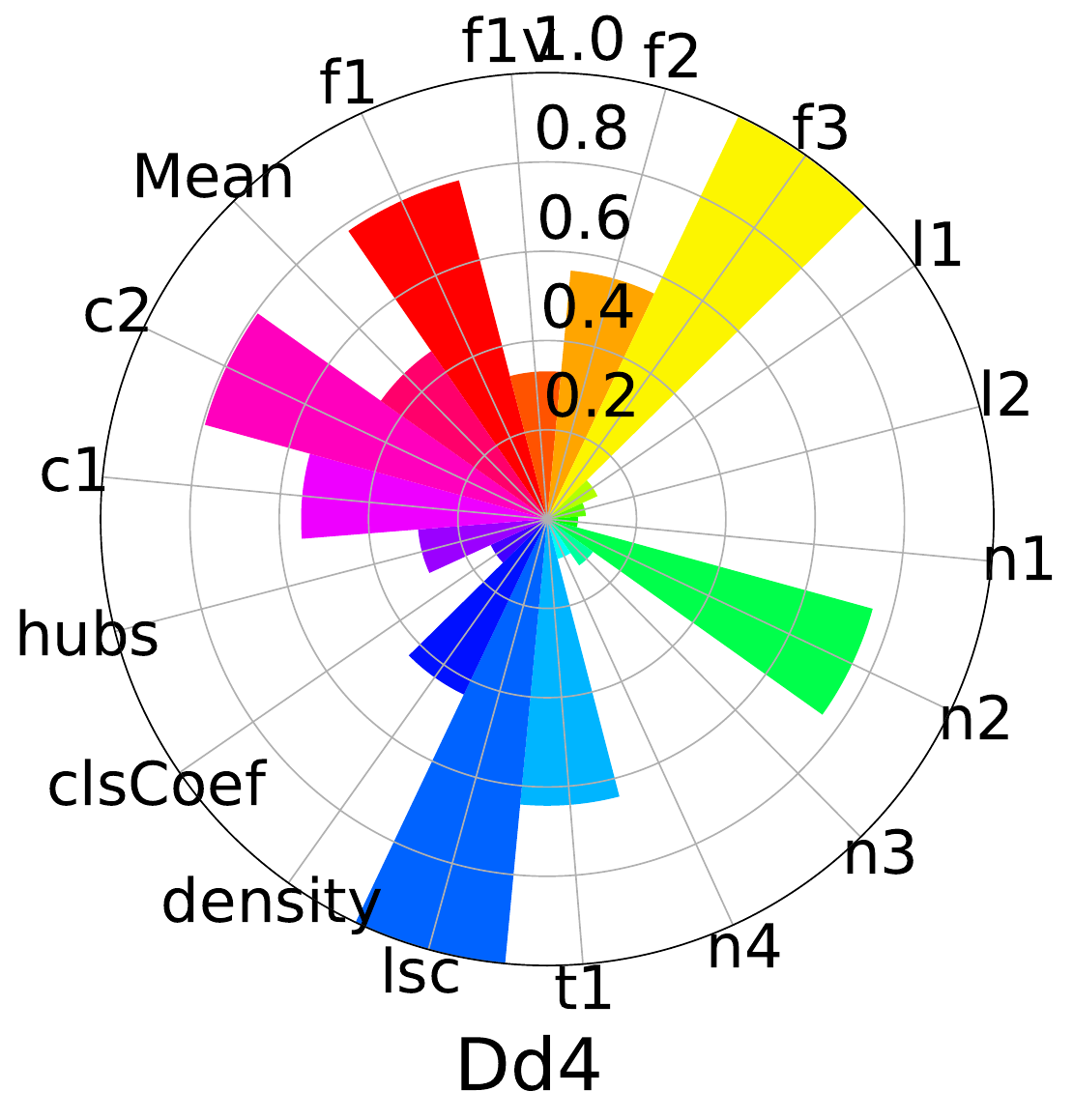}
\includegraphics[width=0.195\textwidth]{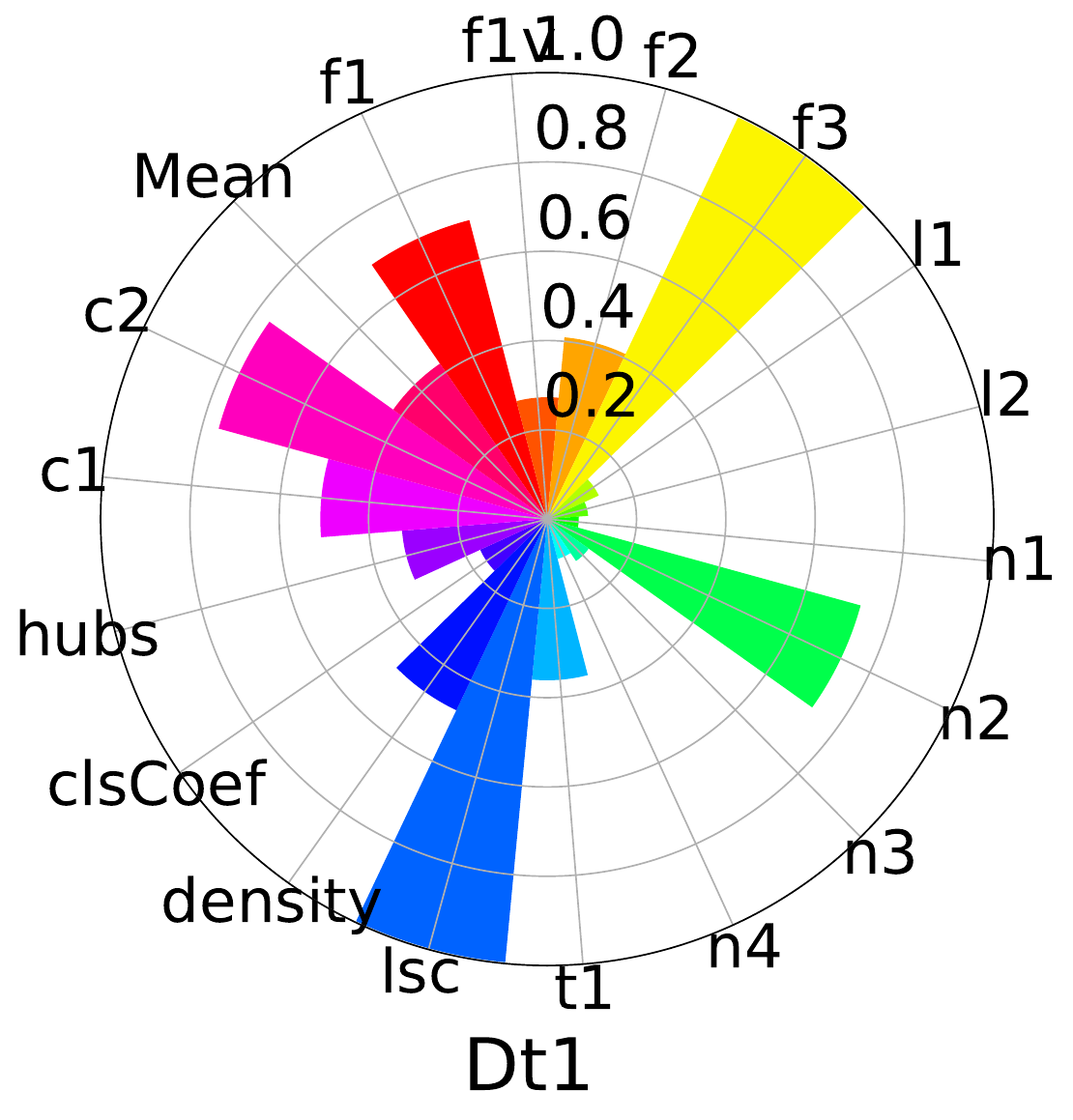}
\includegraphics[width=0.195\textwidth]{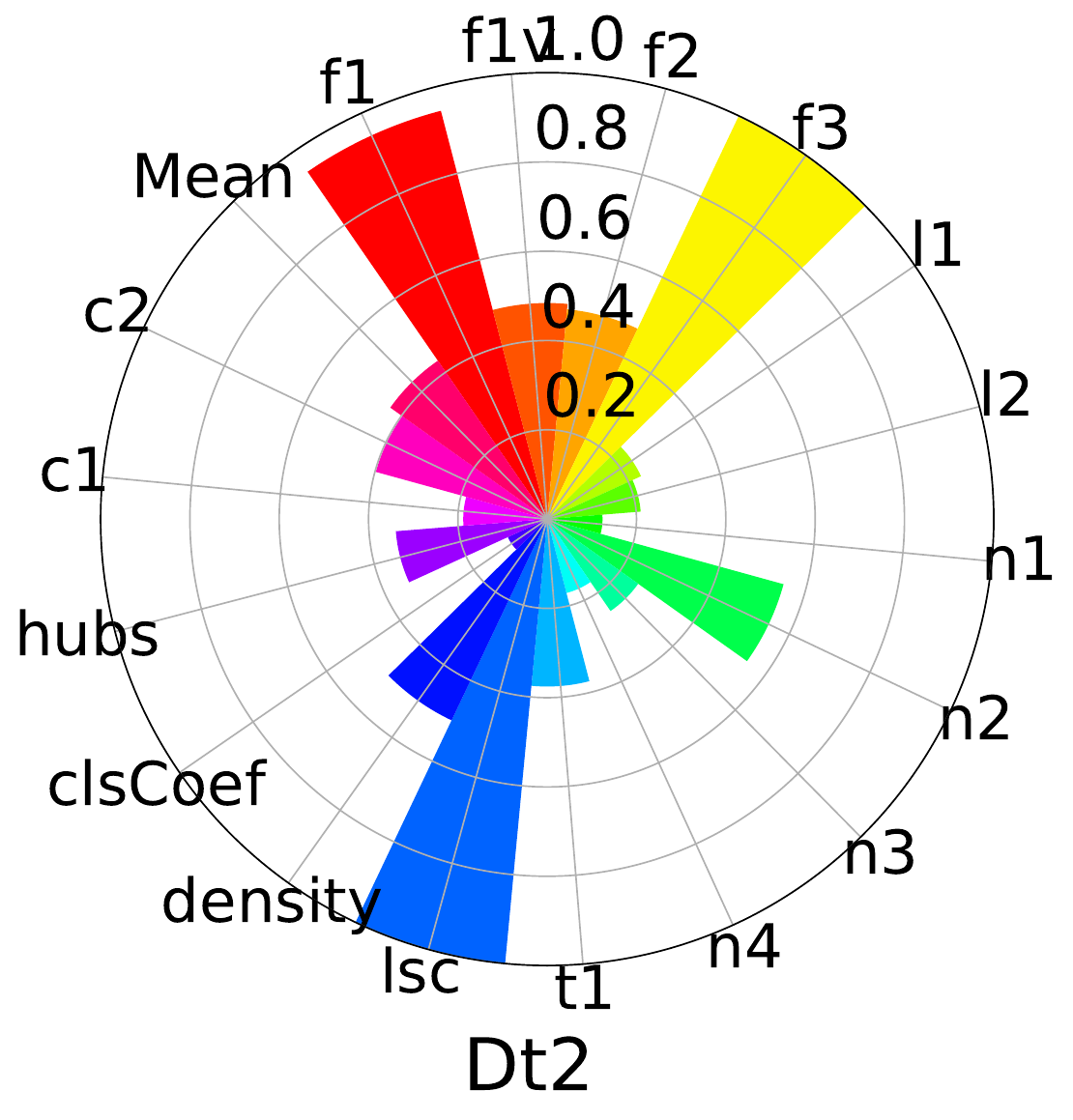}
\caption{Complexity measures per dataset in Table \ref{tb:commonDatasets}.}
\label{fig:complexityFigures}
\end{figure*}

There is a wide deviation between
the thresholds used by the Cosine and Jaccard similarities,
but the actual difference between them is rather low:
$F1_{CS}^{max}$ is higher than $F1_{JS}^{max}$ by just
0.8\%, on average, across the structured and dirty
datasets. In the case of textual datasets, though, Cosine
similarity outperforms Jaccard similarity by 12.3\%  on
average. This is due to the large number of tokens per
record, which significantly reduces the Jaccard scores.

Overall, $F1_{CS}^{max}$ and $F1_{JS}^{max}$ suggest that
\textit{among the structured datasets, only $D_{s3}$,
$D_{s4}$ and $D_{s6}$ are complex enough to call for
non-linear classification models. The same applies only
to $D_{d3}$ and $D_{d4}$ from the dirty datasets, and
to both textual datasets ($D_{t1}$, $D_{t2}$).}

\textbf{Complexity Measures.} We now present the first complexity analysis of the main benchmark datasets for matching algorithms. All measures are
implemented by the \texttt{problexity} Python
package~\cite{komorniczak2022complexity}, using the two-dimensional feature vector defined at the end of Section~\ref{sec:complexityMeasures}.
The results are shown in Figure~\ref{fig:complexityFigures}.

More specifically, datasets identified as rather easy by
the above linearity degree analysis achieve the lowest
scores for the majority of the complexity measures, and
the lowest ones on average. As expected, the lowest average score corresponds to $D_{s7}$ (0.179), since the vast majority of its individual measures falls far below 0.2. Among the remaining easy datasets ($D_{s1}$, $D_{s2}$, $D_{s5}$, $D_{d1}$ and $D_{d2}$), the maximum average score corresponds to $D_{s5}$, amounting to 0.346.

We observe that there are three datasets with an average score close to 0.346: $D_{s3}$ with 0.354, $D_{d2}$ with 0.341 and $D_{d3}$ with 0.355. Comparing them with $D_{s5}$, we observe that they get higher scores for most measures. However, $D_{s3}$ and $D_{d3}$ achieve the minimum value for $f_2$ among all datasets, i.e., the overlapping region between the two classes has the smallest volume among all datasets. They also exhibit comparatively low values for $f_3$, which implies that the two features we have defined are quite effective in separating the two classes.
Most importantly, their scores for both class imbalance measures are close to the minimum score across all datasets. The reason is that they have an unrealistically high portion of positive instances, which, as shown in Table \ref{tb:commonDatasets}, is second only to that of $D_{t2}$ (which gets the lowest scores for $c_1$ and $c_2$ in Figure \ref{fig:complexityFigures}). 
Regarding $D_{d2}$, it exhibits scores close to the minimum ones for $l_1$, which indicates low distances of incorrectly classified instances from a linear classification boundary, as well as for $n_1$ and $n_2$, which indicate a low number of instances
surrounded by examples from the other class. As a result, \textit{the complexity measures indicate that $D_{s3}$, $D_{d2}$ and $D_{d3}$ pose easy classification tasks}, which is in line with Table~\ref{tb:f1PerDataset}, where most matchers achieve high $F1$ over these datasets. 

All other datasets achieve an average score that fluctuates between 0.423
and 0.457. Overall, \textit{a mean complexity score below 0.400 indicates easy classification tasks, with only $D_{s4}$, $D_{s6}$, $D_{d4}$, $D_{t1}$ and $D_{t2}$ being challenging.}

\subsection{Practical Measures}
\label{sec:quantitativeAnalysis}

\begin{table*}[t]\centering
\caption{F1 per method and data set. Hyphen indicates insufficient memory. 
The highest F1 per category and dataset 
is in bold.}
{\small
\begin{tabular}{ | l | ccccccc|cccc|cc | }\cline{1-14}

\multicolumn{1}{|c|}{}& $D_{s1}$ & $D_{s2}$ & $D_{s3}$ & $D_{s4}$ & $D_{s5}$ & $D_{s6}$  & $D_{s7}$ & $D_{d1}$ & $D_{d2}$ & $D_{d3}$ & $D_{d4}$ & $D_{t1}$ & $D_{t2}$\\
\hline
\hline
\multicolumn{14}{|c|}{(a) DL-based matching algorithms}\\
\hline
DeepMatcher (15) & 98.65 & 95.50 & 88.46 & 69.66 & 75.86 & 65.98 & 95.45 & 96.63 & 93.07 & 75.00 & 46.56 & 68.53 & \textbf{94.04} \\
DeepMatcher (40) & 98.76 & 93.70 & 84.62 & 64.42 & 66.67 & 53.73 & 91.67 & 96.54 & 92.73 & 66.67 & 46.99 & 69.21 & - \\
\hline
DeepMatcher \cite{Mudgal2018sigmod} & 98.40 & 94.00 & 88.00 & 66.90 & 72.70 & 69.30 & 100.00 & 98.10 & 93.80 & 74.50 & 46.00 & 62.80 & 92.70  \\
\hline
\hline
DITTO (15) & 51.46 & 88.62 & 67.61 & 51.44 & 42.62 & 70.66 & 28.76 & {42.29} & 91.21 & 61.73 & {44.15} & {38.94} & {54.60} \\
DITTO (40) & {89.43} & {91.18} & {56.82} & {58.02} & {28.00} & 66.94 & 65.67 & {90.16} & 91.05 & 65.06 & 60.80 & {42.09} & {64.77} \\
\hline
DITTO \cite{DBLP:journals/pvldb/0001LSDT20} & 98.99 & 95.60 & 97.06 & 86.76 & 94.37 & 75.58 & 100.00 & 99.03 & 95.75 & 95.65 & 85.69 & 89.33 & {93.85} \\
\hline
\hline
EMTransformer-B (15) & {98.99} & 95.42 & 92.59 & 80.80 & \textbf{82.35} & 68.14 & 97.78 & {98.88} & 95.24 & \textbf{98.04} & 79.59 & 83.94 & {78.31}\\
EMTransformer-B (40) & \textbf{99.21} & 95.38 & 92.31 & 82.72 & \textbf{82.35} & 66.20 & 97.78 & \textbf{98.99} & {95.53} & 94.34 & 82.81 & 85.42 & {77.65}  \\
\hline
EMTransformer-R (15) & 98.87 & \textbf{95.90} & 96.15 & {84.83} & {80.00} & 69.04 & \textbf{100.00} & 98.19 & \textbf{95.78} & 94.12 & \textbf{83.95} & \textbf{89.29} & {77.65} \\
EMTransformer-R (40) & {98.52} & {95.83} & 94.55 & \textbf{85.04} & {80.00} & 68.36 & \textbf{100.00} & 98.30 & 95.22 & 94.34 & 82.69 & 87.11 & {77.12} \\
\hline
EMTransformer \cite{DBLP:journals/corr/abs-2004-00584,DBLP:conf/edbt/BrunnerS20} & N/A & N/A & N/A & N/A & N/A & N/A & N/A & 98.90 & 95.60 & 94.20 & 85.50 & 90.90 & N/A \\
\hline
\hline
GNEM (10) & {98.21} & 95.19 & {96.43} & 84.96 & {77.78} & 70.85 & \textbf{100.00} & 98.87 & 93.93 & {94.74} & 79.19 & {88.66} & - \\ 
GNEM (40) & {98.55} & 94.95 & \textbf{98.18} & {20.45} & {80.00} & \textbf{74.75} & \textbf{100.00} & 98.87 & 93.92 & 89.66 & {83.87} & 86.49 & - \\
\hline
GNEM \cite{DBLP:conf/www/ChenSZ20} & N/A & N/A & N/A & 86.70 & N/A & 74.70 & N/A & N/A & N/A	 & N/A & N/A & 87.70 & N/A \\
\hline
\hline
HierMatcher (10) & - & 94.85 & - & 79.37 & 72.00 & 72.06 & \textbf{100.00} & - & - & - & 58.63 & - & - \\
HierMatcher (40) & - & 94.85 & - & 79.37 & 72.00 & 72.06 & \textbf{100.00} & - & - & - & 58.63 & - & - \\
\hline
HierMatcher \cite{DBLP:conf/ijcai/FuHHS20} & {98.80} & 95.30 & N/A & 81.60 & N/A & 74.90 & N/A & 98.01 & 94.50 & N/A & 68.50 & N/A & N/A \\
\hline
\hline
\multicolumn{14}{|c|}{(b) Non-neural, non-linear ML-based matching algorithms}\\
\hline
Magellan-DT & 97.65 & {86.88} & {88.52} & {62.37} & {84.85} & 54.42 & \textbf{100.00} & 40.07 & 78.76 & 50.00 & 33.89 & 48.46 & \textbf{100.00} \\
Magellan-LR & 97.66 & 88.61 & 84.21 & 65.99 & 80.00 & 44.44 & \textbf{100.00} & \textbf{83.20} & 76.03 & 50.00 & 32.77 & 37.36 & \textbf{100.00} \\
Magellan-RF & 98.32 & \textbf{92.96} & \textbf{89.66} & \textbf{67.76} & \textbf{84.85} & \textbf{56.10} & \textbf{100.00} & 60.47 & \textbf{81.67} & \textbf{52.00} & \textbf{38.06} & \textbf{51.30} & \textbf{100.00} \\
Magellan-SVM & 90.19 & 81.41 & 84.62 & 65.03 & 84.62 & 2.53 & 84.21 & 10.99 & 48.15 & 12.12 & 12.62 & 0.00 & 99.96 \\
\hline
Magellan \cite{Mudgal2018sigmod} & 98.40 & 92.30 & 91.20 & 71.90 & 78.80 & 49.10 & 100.00 & 91.90 & 82.50 & 46.80 & 37.40 & 43.60 & 79.80 \\
\hline
\hline
ZeroER & \textbf{98.80} & 65.67 & 49.81 & 64.41 & 35.90 & 18.50 & 90.91 & 36.53 & 39.23 & 10.42 & 20.00 & 2.56 & - \\
\hline
ZeroER \cite{wu2020sigmod} & 96.00 & 86.00 & N/A & N/A & N/A & 48.00 & 100.00 & N/A & N/A & N/A & N/A & 52.00 & N/A \\
\hline
\hline
\multicolumn{14}{|c|}{(c) Non-neural, linear supervised matching algorithms}\\
\hline
SA-ESDE & 93.06 & 87.57 & 52.94 & 45.27 & 85.71 & 51.58 & \textbf{100.00} & 92.71 & 86.80 & 52.94 & \textbf{45.27} & 37.67 & 43.97 \\
SAQ-ESDE & 93.08 & \textbf{88.62} & 55.81 & 43.91 & 82.76 & \textbf{54.13} & 97.77 & {93.16} & \textbf{88.51} & 49.41 & 42.82 & 37.94 & 58.40 \\
SAF-ESDE & 88.84 & 83.14 & 32.43 & 28.51 & 61.54 & 36.36 & 76.92 & 88.94 & 82.64 & 33.71 & 27.29 & 21.24 & 40.08 \\
SAS-ESDE & \textbf{93.49} & 87.40 & 64.00 & 43.62 & \textbf{87.50} & 48.17 & 95.45 & \textbf{93.35} & 86.79 & \textbf{64.00} & 42.27 & 40.57 & \textbf{79.86} \\
\hline
SB-ESDE & 91.19 & 79.63 & \textbf{92.31} & \textbf{67.81} & 82.76 & 52.65 & 84.44 & 84.27 & 78.18 & 46.43 & 42.94 & 45.63 & 41.23 \\
SBQ-ESDE & 91.44 & 82.71 & 84.21 & 67.55 & 83.33 & 45.20 & \textbf{100.00} & 87.54 & 82.29 & 55.70 & 37.47 & 47.17 & 58.37  \\
SBF-ESDE & 90.89 & 80.69 & 77.27 & 67.34 & 78.57 & 33.88 & 95.65 & 70.36 & 69.91 & 30.43 & 22.27 & 33.91 & 39.12\\
SBS-ESDE & 90.89 & 82.45 & 87.72 & 67.35 & 82.76 & 46.68 & \textbf{100.00} & 85.68 & 80.06 & 43.14 & 41.29 & \textbf{49.15} & \textbf{79.86}\\
\hline
	\end{tabular}
 	}
	\label{tb:f1PerDataset}
\end{table*}

\textbf{Setup.} We conducted all experiments on a server with an
Nvidia GeForce RTX 3090 GPU (24 GB RAM) and a dual AMD EPYC 7282
16-Core CPU (256 GB RAM), with all implementations running in Python. Given that every method depends on different Python
versions and packages, we aggregated all of them into a Docker image, which facilitates the reproducibility of our experiments. Note
that every evaluated method requires a different format for the
input data; we performed all necessary transformations and
will publish the resulting files and the Docker image upon acceptance.

\textbf{Methods Configuration.} 
Following \cite{Mudgal2018sigmod}, DeepMatcher is combined with
fastText~\cite{DBLP:journals/tacl/BojanowskiGJM17}
embeddings in the attribute embedding module, the Hybrid model in the attribute similarity vector module, and a two layer fully connected ReLU HighwayNet~\cite{DBLP:conf/icml/ZillySKS17}
classifier followed by a softmax layer in the classification
module. 

EMTransformer has two different versions:
EMTransformer-B and EMTransformer-R, which use BERT and RoBERTa, respectively. As noted in~\cite{DBLP:journals/corr/abs-2004-00584} (Section E in their Appendix), its original implementation ignores the validation set. Instead, ``it reports the best F1 on the test set among all the training epochs''. To align it with the other methods, we modified its code so that it uses the validation set to select the best performing model that is applied to the testing~set.

For GNEM, we employ a BERT-based embedding
model, because its dynamic nature outperforms the static pre-trained models like
fastText, as shown by the authors in~\cite{DBLP:journals/tacl/BojanowskiGJM17}. In the
interaction module, we apply a single-layer gated graph
convolution network, following the recommendation of the
authors. 

DITTO employs RoBERTa, since it is best performing in~\cite{DBLP:journals/pvldb/0001LSDT20,DBLP:journals/jdiq/LiLSWHT21}. However, we were
not able to run DITTO with part-of-speech tags, because these
tags are provided by a service that was not available. 
Therefore, like all other methods we evaluated, DITTO did not
employ any external knowledge.

HierMatcher employs the pre-trained fastText
model~\cite{DBLP:journals/tacl/BojanowskiGJM17} for embeddings, while the hidden size of each GRU layer is set to 150 in the
representation layer, following \cite{DBLP:conf/ijcai/FuHHS20}.

We also modified the functionality of ZeroER, decoupling it from
the blocking function that is hand-crafted for each dataset. Only
in this way we can ensure that it applies to exactly the same
instances as all other methods, allowing for a fair comparison.

Finally, we combine Magellan with four
different classification algorithms: Magellan-DT uses a Decision Tree,
Magellan-LR Logistic Regression, Magellan-RF a Random Forest, and Magellan-SVM a
Support Vector Machine. Similar to ZeroER, for a fair
comparison, we decoupled the
blocking functionality provided by Magellan, applying it to the same
blocked data sets as all other methods.  

\textbf{Hyperparameters.} Initial experiments showed
that the number of epochs is probably the most important
hyperparameter for most DL-based matching algorithms. To
illustrate this, we report the performance of every DL
algorithm for two different settings: (1) the default number of
epochs as reported in the corresponding paper, and (2) 40 epochs,
which is common in the original papers.
Table~\ref{tb:f1PerDataset} shows the
resulting performance, with the number in parenthesis next to each DL-based algorithm indicating the number of epochs. 

\textbf{Reproducibility Analysis.} To verify the validity of the above configurations, which will also be applied to the new datasets, Table \ref{tb:f1PerDataset} also presents the fine-tuned performance of each non-linear matching algorithm, as reported in the literature. The lower the difference between the best F1 performance we achieved for each method on a specific dataset and the one reported in the literature, the closer we are to the optimal configuration for this method. 

Starting with the DL-based matching algorithms, our experiments with DeepMatcher exceed those of~\cite{Mudgal2018sigmod} in most cases, by 1.5\%, on average.
For EMTransformer, we consider the results reported in \cite{DBLP:journals/corr/abs-2004-00584}, because the original experiments in \cite{DBLP:conf/edbt/BrunnerS20} show the evolution of F1 across the various epochs, without presenting exact numbers. 
The average difference with our F1 is just 0.15\% on average.
Slightly higher, albeit negligible, just 0.3\%, is the mean difference between our results and GNEM's performance in 
\cite{DBLP:conf/www/ChenSZ20}.
Our results for HierMatcher are consistently lower
than those in~\cite{DBLP:conf/ijcai/FuHHS20}, 
with an average difference of 5.4\%.
HierMatcher constitutes a relatively reproducible algorithm.
Finally, DITTO's performance
in~\cite{DBLP:journals/pvldb/0001LSDT20} is consistently higher than our
experimental results to a significant extent -- on average
by 25\%. This is caused by the lack of external
knowledge and the absence of two optimizations (see Section 3 in~\cite{DBLP:journals/pvldb/0001LSDT20}).
Yet, our best performance among all DL-based matching algorithms per dataset is very close or even higher than DITTO's performance in \cite{DBLP:journals/pvldb/0001LSDT20} in all datasets, but $D_{s5}$.

Regarding the ML-based matching algorithms, Magellan underperforms the results in \cite{Mudgal2018sigmod} in $D_{s4}$ and $D_{d1}$, while for $D_{s1}$, $D_{s2}$, $D_{s3}$, $D_{d2}$ and $D_{d4}$, the differences are minor ($\le1.8\%$ in absolute terms). For the remaining five datasets, though, our results are significantly higher than~\cite{Mudgal2018sigmod}, by 13\% on average. For ZeroER, we get slightly better performance in $D_{s1}$, while for the other four datasets examined in \cite{wu2020sigmod}, our results are lower by 60\%, on average. The reason is that~\cite{wu2020sigmod} combines ZeroER with custom blocking methods and configurations in
each case, whereas we use the same configuration in all datasets. Yet, the best performance per dataset that is achieved by one of Magellan's variants consistently outperforms ZeroER's performance in \cite{wu2020sigmod}, except for $D_{t1}$, where its top F1 is 0.7\%~lower.

On the whole, the selected configurations provide an overall performance close to or even better than the best one for non-linear matching algorithms in the literature. 

\begin{figure}[t]
\centering
\includegraphics[width=0.47\textwidth]{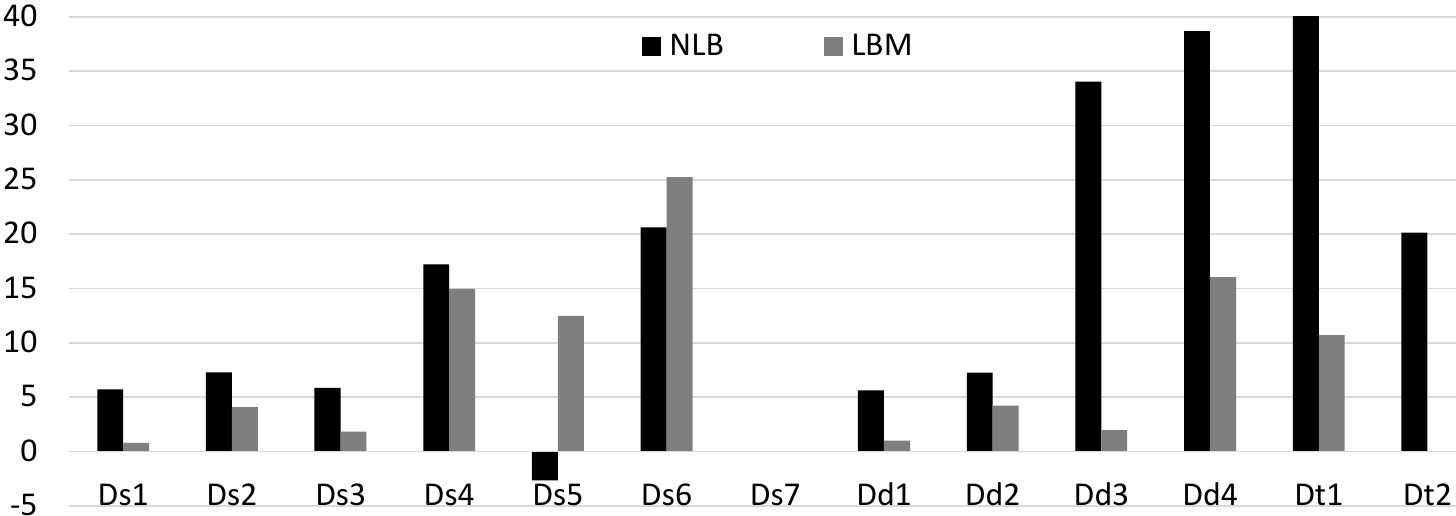}
\vspace{-10pt}
\caption{Practical measures per dataset in Table \ref{tb:commonDatasets}.}
\vspace{-15pt}
\label{fig:practicalMeasures}
\end{figure}

\textbf{Aggregate Practical Measures.} Figure \ref{fig:practicalMeasures} presents the non-linear boost (NLB) and the learning-based margin (LBM),
as we described in Section~\ref{ref:aggmeas}, per dataset. Both measures should exceed 5\%, ideally 10\%, in a dataset that is considered as challenging. Only in such a dataset are the two classes linearly inseparable to a large extent, while there is a significant room for improvements. 

\textit{Among the structured datasets, this requirement is met only by $D_{s4}$ and $D_{s6}$.} The first three datasets exhibit a high NLB, demonstrating their non-linearity, but have a very low LBM, as many algorithms achieve a practically perfect performance. In $D_{s5}$, there is room for improvements, but the two classes are linearly separable to a large extent, as the best linear algorithms outperform the best non-linear ones. Finally, both measures are reduced to 0 over $D_{s7}$, because both linear and non-linear algorithms achieve perfect F1. 

All dirty datasets have a higher degree of non-linearity, as indicated by their NLB, which consistently exceeds 5\%. However, the first three datasets are ideally solved by the DL-based matching algorithms, especially EMTransformer, \textit{leaving $D_{d4}$ as the only challenging dataset of this type.}

Finally, both textual datasets exhibit high non-linearly, but on $D_{t2}$, all Magellan variants achieve perfect performance (outperforming the DL-based matchers). Hence, \textit{only $D_{t1}$ is challenging.}

\section{Methodology for New Benchmarks} 
\label{sec:methodology}

\begin{table}[t]\centering
\renewcommand{\tabcolsep}{5pt}
\caption{{\small Existing Vs New Benchmarks}}
\vspace{-7pt}
{\small
\begin{tabular}{ | l | r | r | r || l | r | r | r |}
\hline
$D_{t1}$ & 0.955 & 0.120 & 12.03\% & $D_{n1}$ & 0.899 & 0.029 & 2.90\% \\
$D_{s1}$ & 0.998 & 0.137 & 13.68\% & $D_{n3}$ & 0.983 & 0.953 & 95.30\% \\
$D_{s2}$ & 1.000 & 0.229 & 22.89\% & $D_{n8}$ & 0.906 & 0.166 & 16.60\% \\
$D_{s4}$ & 1.000 & 0.104 & 10.37\% & $D_{n7}$ & 0.894 & 0.018 & 1.80\% \\
$D_{s6}$ & 0.898 & 0.247 & 24.66\% & $D_{n2}$ & 0.910 & 0.074 & 7.40\% \\
\hline
	\end{tabular}
	}
	\vspace{-14pt}
	\label{tb:existingVsNew}
\end{table}

\begin{table*}[ht!]\centering
\renewcommand{\tabcolsep}{2.8pt}
{\small
\caption{{
Characteristics of the new datasets generated by
DeepBlocker~\cite{DBLP:journals/pvldb/Thirumuruganathan21}. The
blocking performance is reported with recall ($PC$),
precision ($PQ$), the total number of candidates ($|C|$) and the
number of matching candidates ($|P|$). DeepBlocker applies
Autoenconder to the selected attribute(s) ($attr.$) using
stemming and stop-word removal 
or not ($cl.$), and $K$ candidates per query record of the indexed
dataset ($ind.$).}}
	\begin{tabular}{ | l | c |  c | r | r | c || c | c | r | r || c | c | r | c || r | r |  r | r | r | r | r |}
		\cline{1-21}
		 \multicolumn{1}{|c|}{} & 
		 \multirow{2}{*}{$D_1$} & 
		 \multirow{2}{*}{$D_2$} & 
         \multicolumn{1}{c|}{\multirow{2}{*}{$|D_1|$}} & 
         \multicolumn{1}{c|}{\multirow{2}{*}{$|D_2|$}} & 
         \multirow{2}{*}{$|A|$} &
         \multicolumn{4}{c||}{Blocking performance} & 
         \multicolumn{4}{c||}{DeepBlocker config.} & 
         \multicolumn{1}{c|}{\multirow{2}{*}{$|I_{tr}|$}} & 
         \multicolumn{1}{c|}{\multirow{2}{*}{$|I_{te}|$}} & 
         \multicolumn{1}{c|}{\multirow{2}{*}{$|P_{tr}|$}} &
         \multicolumn{1}{c|}{\multirow{2}{*}{$|P_{te}|$}} & 
         \multicolumn{1}{c|}{\multirow{2}{*}{$|N_{tr}|$}} & 
         \multicolumn{1}{c|}{\multirow{2}{*}{$|N_{te}|$}} &
         \multicolumn{1}{c|}{\multirow{2}{*}{IR}} \\
         \multicolumn{1}{|c|}{} & & & & & & 
         \multicolumn{1}{c|}{$PC$} & 
         \multicolumn{1}{c|}{$PQ$} & 
         \multicolumn{1}{c|}{$|C|$} &
         \multicolumn{1}{c||}{$|P|$} &
         \multicolumn{1}{c|}{$attr.$} & 
         \multicolumn{1}{c|}{$cl.$} & 
         \multicolumn{1}{c|}{$K$} & 
         \multicolumn{1}{c||}{$ind.$} & & & & & & & \\
        \hline
        \hline
        $D_{n1}$ & Abt & Buy & 1,076 & 1,076 & 3 & 0.899 & 0.029 & 33,356 & 967 & name & $\times$ & 31 & $D_2$ & 20,014 & 6,671 & 580 & 193 & 19,433 & 6,478 & 2.9\%\\
        $D_{n2}$ & Amazon & GP & 1,354 & 3,039 & 4 & 0.910 & 0.074 & 13,540 & 1,005	 & title & $\times$ & 10 & $D_1$ & 8,124 & 2,708 & 603 & 201 & 7,521 & 2,507 & 7.4\%\\
        $D_{n3}$ & DBLP & ACM & 2,616 & 2,294 & 4 & 0.983 & 0.953 & 2,294 & 2,186 & all & \checkmark & 1	 & $D_2$ & 1,376 & 459 & 1,312 & 437 & 65 & 22 & 95.3\%\\
        $D_{n4}$ & IMDB & TMDB & 5,118 & 6,056 & 5 & 0.898 & 0.011 & 158,658 & 1,768 & all & \checkmark	 & 31 & $D_1$ & 95,195 & 31,732 & 1,061 & 354 & 94,134 & 31,378 & 1.1\%\\
        $D_{n5}$ & IMDB & TVDB & 5,118 & 7,810 & 4 & 0.891 & 0.003 & 322,434 & 955 & all & $\times$	 & 63 & $D_1$ & 193,460 & 64,487 & 573 & 191 & 192,887 & 64,296 & 0.3\%\\
        $D_{n6}$ & TMDB & TVDB & 6,056 & 7,810 & 6 & 0.927 & 0.130 & 7,810 & 1,015 & all & \checkmark & 1 & $D_2$ & 4,686 & 1,562 & 609 & 203 & 4,077 & 1,359 & 13.0\%\\
        $D_{n7}$ & Walmart & Amazon & 2,554 & 22,074 & 6 & 0.894 & 0.018 & 43,418 & 763 & all	 & \checkmark & 17 & $D_1$ & 26,051 & 8,684 & 458 & 153 & 25,593 & 8,531 & 1.8\%\\
        $D_{n8}$ & DBLP & GS & 2,516 & 61,353 & 4 & 0.906 & 0.166 & 12,580 & 2,091 & all & \checkmark & 5 & $D_1$ & 7,548 & 2,516 & 1,255 & 418 & 6,293 & 2,098 & 16.6\%\\
	\hline
	\end{tabular}
	\label{tb:newDatasets}
}
\end{table*}

Our methodology for creating new benchmarks consists of the following three steps:
\begin{enumerate}[leftmargin=*]
    \item Given a dataset with a complete ground truth, apply a state-of-the-art blocking method 
    that is suitable for the data at hand. Blocking is indispensable for reducing the search space to the most likely duplicates, which can be processed by a matching algorithm within a reasonable time frame.
    \item Based on the available ground truth, fine-tune the selected blocking method for a minimum level of recall. In practical situations, recall should be very high (e.g., 90\%), because most learning-based matchers take decisions at the level of individual record pairs and, thus, they cannot infer duplicates not included in the candidate pairs. The fine-tuning maximizes precision for the selected recall so as to minimize the class imbalance. In this process, the selected recall level determines the  difficulty of the labeled instances. The higher the recall levels are, the more difficult to classify positive instances (true matches) are included at the expense of including more and easier negative instances (true non-matches), and vice versa for low recall levels. We term ``easy positive instances'' the duplicate entities whose similarity is higher than most non-matching pairs, whereas ``easy negative instances'' involve non-matching entities with a similarity lower than most matching ones.
    \item Randomly split the candidates pairs into training, validation and testing sets with a typical ratio, using the ground truth.
    \item Apply all
    difficulty measures from Section \ref{sec:theoreticalMeasures} to decide whether the resulting benchmark is challenging enough.
\end{enumerate}

To put this methodology into practice, we use the eight, publicly available, established datasets for RL in Table \ref{tb:newDatasets}. They cover a wide range of domains, from product matching ($D_{n1}$, $D_{n2}$, $D_{n7}$) to bibliographic data ($D_{n3}$, $D_{n8}$) and movies ($D_{n4}$-$D_{n6}$). 
We apply DeepBlocker~\cite{DBLP:journals/pvldb/Thirumuruganathan21}
to these datasets, a generic state-of-the-art approach leveraging Autoenconder, self-supervised learning and fastText embeddings.
Through grid search, DeepBlocker is configured so that 
its recall, also known as pair completeness ($PC$)~\cite{Christen2012springer}, exceeds 90\%.
Note that \textit{our methodology is generic enough to support any other blocking method and recall limit}.

For every dataset, DeepBlocker generates the candidate pair set $C$ by indexing one of the two data sources ($D_1$ or $D_2$ in Table \ref{tb:newDatasets}), while every record of the other source is used as a query that retrieves the $K$ most likely matches. To maximize precision, we consider the lowest $K$ that exceeds the minimum recall. In each dataset, we use both combinations of indexing and query sets and select the one yielding the lowest number of candidates for the required recall. 

\begin{figure*}[t]
\centering
\includegraphics[width=0.485\textwidth]{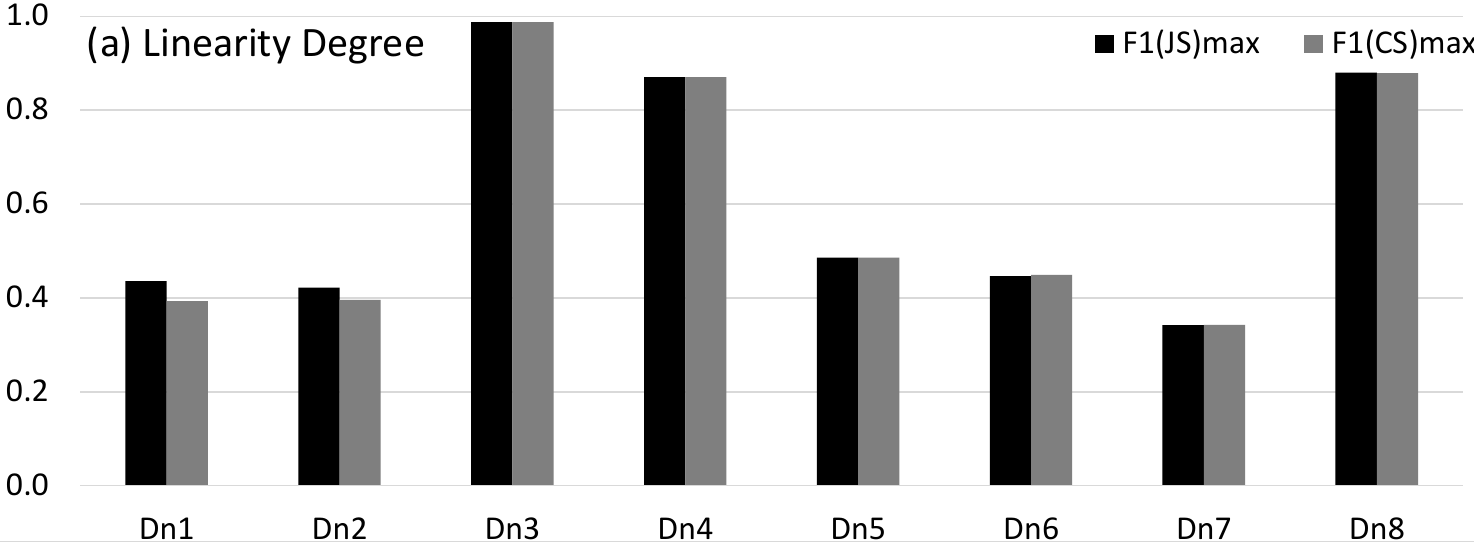}
\hspace{5pt}
\includegraphics[width=0.47\textwidth]{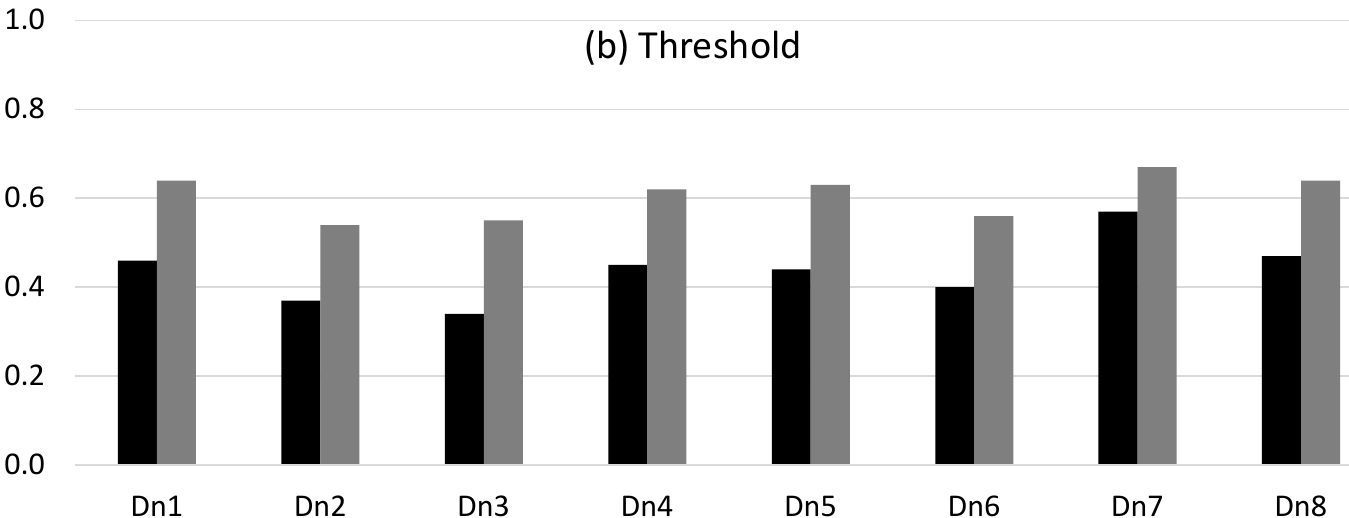}
\caption{Degree of linearity per dataset in Table \ref{tb:newDatasets} (left) and the respective threshold (right) with respect to Equations \ref{eq:cosine} and \ref{eq:jaccard}.}
\label{fig:theoreticalLnearityNew}
\end{figure*}

We also fine-tune two more hyperparameters: (1) whether cleaning is used or not (if it does, stop-words are removed and stemming is applied to all words), and (2) the attributes providing the values to be blocked. We consider all individual attributes as well as a schema-agnostic setting that concatenates all attributes into a sentence. Note that DeepBlocker converts these attributes into embedding vectors using fastText and then applies self-supervised learning to boost its accuracy without requiring any manually labelled instances; fastText's static nature ensures that the order of words in the concatenated text does not affect the resulting vector. For every hyperparameter, we consider all possible options and select the one minimizing the returned set of candidates. This means that we maximize precision, also known as pairs quality ($PQ$)~\cite{Christen2012springer}, so as to maximize the portion of matching pairs in the resulting candidate set, minimizing the class imbalance.

The exact configuration of DeepBlocker per dataset is shown in Table~\ref{tb:newDatasets}. Column \textit{attr.} indicates that the schema-agnostic setting yields the best performance in most cases, column \textit{cl.} suggests that cleaning is typically required, and column \textit{ind.} shows that the smallest data source is typically indexed. Finally, the number of candidates per query entity, $K$, differs widely among the datasets, depending heavily on the data at hand.

Given that DeepBlocker constitutes a stochastic approach, the performance reported in Table~\ref{tb:newDatasets} corresponds to the average after 10 repetitions. For this reason, in some cases, $PC$ drops slightly lower than 0.9. The resulting candidate pairs are randomly split into training, validation, and testing sets with the same ratio as the benchmarks in Table~\ref{tb:commonDatasets} (3:1:1). These settings simulate the realistic scenarios, where blocking is applied to exclude obvious non-matches, and then a subset of the generated candidate pairs is labelled to train a matching algorithm that resolves the rest of the candidates. The instances per class and set are reported in Table~\ref{tb:newDatasets} (note that the testing and validation sets have the same size). 

At this point, it is worth juxtaposing the existing and the new benchmarks that have the same origin. These datasets are compared in Table \ref{tb:existingVsNew}. We observe that $D_{t1}$ and $D_{s2}$ outperform $D_{n1}$ and $D_{n8}$, respectively, both in terms of recall and precision, even though DeepBlocker outperforms Magellan's blocking methods \cite{DBLP:journals/pvldb/Thirumuruganathan21}. Therefore, the higher precision of $D_{t1}$ and $D_{s2}$ is most likely achieved due to the removal of negative pairs. Moreover, the recall of $D_{s4}$ is lower than $D_{s7}$ by just 6\%, but its precision is higher by a whole order of magnitude, whereas the recall of $D_{s6}$ is lower than $D_{n2}$ by just 1.2\%, but its precision is higher by 3.3 times. These tradeoffs are not common in blocking over these two particular datasets \cite{DBLP:journals/corr/abs-2202-12521} and could be caused by removing a large portion of negative pairs. Finally, $D_{n3}$ exhibits much higher precision (almost by 7 times) than $D_{s1}$, even though their difference in recall is just 1.5\%. Given that a wide range of blocking methods achieves exceptionally high precision in this bibliographic dataset \cite{DBLP:journals/corr/abs-2202-12521}, the low precision of the existing benchmark could be caused by including a large number of easy, negative pairs, i.e. obvious non-matches. Overall, \textit{the five existing benchmarks in Table \ref{tb:existingVsNew} seem to involve an undocumented approach for inserting or removing an arbitrary number of negative pairs}.

The only exception is $D_{n3}$, which is dominated by positive candidate pairs after blocking. This indicates a rather easy RL dataset, where blocking suffices for detecting the duplicate records, rendering matching superfluous. This is due to the low levels of noise and the distinguishing information in its bibliographic data. Yet, the imbalance ratio in the respective existing benchmarks, $D_{s1}$ and $D_{d1}$, is 81\% lower, which implies that \textit{they contain a large portion of obvious non-matches. This explains why $D_{s1}$ and $D_{d1}$ have been marked as easy classification tasks in the analysis of Section \ref{sec:datasetAnalysis}.}

\subsection{Analysis of new benchmarks}

The above process is not guaranteed to yield challenging
classification tasks for learning-based matching algorithms. 
For this reason, our methodology includes a third step, which assesses the difficulty of each new benchmark through the theoretical and practical measures defined in Sections \ref{sec:theoreticalMeasures} and \ref{sec:practicalMeasures}, respectively.
We should stress, though, that our methodology allows analysts to tune the difficulty level of the generated datasets by changing the level of blocking recall (90\% in our case) and by replacing DeepBlocker with another state-of-the-art blocking algorithm.

\subsubsection{Theoretical Measures}

\textbf{Degree of Linearity.} In Figure \ref{fig:theoreticalLnearityNew}(a), we show the values of $F1_{CS}^{max}$ and $F1_{JS}^{max}$ per dataset along with the respective thresholds. We observe that both measures exceed 0.87 for $D_{n3}$, $D_{n4}$ and $D_{n8}$, while remaining below 0.49 for all other datasets. For $D_{n3}$, this is expected, because, as explained above, it involves quite unambiguous duplicates, due to the low levels of noise. The same is also true for $D_{n8}$, which also conveys bibliographic data, with its duplicates sharing clean and distinguishing information. The opposite is true for $D_{n4}$: its low precision after blocking indicates high levels of noise and missing values, requiring many candidates per entity to achieve high recall. The fastText embeddings may add to this noise, as the attribute values are dominated by movie titles and the names of actors and directors, which are underrepresented in its training corpora. In contrast, the traditional textual similarity measures at the core of $F1_{CS}^{max}$ and $F1_{JS}^{max}$ are capable of separating linearly the two matching classes.

On the whole, these a-priori measures suggest that \textit{$D_{n1}$, $D_{n2}$, and $D_{n5}$ to $D_{n7}$ pose challenging matching tasks.}

\begin{figure*}[t]
\centering
\includegraphics[width=0.24\textwidth]{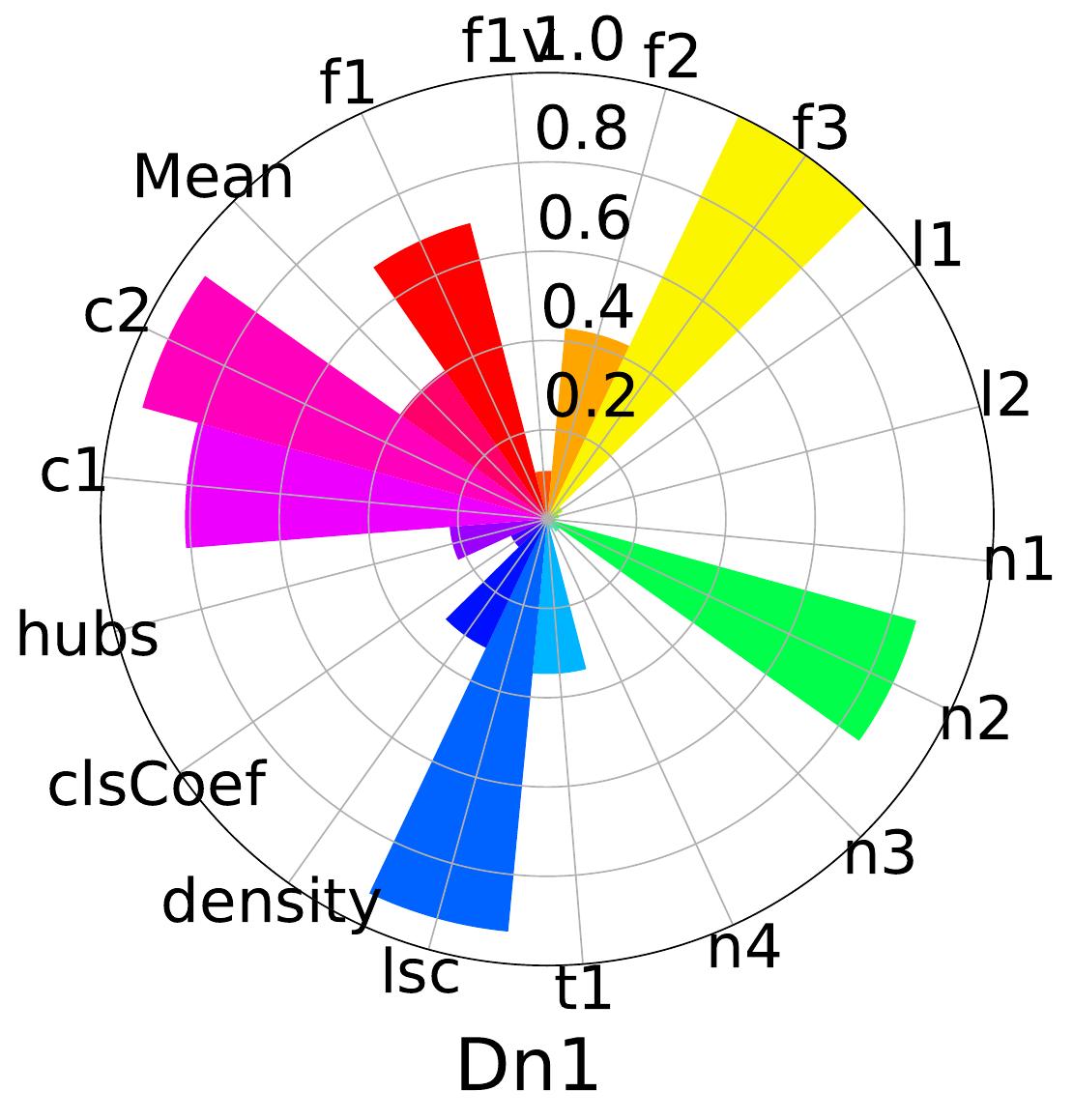}
\includegraphics[width=0.24\textwidth]{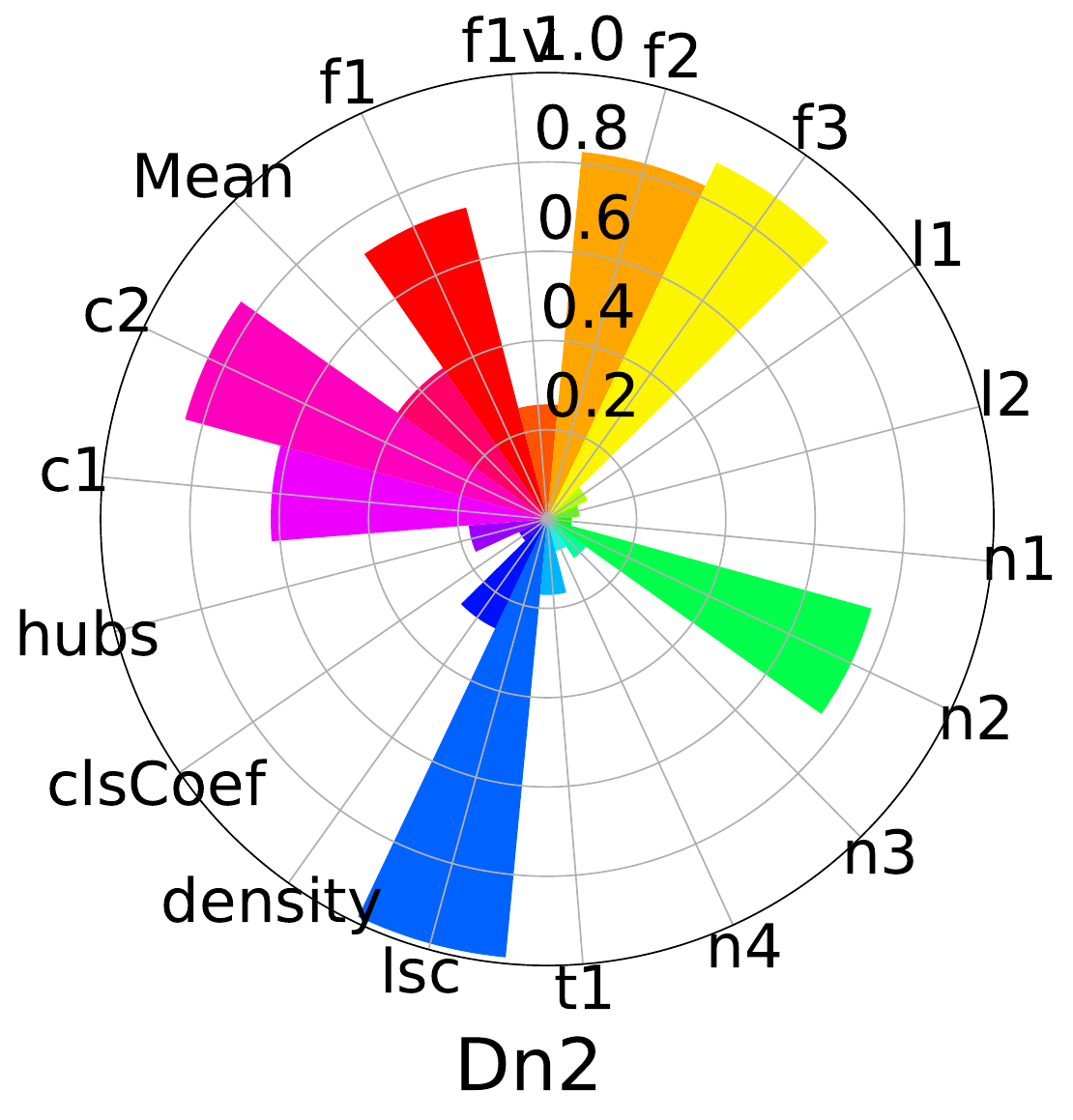}
\includegraphics[width=0.24\textwidth]{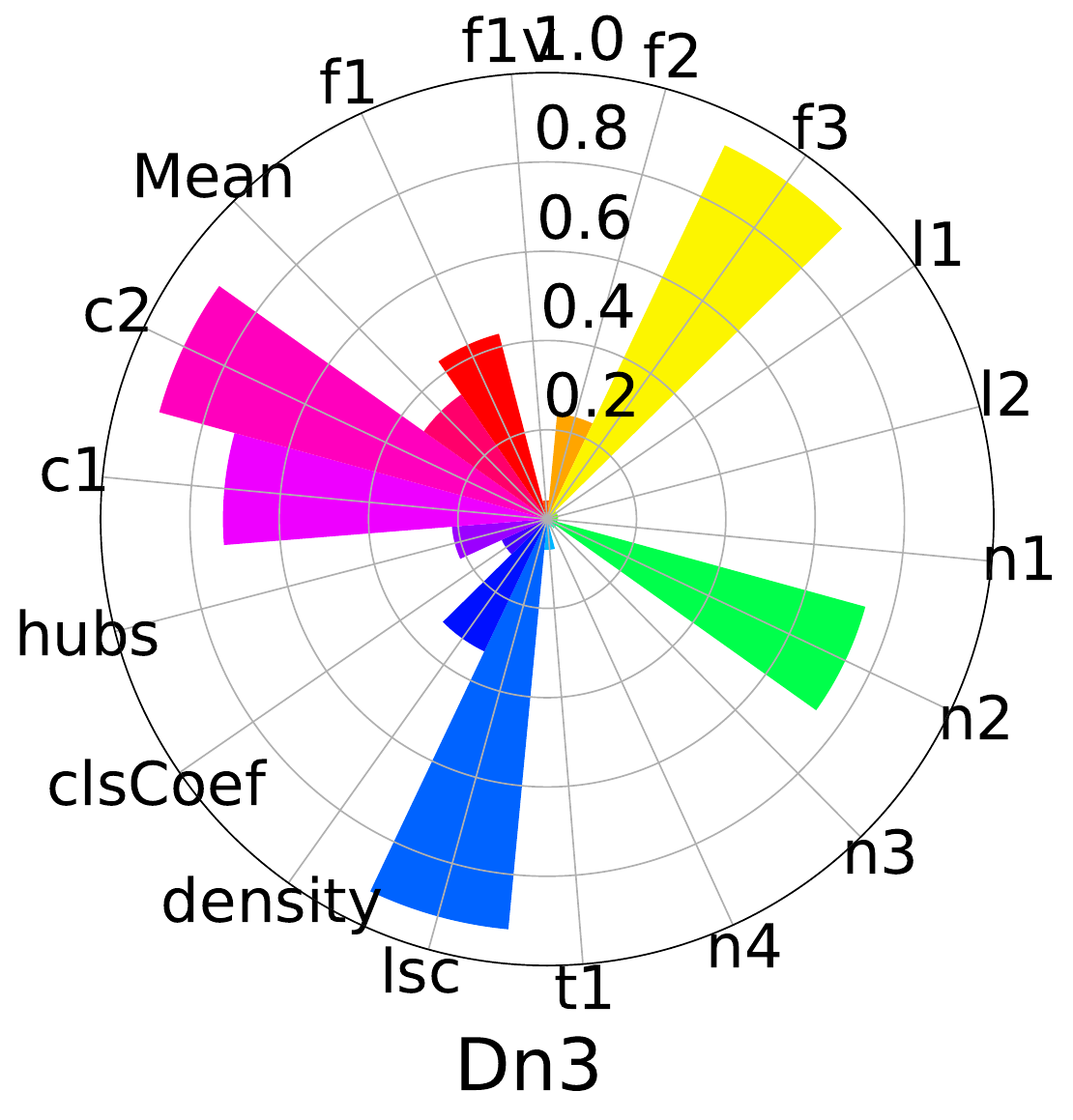}
\includegraphics[width=0.24\textwidth]{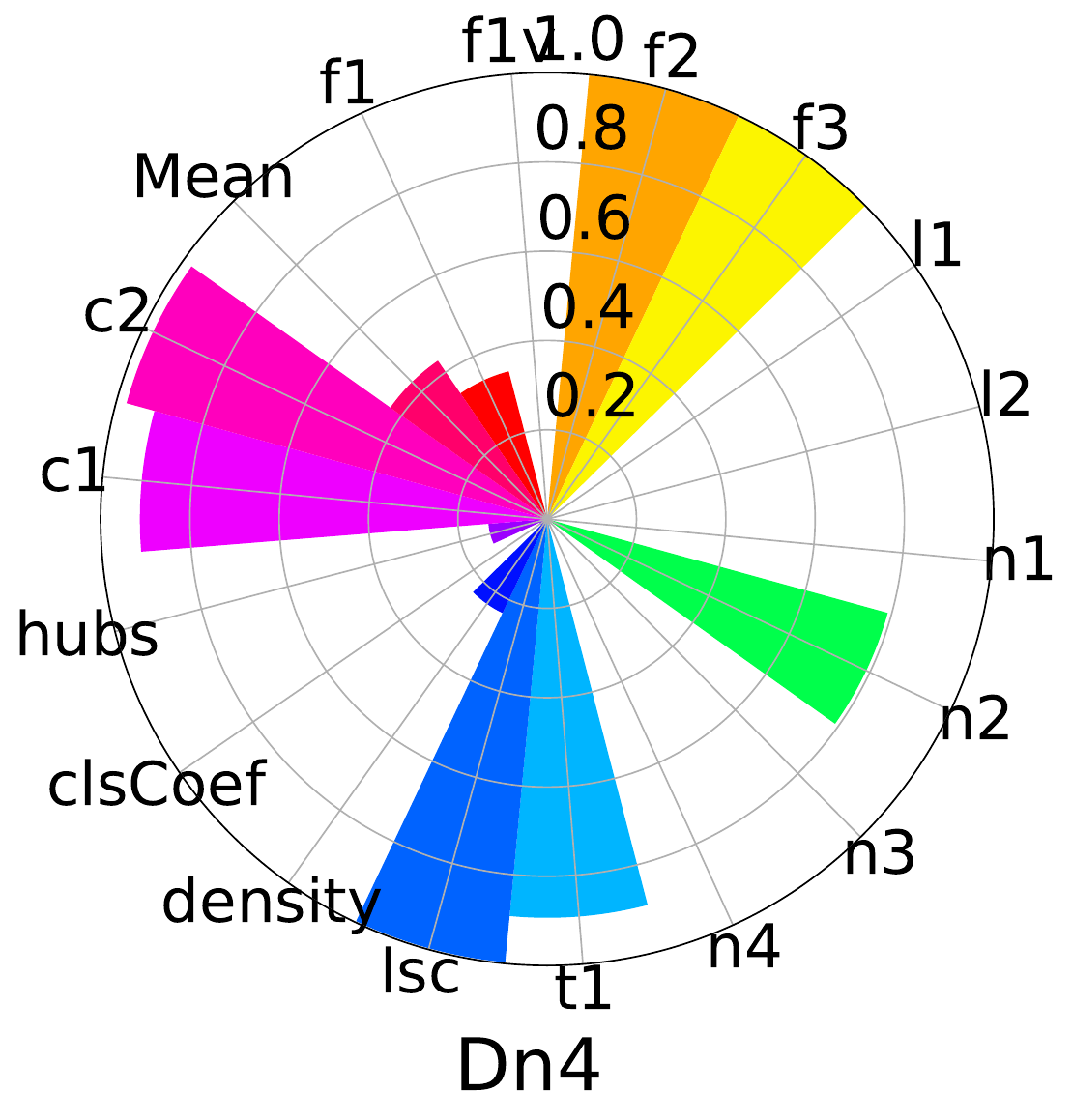}
\includegraphics[width=0.24\textwidth]{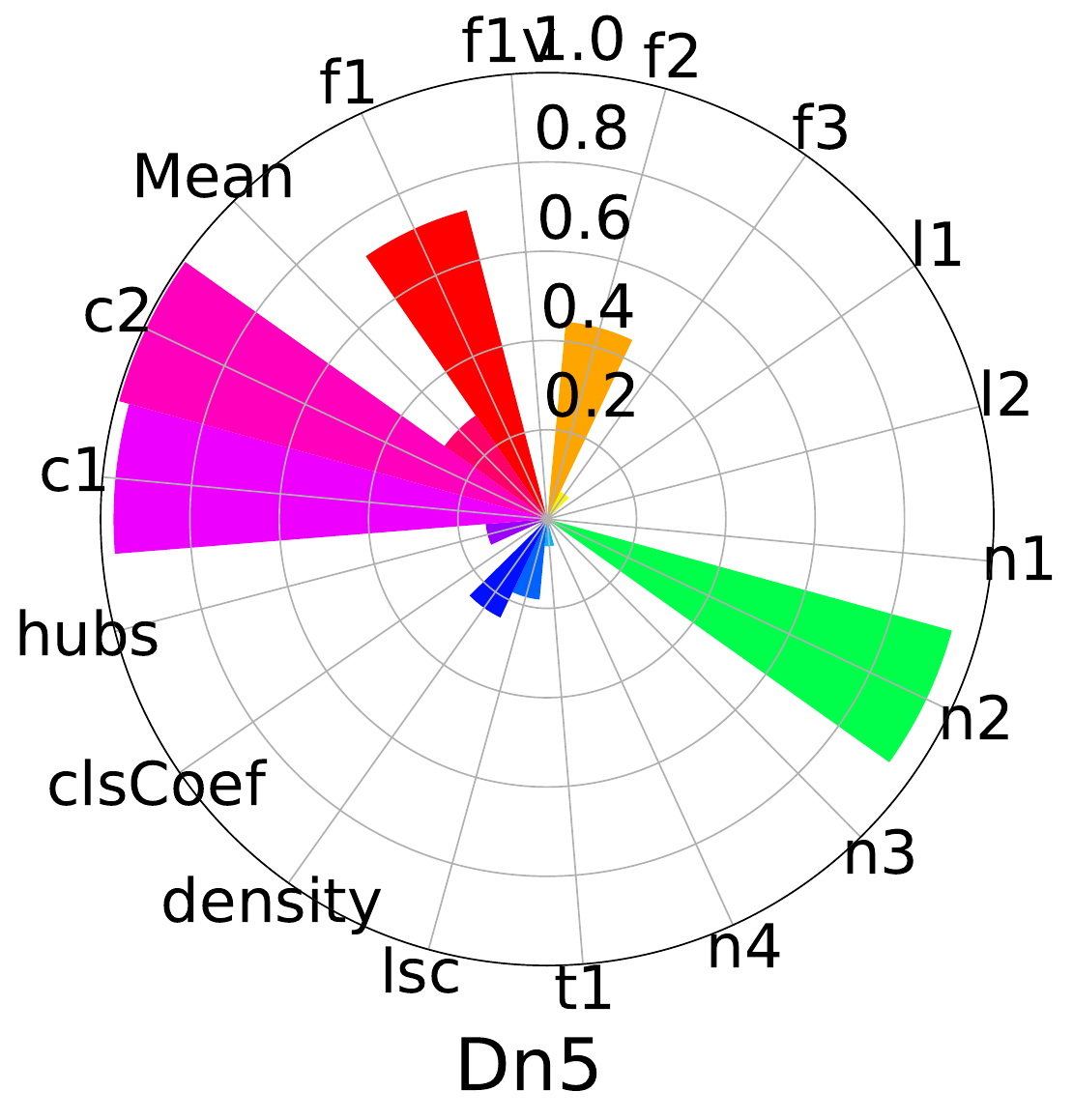}
\includegraphics[width=0.24\textwidth]{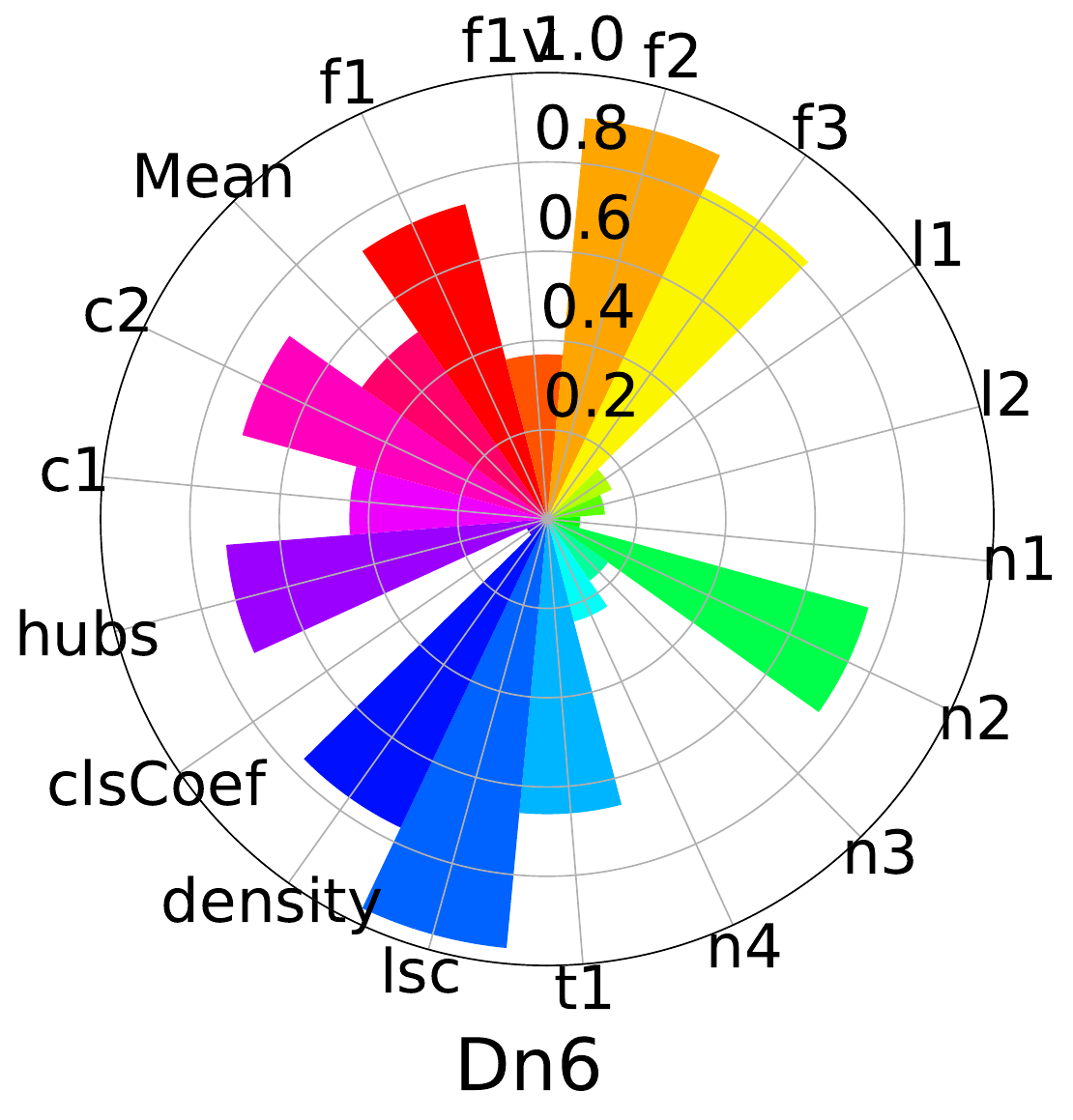}
\includegraphics[width=0.24\textwidth]{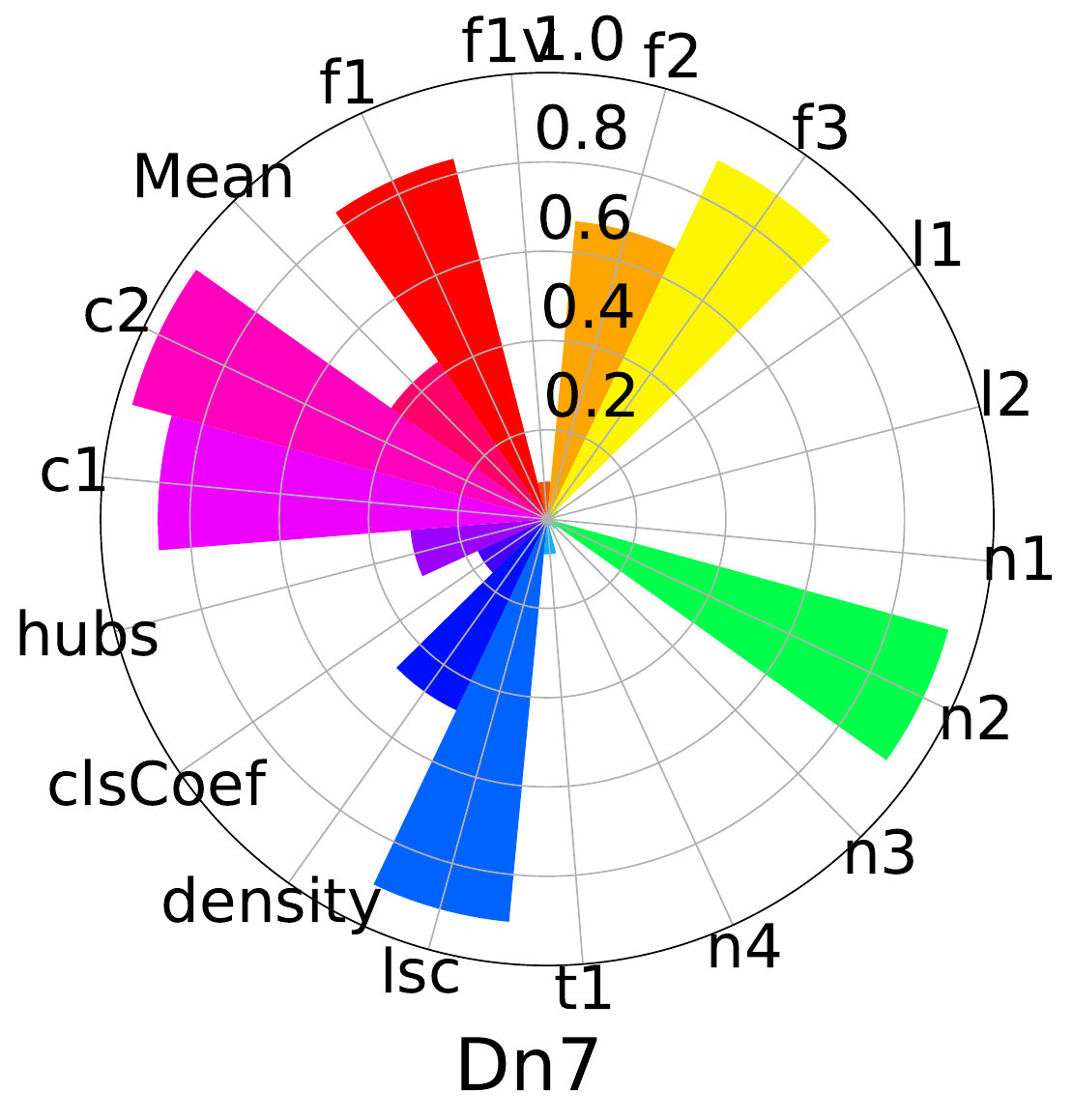}
\includegraphics[width=0.24\textwidth]{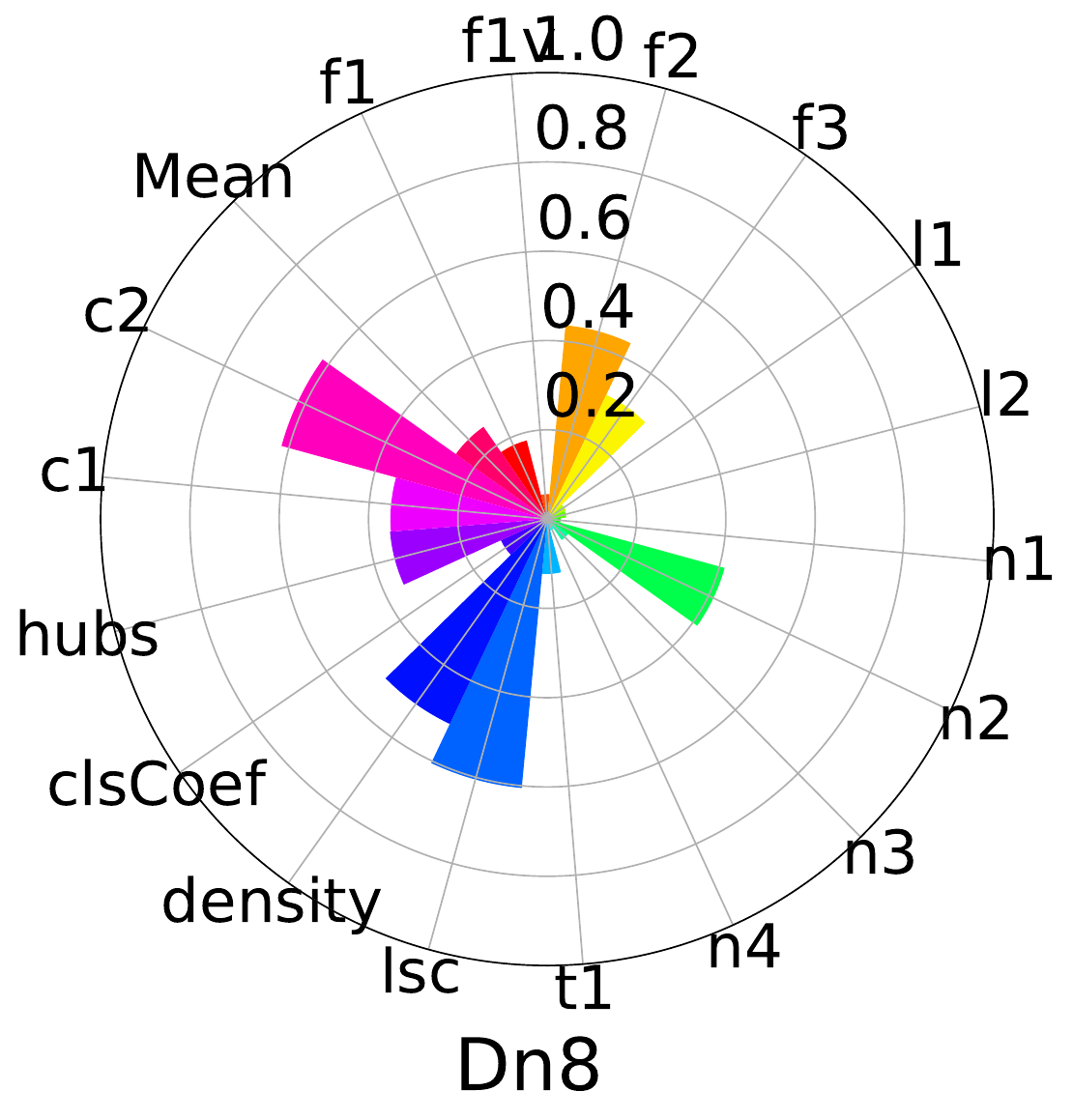}
\caption{Complexity measures per dataset in Table \ref{tb:newDatasets}.}
\label{fig:complexityFiguresNew}
\end{figure*}

\begin{table}[t]\centering
\renewcommand{\tabcolsep}{3pt}
\caption{{
F1 per method and dataset in Tb. \ref{tb:newDatasets}. Hyphen indicates insufficient memory. 
The highest F1 per category and dataset  is in bold.}}
{\small
\begin{tabular}{ | l | cccccccc|}\cline{1-9}

\multicolumn{1}{|c|}{}& $D_{n1}$ & $D_{n2}$ & $D_{n3}$ & $D_{n4}$ & $D_{n5}$ & $D_{n6}$ & $D_{n7}$ & $D_{n8}$ \\
\hline
\hline
\multicolumn{9}{|c|}{(a) DL-based matching algorithms}\\
\hline
DM(15) & 70.49 & 52.01 & 99.32 & 90.50 & 59.88 & 69.95 & 56.57 & 95.10 \\
DM(40)	& 71.43& 56.15 & 99.32 & 89.73 & 63.18 & 67.28 & 57.14 & 93.51 \\
\hline
\hline
DITTO(15) & 86.43 & 38.10 & - & 86.50 & 66.82 & - & \textbf{71.73} & 95.31 \\
DITTO(40) & - & 67.95 & - & 86.84 & 0.59 & - & 63.91 & 95.04 \\
\hline
\hline
EM-B(15) & 84.68 & 64.39 & 99.43 & 91.91 & \textbf{67.14} & 77.78 & 67.56 & 93.16 \\
EM-B(40) & 85.88 & 65.38 & 99.54 & 91.26 & - & 78.54 & 62.86 & 92.98 \\
\hline
EM-R(15) & \textbf{91.35} & 65.49 & 99.43 & \textbf{92.51} & - & \textbf{79.28} & 67.55 & 94.81 \\
EM-R(40) & - & \textbf{70.12} & \textbf{99.54} & - & - & 77.56 & 63.29 & 93.21 \\
\hline
\hline
GNEM(10) & - & - & 99.43 & - & - & - & 62.89 & \textbf{95.53} \\
GNEM(40) & - & - & 99.43 & - & - & - & 60.05 & 95.34 \\
\hline
\hline
HM(10)	& - & - & - & 91.39 & 58.52 & - & 63.31 & - \\
HM(40) & -	 & - & - & 91.39 & 58.52 & - & 63.31 & - \\
\hline
\hline
\multicolumn{9}{|c|}{(b) Non-neural, non-linear ML-based matching algorithms}\\
\hline
MG-DT & 52.55 & 41.67 & 99.54 & 91.69 & 59.72 & 56.84 & 50.00 & 91.73 \\
MG-LR & 43.84 & 39.19 & 99.66 & 91.25 & 59.64 & \textbf{61.10} & 55.65 & 91.06 \\
MG-RF & \textbf{57.42} & \textbf{44.44} & \textbf{99.66} & \textbf{92.64} & \textbf{61.11} & 59.74 & 61.18 & \textbf{93.82} \\
MG-SVM & - & - & 98.20 & 91.01 & 59.34 & 61.01 & \textbf{61.67} & 88.70 \\
\hline
\hline
ZeroER & 32.66 & 22.14 & 99.32 & 43.32 & 0.50 & 53.76 & 61.52 & 84.14 \\
\hline
\hline
\multicolumn{9}{|c|}{(c) Non-neural, linear supervised matching algorithms}\\
\hline
SA-ESDE & 47.79 & 40.35 & 98.64 & \textbf{85.75} & 47.86 & 43.98 & 34.41 & 88.24\\
SAQ-ESDE & 44.59 & 41.41 & 98.64 & 82.80 & 49.93 & 43.96 & 37.77 & 88.57\\
SAF-ESDE & 20.27 & 28.09 & 98.75 & 82.69 & 49.25 & 44.20 & 29.08 & 76.38\\
SAS-ESDE & 47.97 & 39.58 & 98.75 & 77.41 & 49.53 & 44.22 & 35.19 & 87.47 \\
\hline
SB-ESDE & 49.62 & 46.87 & \textbf{99.66} & 61.95 & \textbf{58.87} & \textbf{60.50} & \textbf{66.13} & 89.95\\
SBQ-ESDE & 52.95 & \textbf{49.79} & \textbf{99.66} & 20.00 & 7.61 & 54.26 & 34.07 & \textbf{91.36} \\
SBF-ESDE & 36.64 & 36.52 & \textbf{99.66} & 20.00 & 7.61 & 53.81 & 33.40 & 85.41\\
SBS-ESDE & \textbf{53.65} & 45.39 & \textbf{99.66} & 20.00 & 7.61 & 53.60	 & 33.43 & 88.29
\\
\hline
\end{tabular}
}
	\label{tb:f1PerNewDataset}
\end{table}

\textbf{Complexity Measures.} TThese are presented in Figure \ref{fig:complexityFiguresNew}.
The average complexity score is lower than 0.40 for $D_{n3}$ and $D_{n8}$ (0.339 and 0.251, respectively), in line with the degree of linearity, but it exceeds this threshold for $D_{n4}$ (0.431). This is caused by its very low imbalance ratio (see also Table \ref{tb:newDatasets}), which results in high scores for the class imbalance measures and some of the feature overlapping ones. In all other cases, though, it exhibits the (second) lowest score among all datasets, including the established ones. Note also that $D_{n5}$ yields a very low average score (0.282), that surpasses only $D_{n8}$. This indicates a rather easy classification task, because of the very low values ($\ll$0.2) for 9 out of the 17 complexity measures. Hence, only \textit{$D_{n1}$, $D_{n2}$, $D_{n6}$ and $D_{n7}$ correspond to challenging, non-linearly separable matching benchmarks}.

\subsubsection{Practical Measures}
For the DL-based matching algorithms, we use the same configurations as for the existing benchmarks in Table \ref{tb:f1PerDataset}, due to their high performance, which matches or surpasses the literature. The results appear in Table \ref{tb:f1PerNewDataset}, while Figure \ref{fig:complexityFiguresNew} reports the non-linear boost (NLB) and the learning-based margin (LBM). 

For the datasets marked as challenging by the theoretical measures ($D_{n1}$, $D_{n2}$, $D_{n6}$ and $D_{n7}$), both practical measures take values well above 5\%. LBM takes its minimum value (8.7\%) over $D_{n1}$, as EMTransformer with RoBERTa performs exceptionally well, outperforming all other DL-based algorithms by at least 5\% and all others by at least 34\%. NLB takes its minimum value over $D_{n7}$, because the F1 for SB-ESDE is double as that of all other linear algorithms, reducing is distance from the top DL-based one to 5.6\%.

Among the remaining datasets, all algorithms achieve perfect performance over $D_{n3}$, thus reducing both practical measures to 0. The same applies to a lesser extent to $D_{n8}$, where both measures amount to $\sim$4.3\%. In $D_{n4}$ and $D_{n5}$, both practical measures exceed 5\% to a significant extent. The reason for the former is that the best DL- and ML-matchers lie in the middle between the perfect F1 and the best linear algorithm, whose performance matches the degree of linearity. For $D_{n5}$, the practical measures are in line with the degree of linearity, unlike the complexity measures, which suggest low levels of difficulty.

Overall, the practical measures suggest that, with the exception of $D_{n3}$ and $D_{n8}$ (which exhibit linear separability of their classes), \emph{all other datasets are challenging enough for assessing the relative performance of DL-based matchers}.

\begin{figure}[t]
\centering
\includegraphics[width=0.47\textwidth]{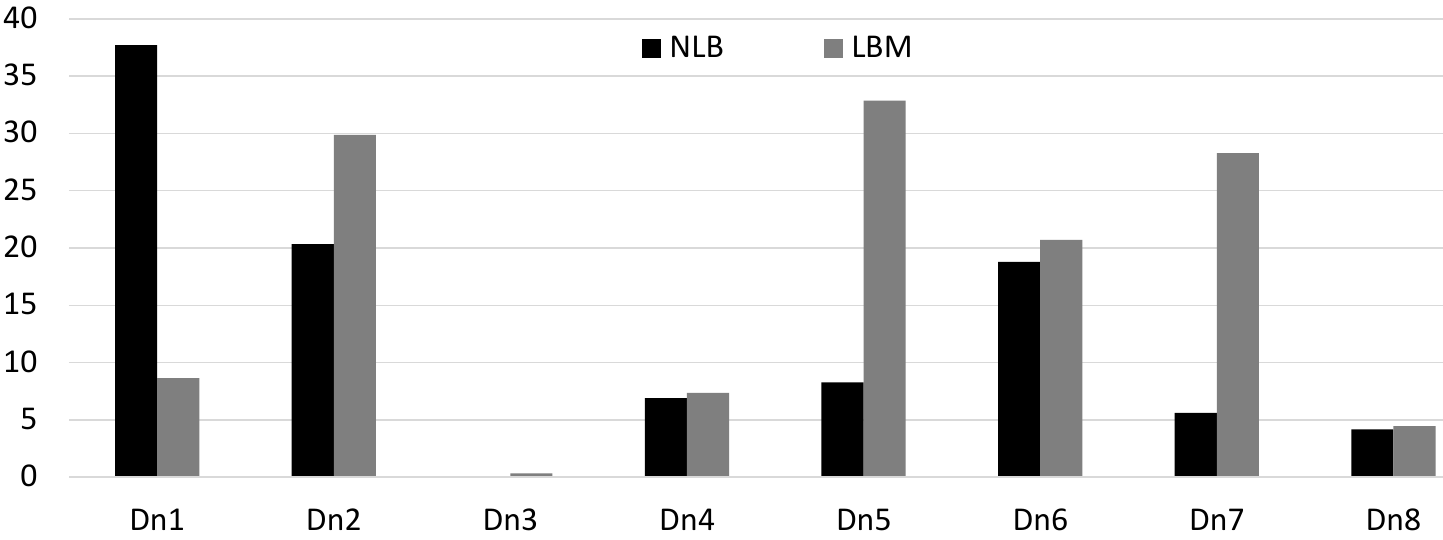}
\caption{Practical measures per dataset in Table \ref{tb:newDatasets}.}
\label{fig:practicalMeasuresNew}
\end{figure}

\subsection{Memory Consumption and Run-times.} 
\begin{figure*}[t]
    \centering
    \includegraphics[width=0.3\textwidth]{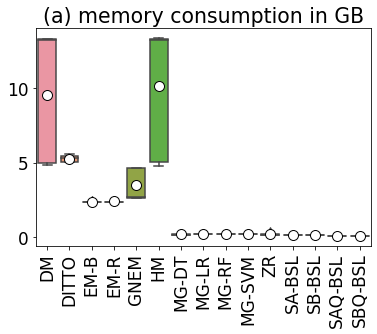}
    \includegraphics[width=0.3\textwidth]{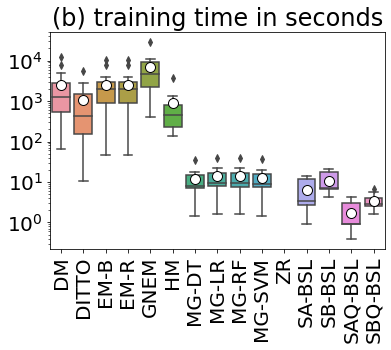}
    \includegraphics[width=0.315\textwidth]{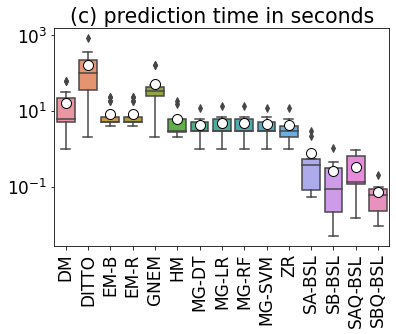}
    \caption{Memory and time requirements per algorithm across the datasets $D_{s4}$, $D_{s6}$, $D_{d3}$, $D_{d4}$, $D_{t1}$, $D_{n1}$, $D_{n2}$, $D_{n6}$, $D_{n7}$. 
    }
    \label{fig:spaceAndTimeReqs}
\end{figure*}

We now investigate the space and time requirements of all matching algorithms across the nine selected datasets. The results are shown in Figure \ref{fig:spaceAndTimeReqs}.

Starting with Figure \ref{fig:spaceAndTimeReqs}(a), we observe that all baseline methods occupy $\leq$200 MB, on average, because their space complexity is dominated by the input data. In contrast, \textit{the space requirements of the DL-based algorithms is determined by their embeddings and their learned model, which together increase the memory consumption by an entire order of magnitude}. The most frugal among them is EMTransformer, which occupies $\sim$2.5 GB on every dataset, regardless of the underlying language model. The reason is that it involves a simple local, heterogeneous operation, which leverages language models in a straightforward manner. Hence, its space complexity is dominated by its 768-dimensional embeddings vectors.

The next most memory efficient approach is DeepMatcher, which requires 10 GB, on average ($\sim$14 GB on every existing and $\sim$5.3 GB on every new dataset). This is determined by the 300-dimensional fastText vectors and the hybrid attribute summarization.

The remaining DL-based algorithms run out of memory in at least one case, making it hard to assess their actual memory consumption. DITTO fails in just three cases, while requiring $\sim$5.5 GB in all other cases. This is determined by the 768-dimensional RoBERTa vectors and the additional records generated during the data augmentation process. GNEM fails in three new datasets for both number of epochs, while requiring $\sim$3.7 GB, on average, in all other cases. The higher memory consumption of GNEM should be attributed to its global operation, which uses a graph to model the relations between all candidate pairs. Finally, HierMatcher has the highest memory consumption among all DL-based algorithms, as it is able to process only four of the nine benchmark datasets. In these cases, it requires 10.6 GB, on average. Its high space complexity should be attributed to its complex, hierarchical neural network, which computes contextual fastText vectors for each token and performs cross-attribute token alignment in one of its layers.

Regarding the training times, their distribution is shown in Figure \ref{fig:spaceAndTimeReqs}(b). As indicated by the log-scale of the vertical axis, the baseline methods are more efficient by at least two orders of magnitude. Magellan requires just 13 seconds, on average, with minor variations among the various classifiers. Our threshold-based baselines are even faster, fluctuating between 1 (SAQ-BSL) and 10 (SB-BSL) seconds per dataset, on average.

For the DL-based algorithms, the training time is determined to a large extent by the number of epochs. The most efficient training corresponds to HierMatcher: 8 and 22 min for 10 and 40 epochs, respectively, on average. However, these measurements probably overestimate its time efficiency, as they capture only the four datasets where HierMatcher is applicable. The second shortest training time is achieved by DITTO, with an average of 12 and 24 min for 15 and 40 epochs, resp. Both times would probably be higher if DITTO was fully applicable to all datasets. The third place involves all methods that apply to all datasets, namely DeepMatcher and EMTransformer with both language models. Despite their fundamentally different functionality, their training takes 23 (61) min for 15 (40) epochs, on average. The most time consuming training corresponds to GNEM: 65 (167) min for 10 (40) epochs, but is most likely even higher over the datasets, where it ran out of memory. Again, this is caused by its global operation.

The most important aspect of time efficiency is the prediction time in Figure  \ref{fig:spaceAndTimeReqs}(c). The difference between the non-neural and the DL-based approaches is again high, but does not exceed the two orders of magnitude. The fastest approaches are our threshold-based baselines, which process every dataset in (far) less than 1 second. They are followed by ZeroER and Magellan, which both require $\sim$4 seconds per dataset, on average. The fastest DL-based approach is HierMatcher, with a mean of $\sim$6 seconds. This is overestimated, though, due to the missing cases. The next fastest approach is EMTransformer, which takes $\sim$8 seconds per dataset, in the average case. This high efficiency stems from the simple, local operation that applies the selected language model in a straightforward manner. DeepMatcher requires twice as much time (i.e., $\sim$16s), on average, due to the higher complexity of its hybrid attribute summarization model. GNEM is much slower, with $\sim$40 seconds per dataset, on average, due to its global operation that examines the relations between all candidate pairs. Yet, the most time-consuming prediction corresponds to DITTO, which depends on the number of epochs (unlike all other DL-based matchers): it requires $\sim$109 and $\sim$219 seconds for 15 and 40 epochs, respectively. This should be attributed to the attribute normalization and the TF-IDF-based summarization of long attribute values, due to its heterogeneous functionality.
\section{Conclusions}
\label{sec:conclusions}

We make the following observations:
(1) The datasets used for benchmarking matching algorithms should be evaluated both a-priori, through their degree of linearity and complexity, and a-posteriori, through the aggregate measures summarizing the performance of linear and non-linear matchers. Excelling in all these respects is necessary for datasets that leave enough room for improvements by complex, DL-based classifiers.
(2) Most of the popular datasets used as benchmarks for DL-based matchers involve almost linearly separable candidate pairs, or are perfectly solved by most existing matching algorithms (therefore, leaving no room for improvement). 
These characteristics render these datasets \emph{unsuitable} for benchmarking matching algorithms.
(3) We experimentally demonstrate that the proposed methodology for creating new ER benchmarks leads to datasets that are better suited for assessing the benefits of DL-based matchers. 
Moreover, our methodology is flexible, and can be used to produce benchmark datasets with a tunable degree of difficulty.
In order to promote further progress, we make all our code and datasets available through a Docker image\footnote{\url{https://github.com/gpapadis/DLMatchers/tree/main/dockers/mostmatchers}}.

In the future, we plan to examine whether different configurations for DeepBlocker, or different blocking methods can extract challenging benchmarks from one of the easy datasets, like the bibliographic ones (DBLP-ACM and DBLP-Google Scholar). We actually intend to create a series of datasets that cover the entire continuum of benchmark difficulty.

\balance

\bibliographystyle{ACM-Reference-Format}
\bibliography{references}

\section*{Appendix}

\begin{figure*}[t]
\centering
\includegraphics[width=0.24\textwidth]{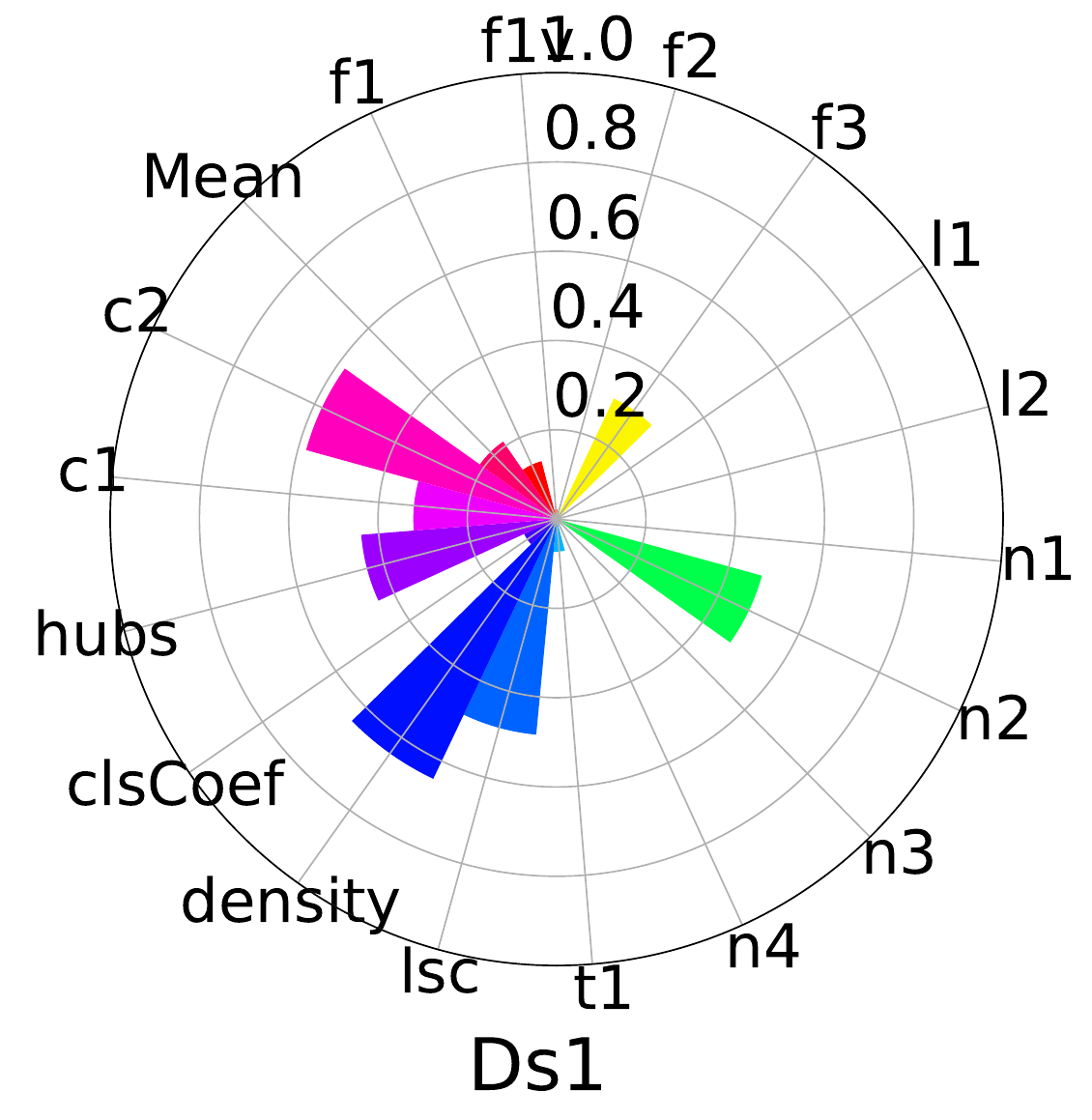}
\includegraphics[width=0.24\textwidth]{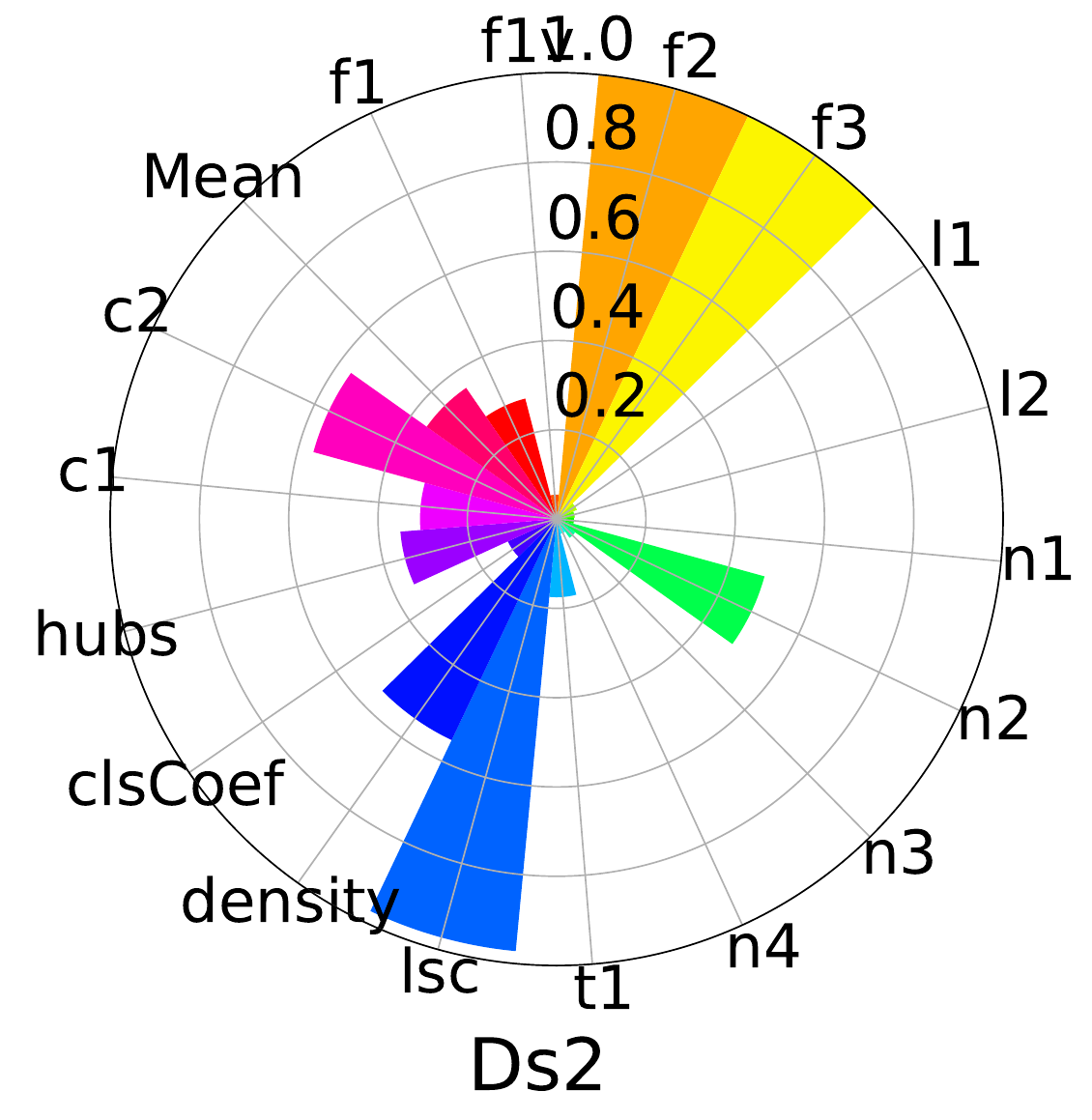}
\includegraphics[width=0.24\textwidth]{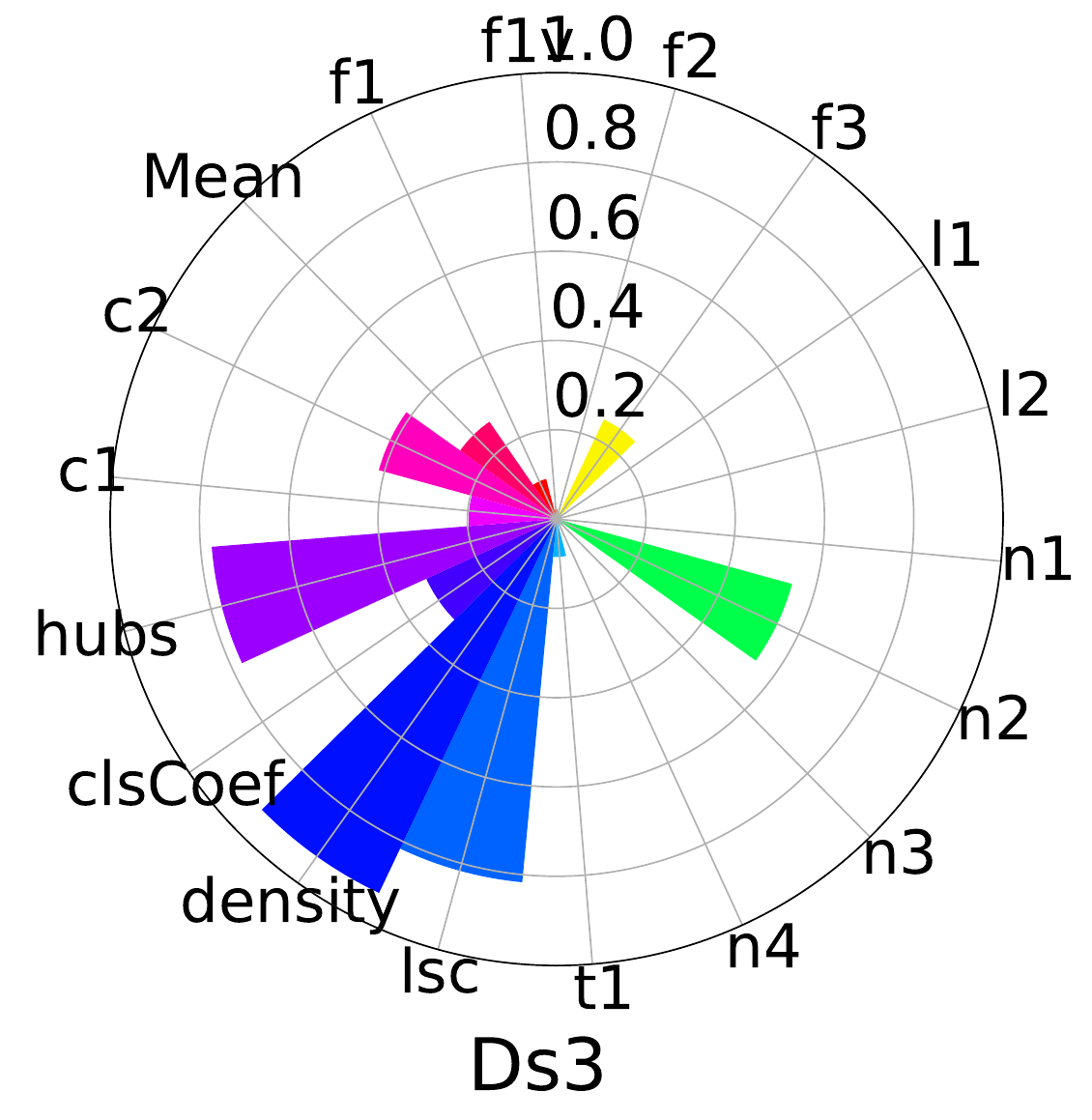}
\includegraphics[width=0.24\textwidth]{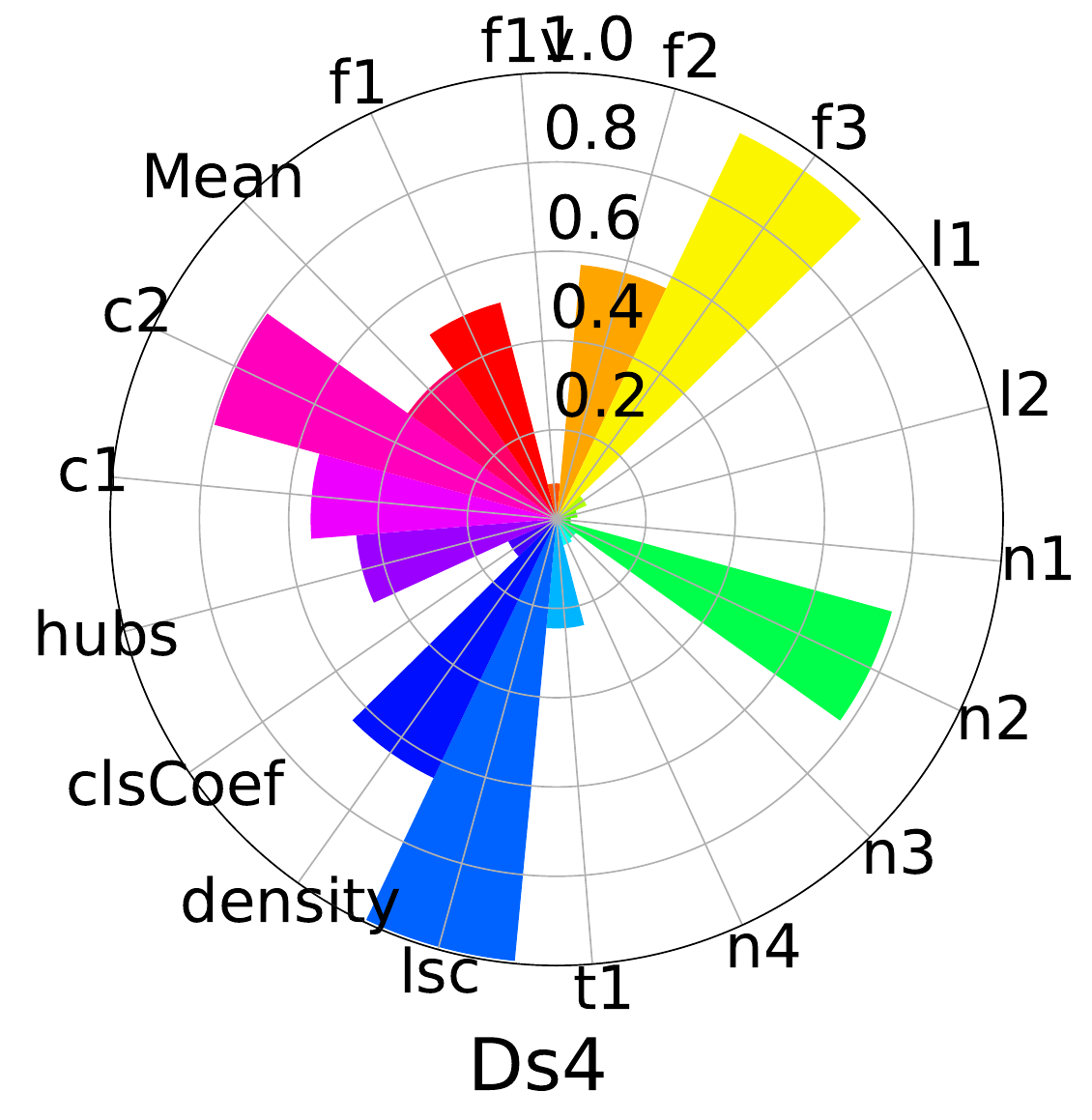}
\includegraphics[width=0.24\textwidth]{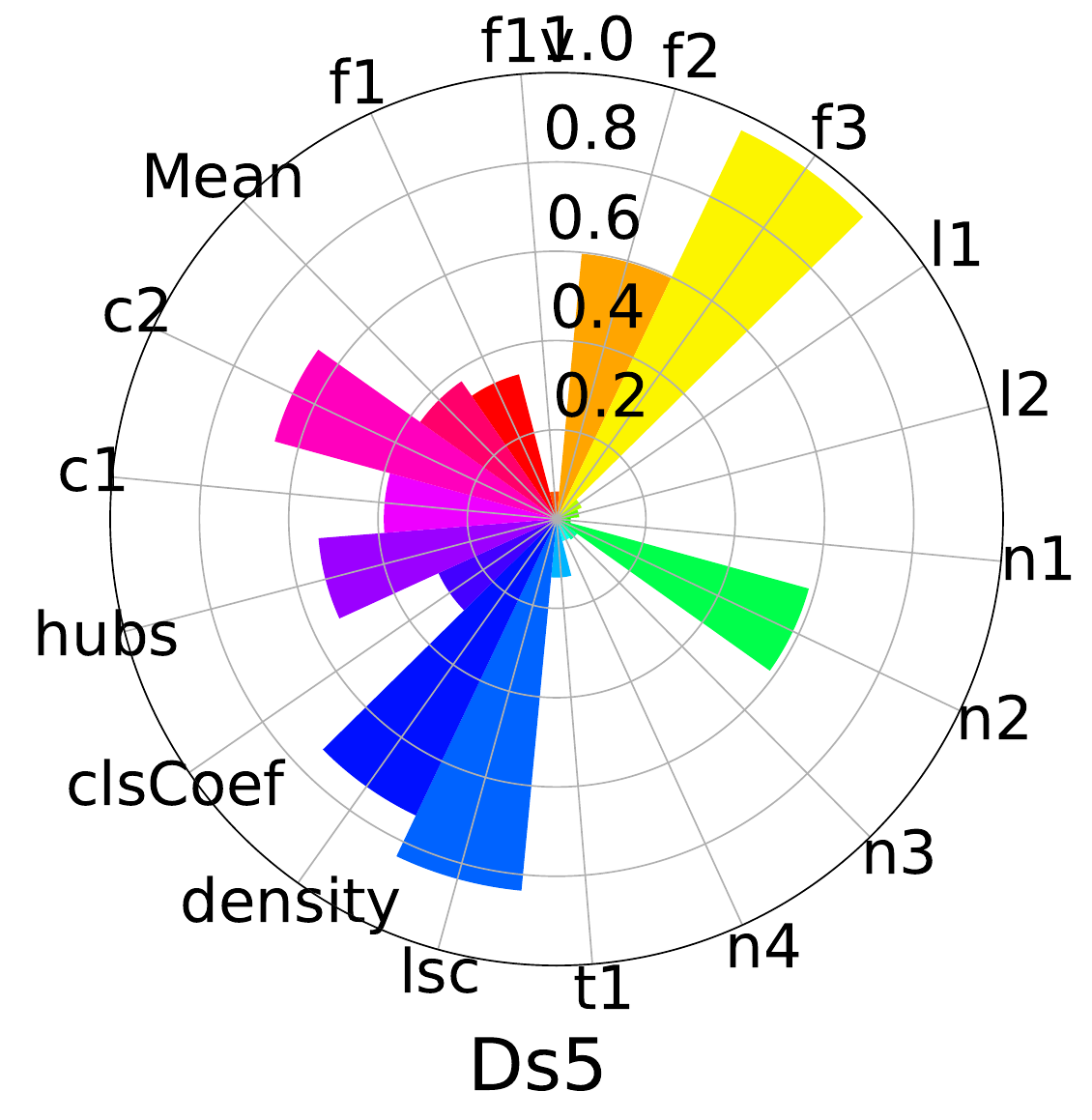}
\includegraphics[width=0.24\textwidth]{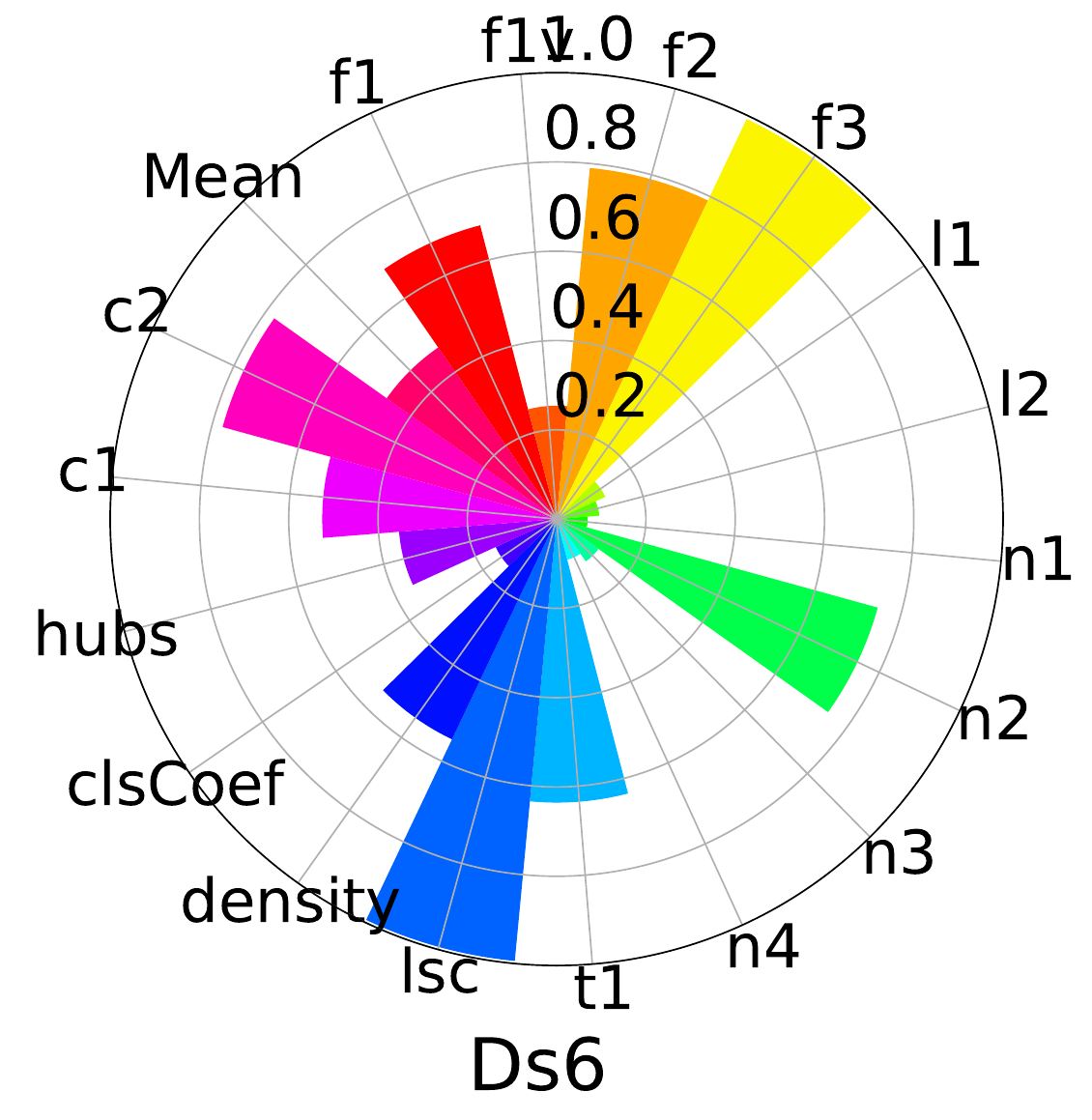}
\includegraphics[width=0.24\textwidth]{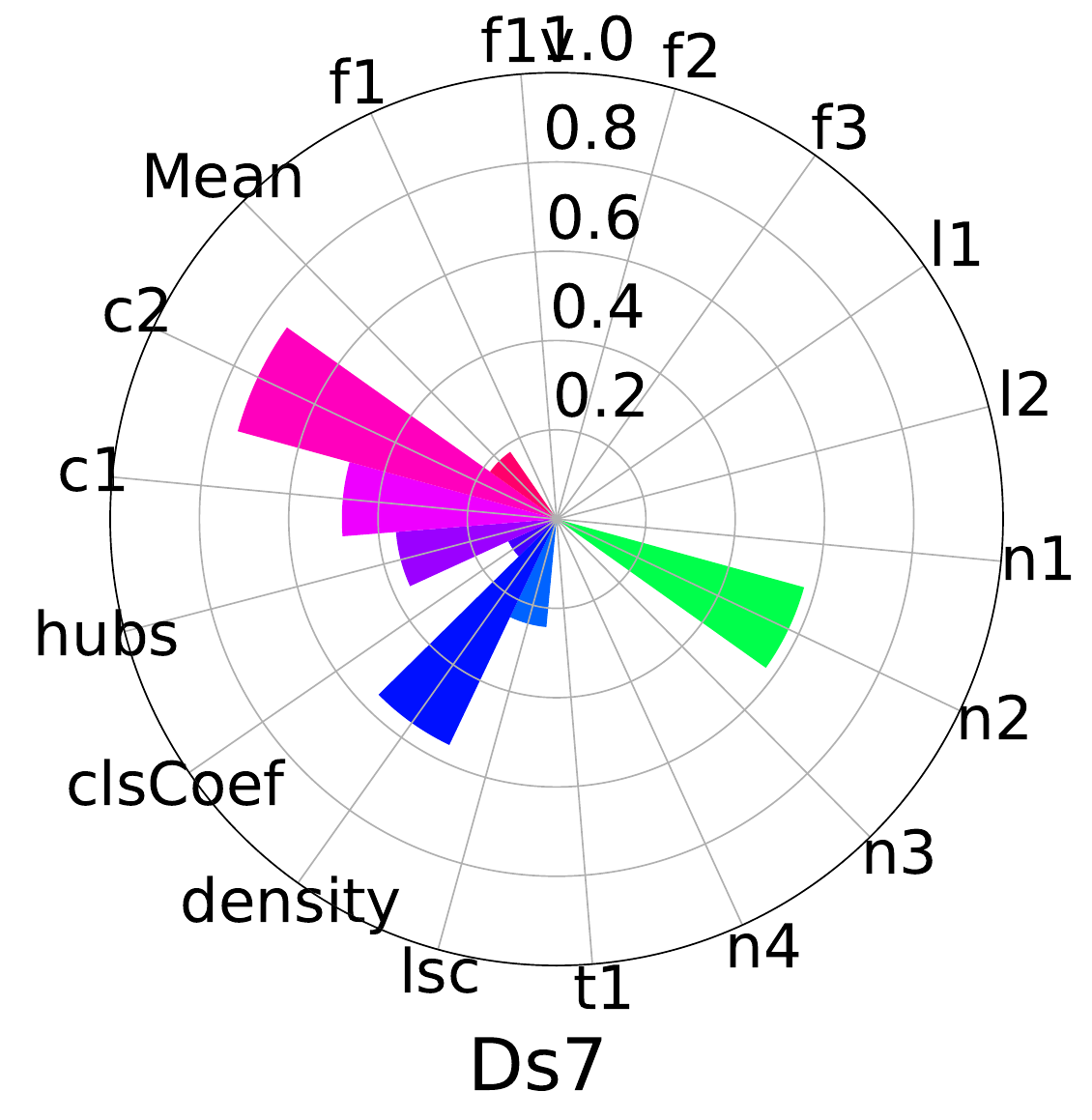}
\includegraphics[width=0.24\textwidth]{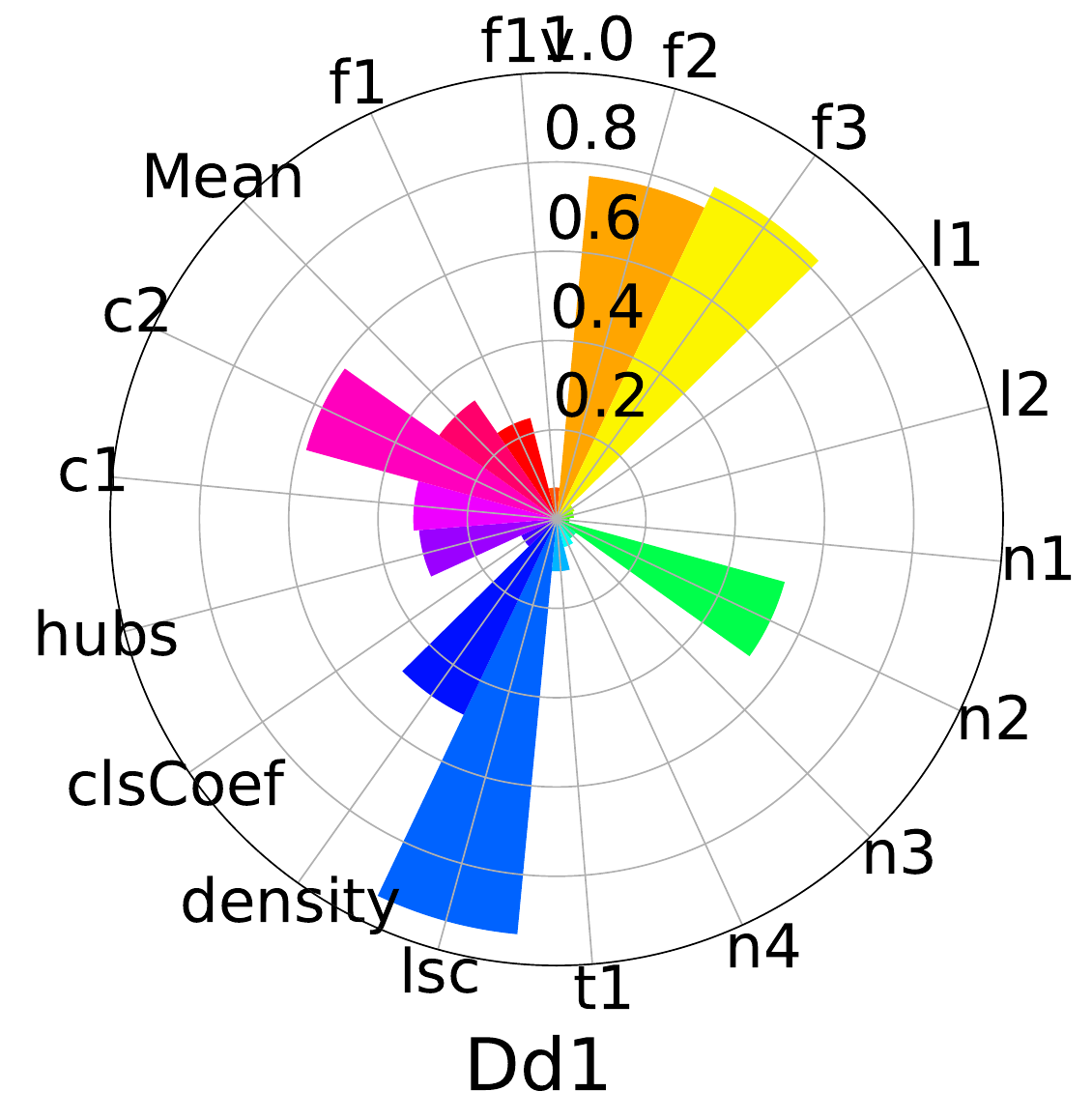}
\includegraphics[width=0.195\textwidth]{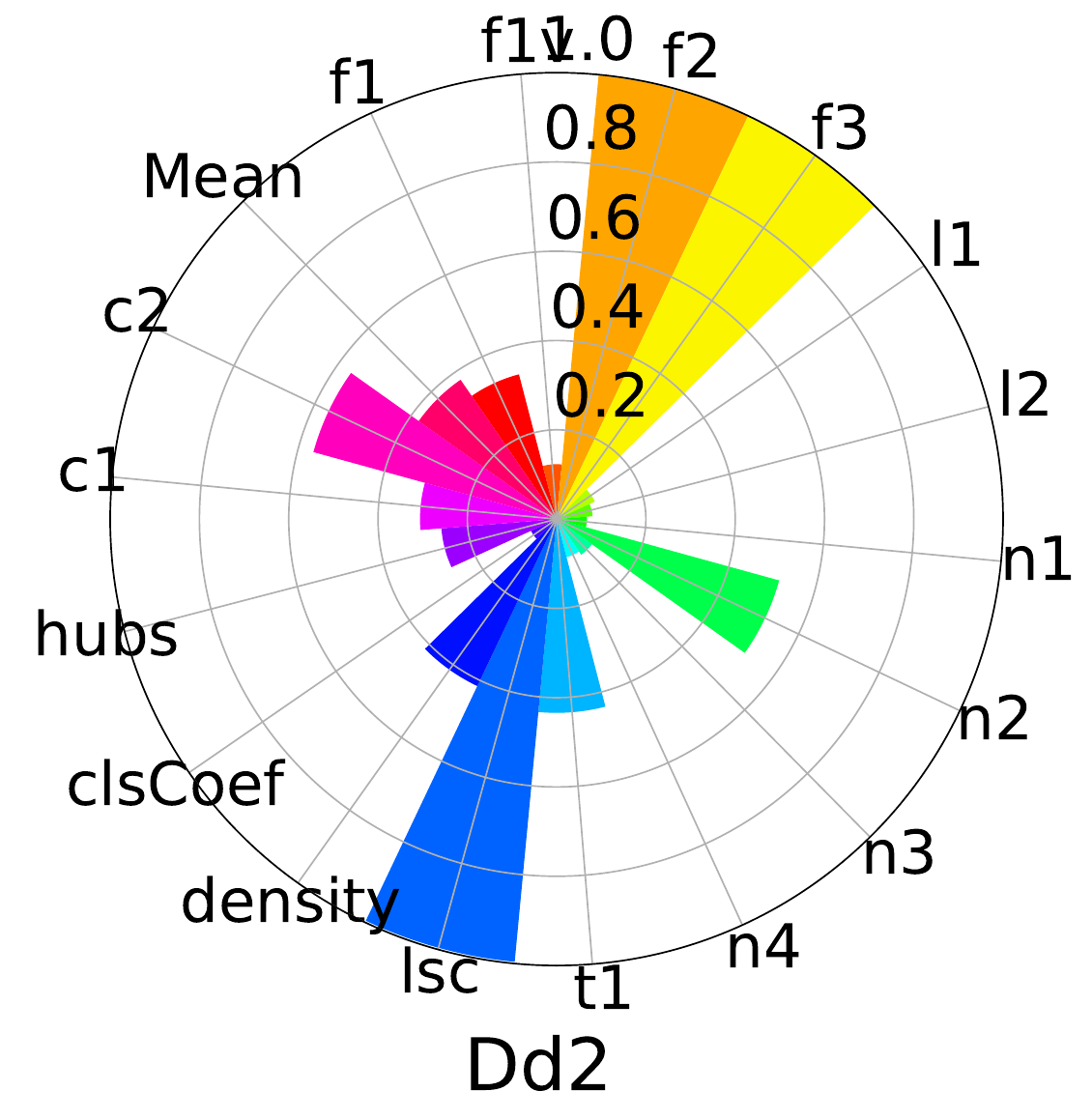}
\includegraphics[width=0.195\textwidth]{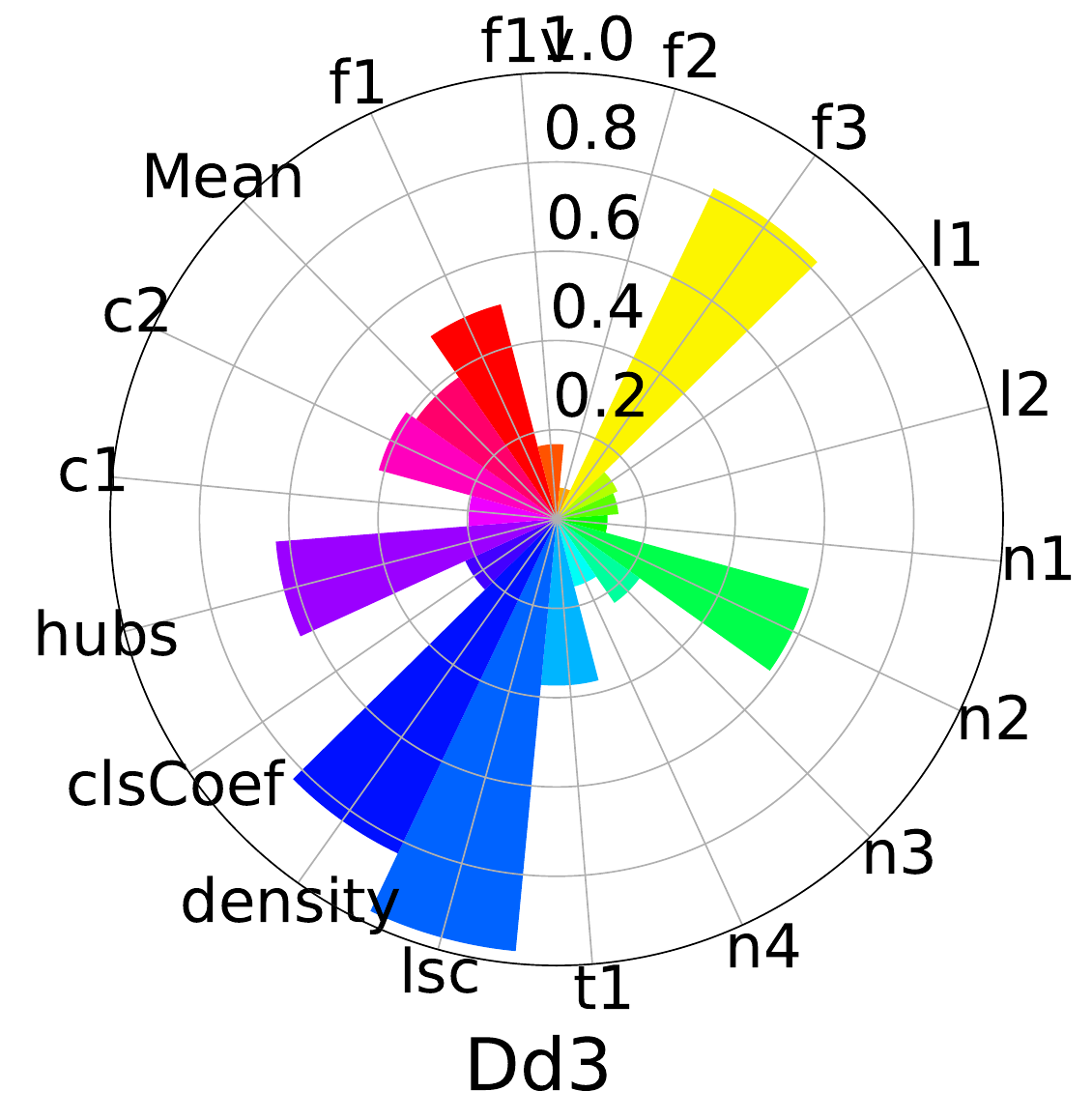}
\includegraphics[width=0.195\textwidth]{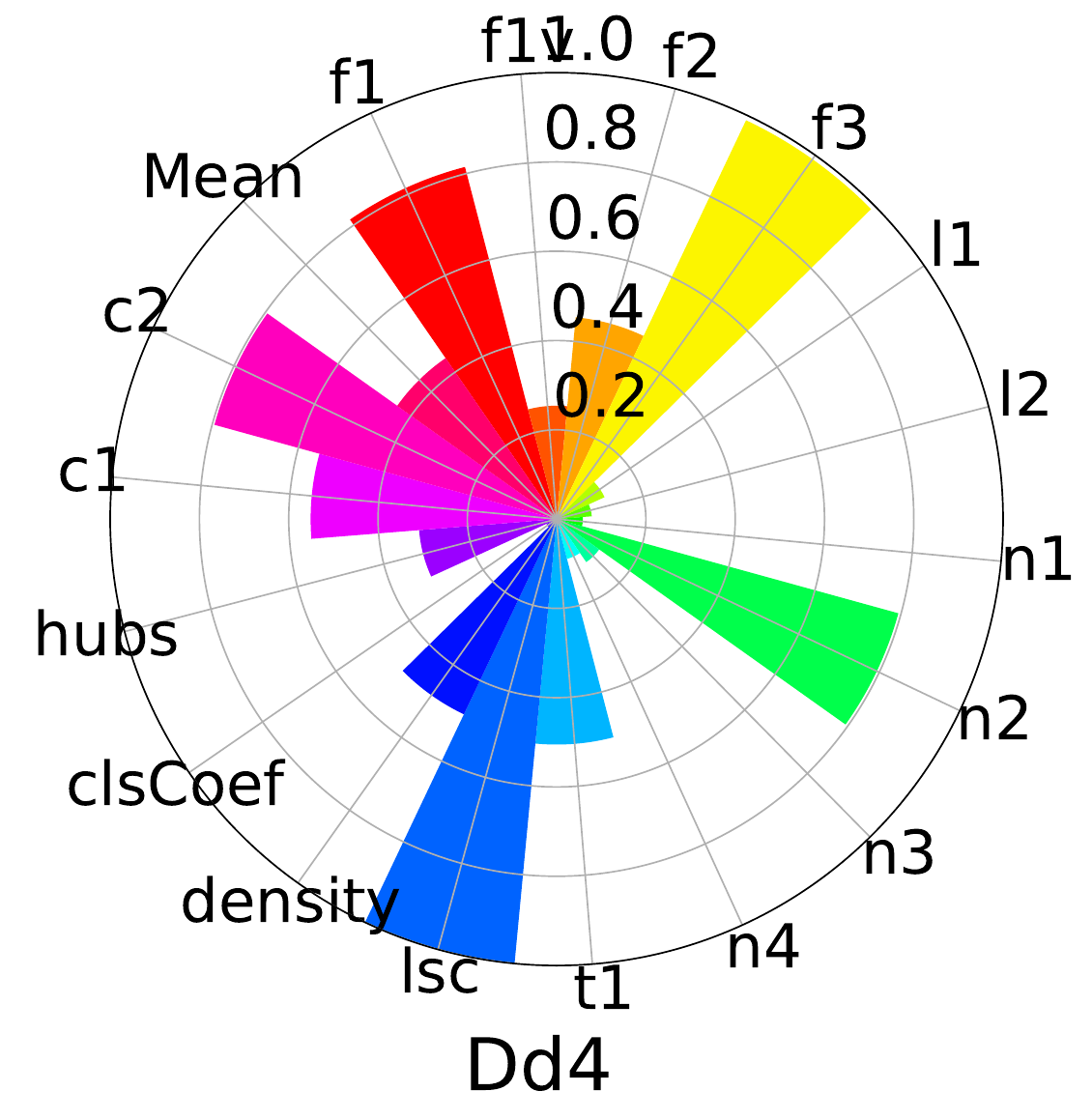}
\includegraphics[width=0.195\textwidth]{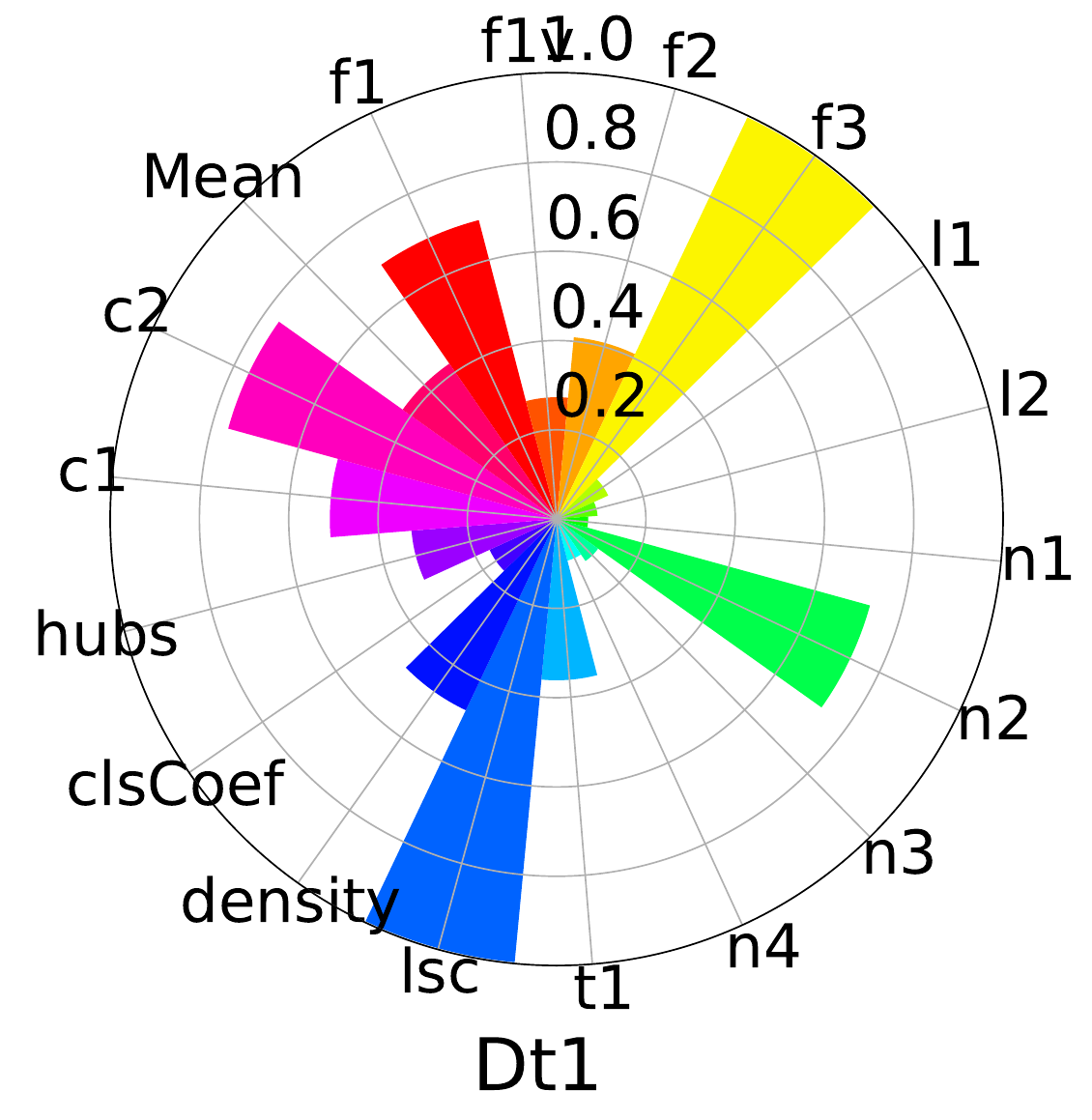}
\includegraphics[width=0.195\textwidth]{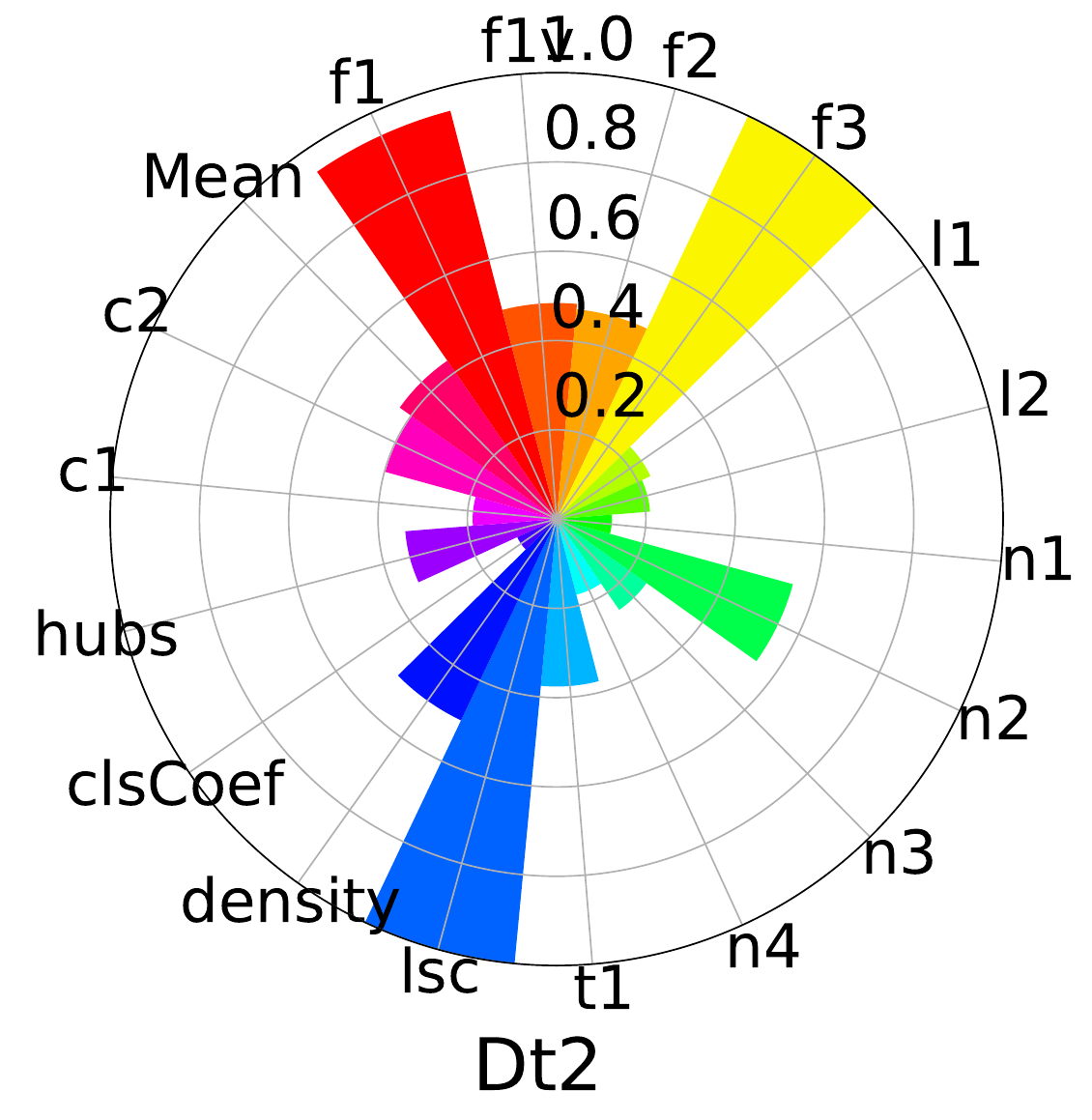}
\caption{Schema-based complexity measures per dataset in Table \ref{tb:commonDatasets}.}
\label{fig:sbComplexityFigures}
\end{figure*}

\begin{figure*}[t]
\centering
\includegraphics[width=0.24\textwidth]{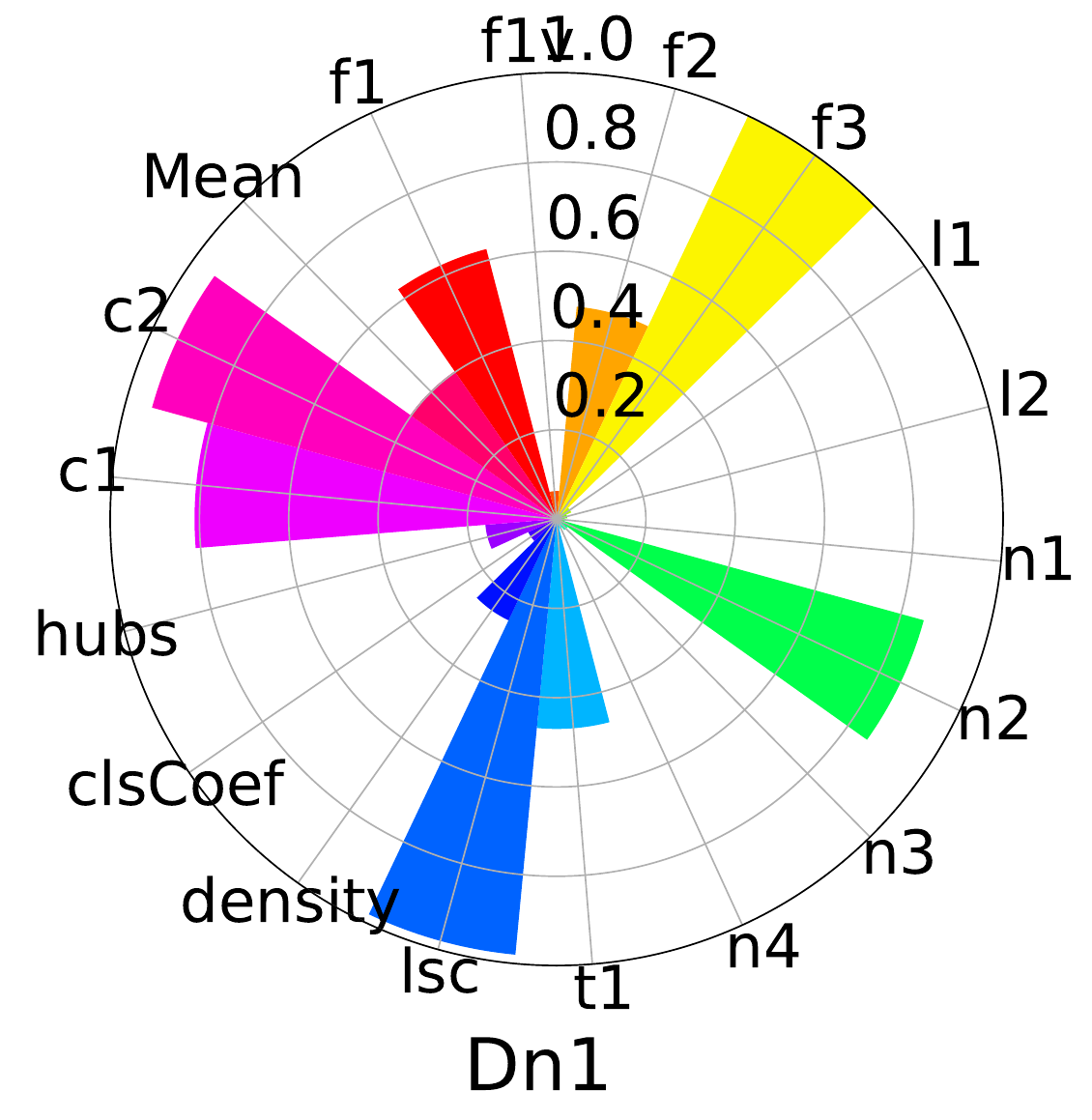}
\includegraphics[width=0.24\textwidth]{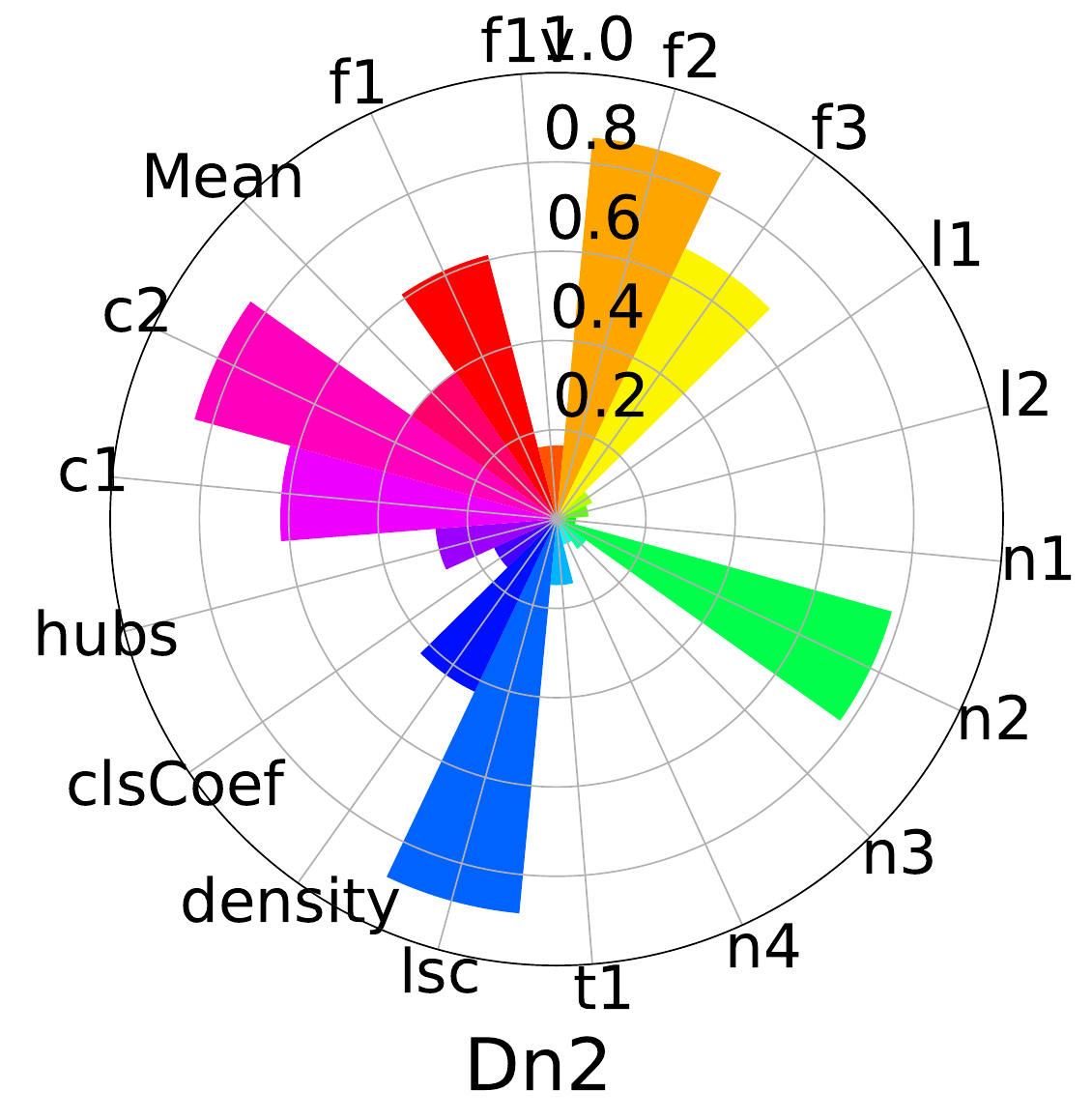}
\includegraphics[width=0.24\textwidth]{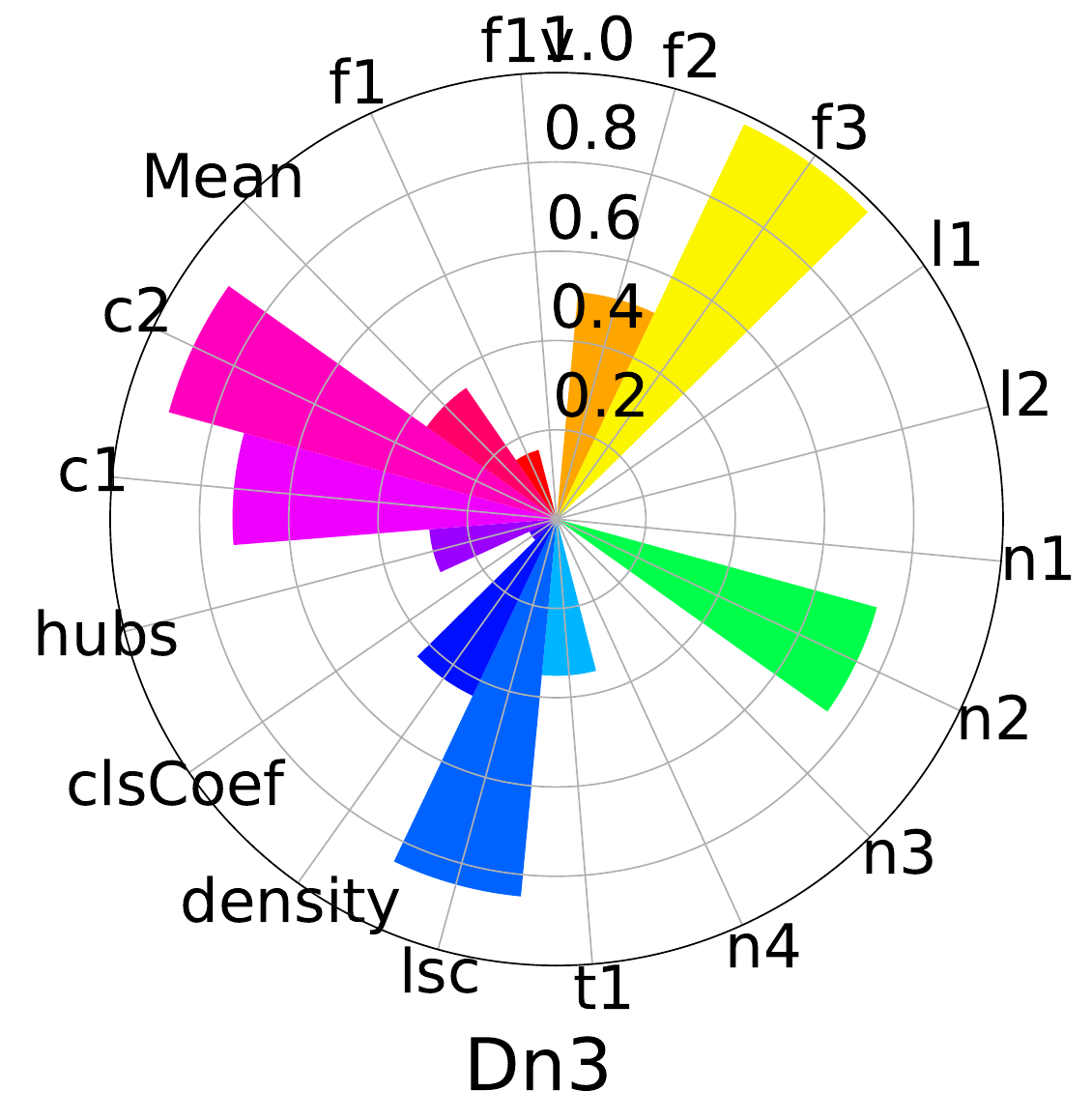}
\includegraphics[width=0.24\textwidth]{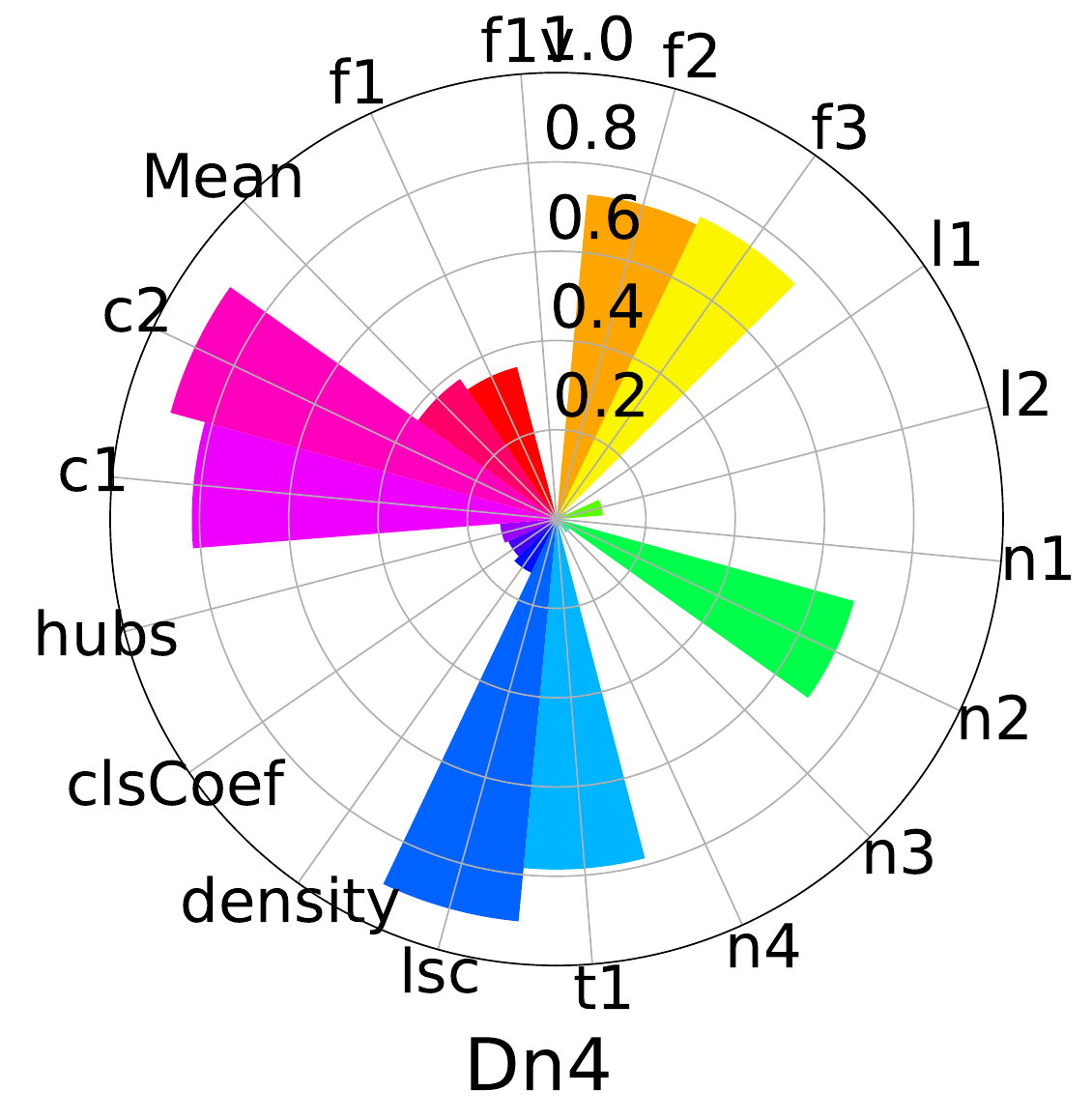}
\includegraphics[width=0.24\textwidth]{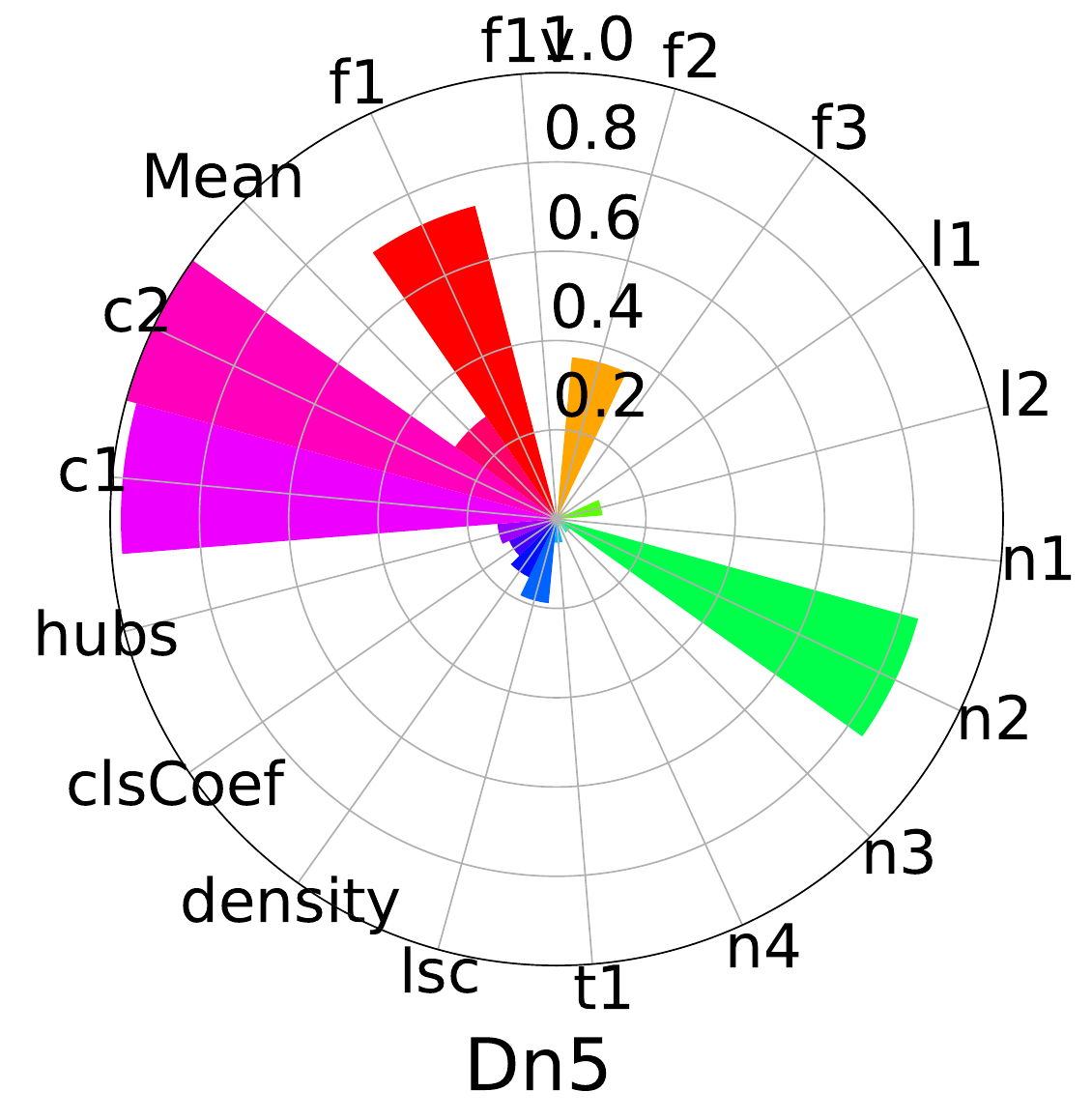}
\includegraphics[width=0.24\textwidth]{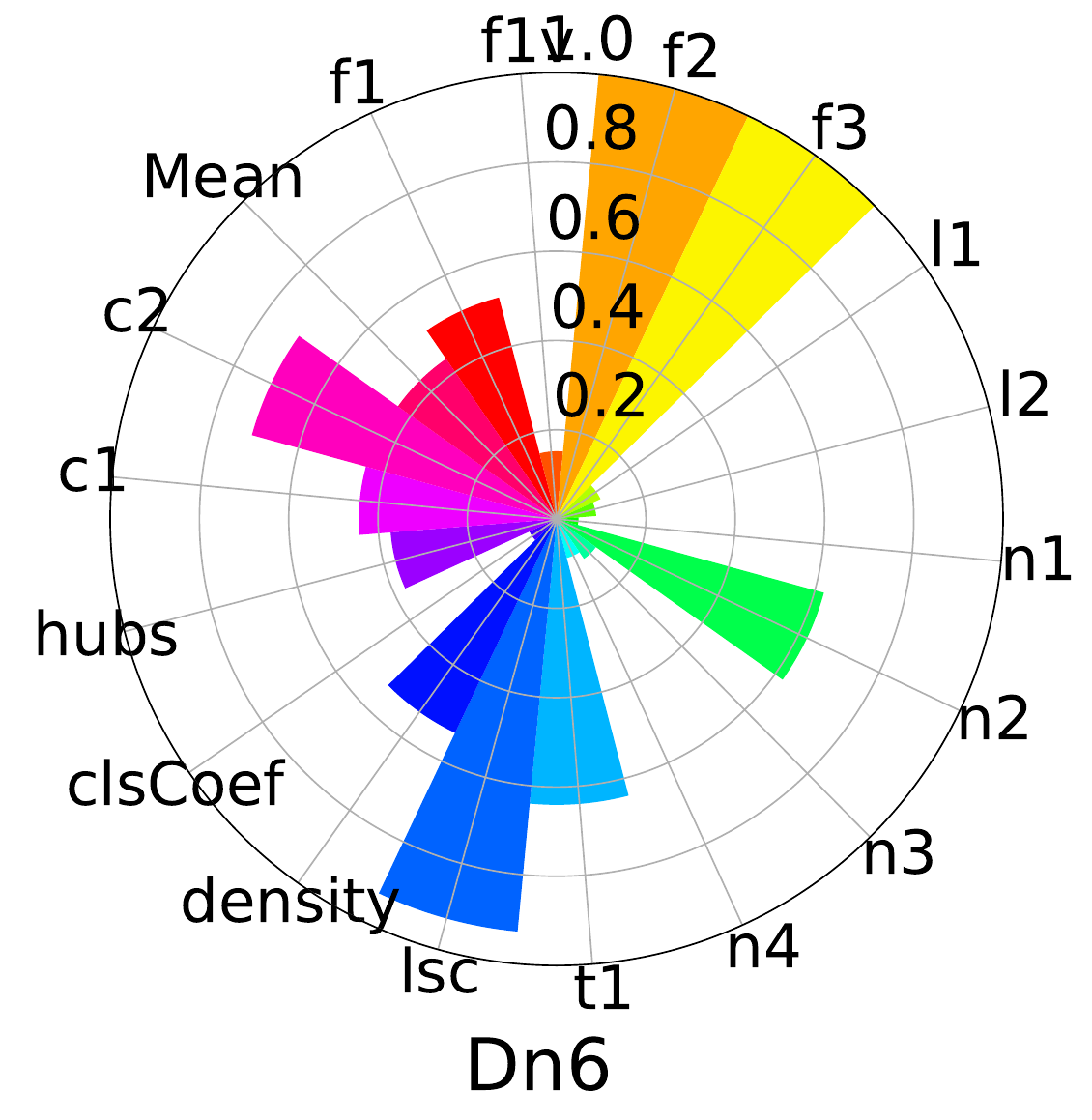}
\includegraphics[width=0.24\textwidth]{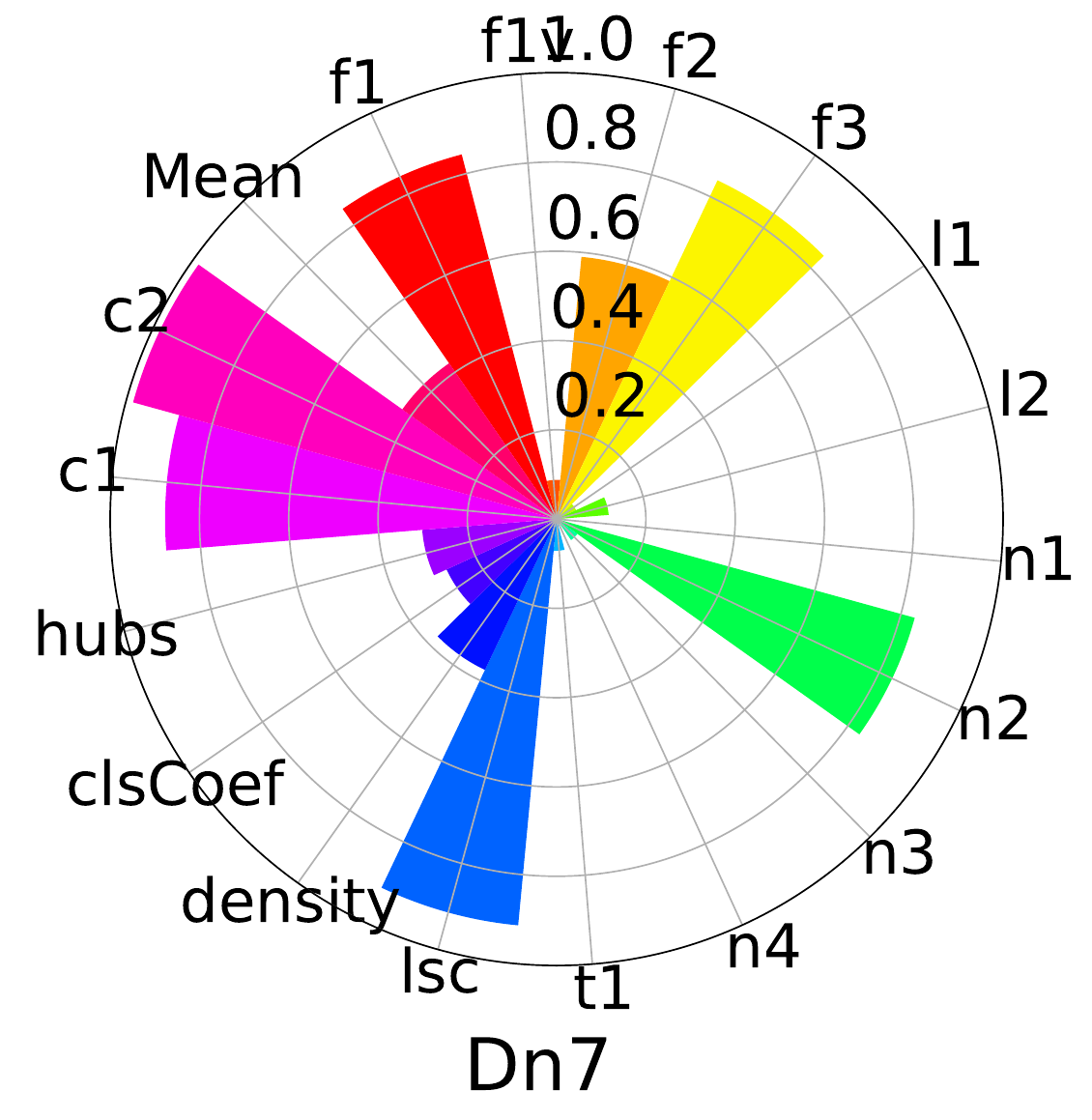}
\includegraphics[width=0.24\textwidth]{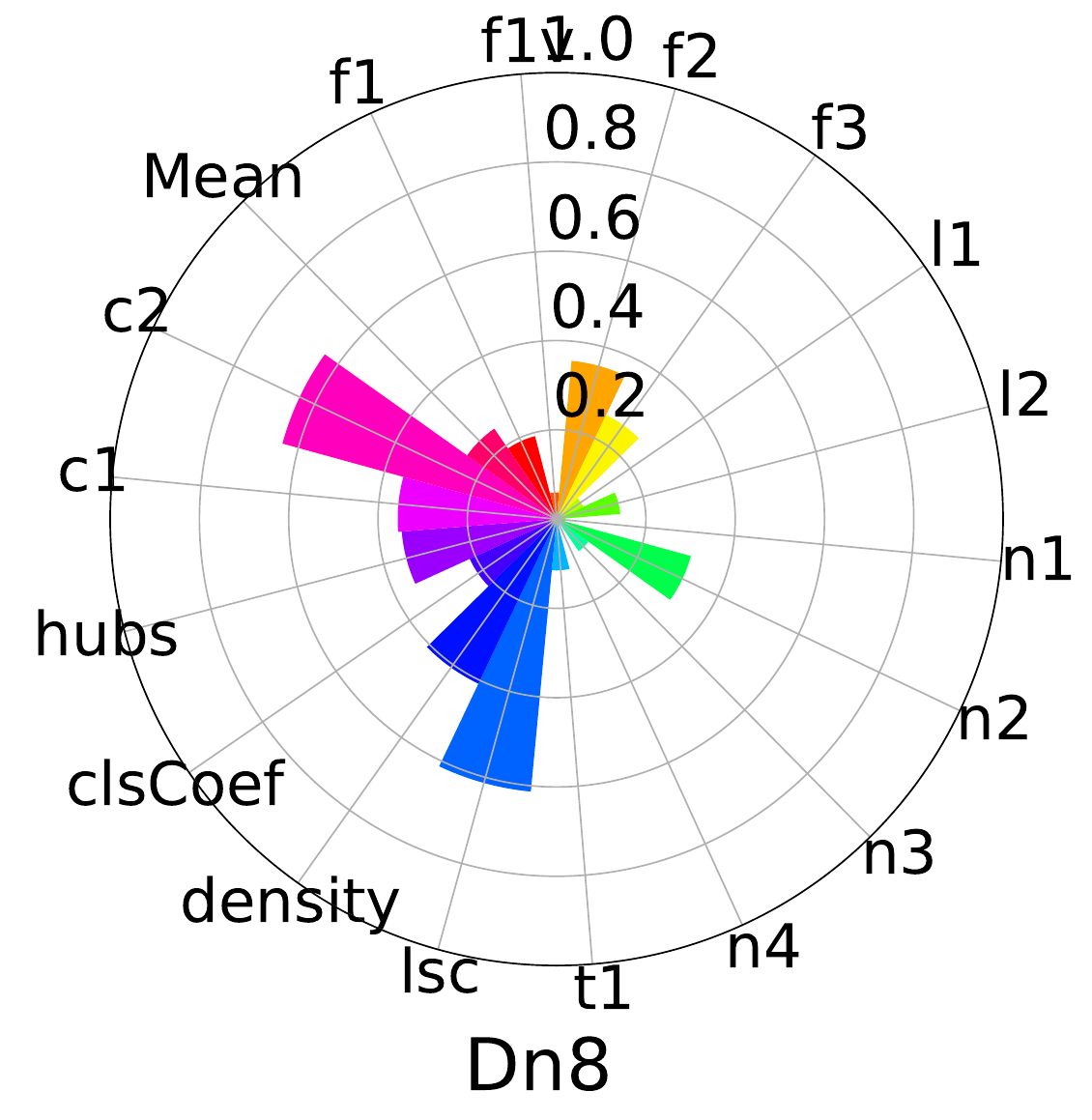}
\caption{Schema-based complexity measures per dataset in Table \ref{tb:newDatasets}.}
\label{fig:scBomplexityFiguresNew}
\end{figure*}

\subsection*{A. Schema-based Theoretical Measures}

The a-priori measures in Section \ref{sec:theoreticalMeasures} leverage schema-agnostic settings, which disregard the attribute values associated with every record. It is possible, though, to adapt them to a schema-aware functionality. Algorithm \ref{algo:linearityDegree} can actually be restricted to the tokens of a specific attribute in a straightforward way. 

Moreover, the complexity measures in Section \ref{sec:complexityMeasures} can be coupled with the Cosine and Jaccard similarities that are calculated for each attribute, representing every candidate pair $c_{i,j}$ through the multi-dimensional feature vector: \[f_{i,j}=[CS^{a_1}(c_{i,j}), JS^{a_1}(c_{i,j}),
\ldots, CS^{|A|}(c_{i,j}), JS^{|A|}(c_{i,j}),]\] where $x^{a_k}(c_{i,j})$ denotes the $x \in \{CS,JS\}$ similarity over the values of attribute $a_k \in A$, while $A$ stands for the set of all attributes describing the records in the input dataset $D$, with $|A|$ being the number of attributes in $D$. The actual value of the complexity measures when using this multi-dimensional feature vector appear in Figures \ref{fig:sbComplexityFigures} and \ref{fig:scBomplexityFiguresNew} for the existing and the new benchmarks in Tables \ref{tb:commonDatasets} and \ref{tb:newDatasets}, respectively.

Comparing Figure \ref{fig:sbComplexityFigures} with  Figure \ref{fig:complexityFigures}, we observe significant variation among the individual measures. Inevitably, the same applies to the mean score across all measures. Yet, using a mean value of 0.4 as the threshold for distinguishing between easy and difficult benchmarks, we are led to the same conclusion: only datasets $D_{s4}$ (0.409), $D_{s6}$ (0.467), 
$D_{d4}$ (0.438), $D_{t1}$ (0.423) and $D_{t2}$ (0.431) are challenging, as all others have an average score well below this threshold, fluctuating between 0.183 ($D_{s7}$) and 0.386 ($D_{d3}$). 

The same pattern appears in Figure \ref{fig:scBomplexityFiguresNew}; there is a significant fluctuation between the individual measures, but when comparing the average one to the cut-off threshold (0.4), the challenging datasets are the , same as the schema-agnostic settings in Figure \ref{fig:complexityFiguresNew}: $D_{n1}$, $D_{n2}$, $D_{n6}$ and $D_{n7}$.

\subsection*{B. Existing Benchmark Datasets}

This section compares in more detail the characteristics of the existing benchmark datasets with those created by following our proposed approach in Section \ref{sec:methodology}.

Starting with the generation process, the datasets typically used for benchmarking learning-based matching algorithms (i.e., the datasets in Table \ref{tb:commonDatasets}) stem from the Magellan repository (note that some publications on the topic use additional datasets, but only the ones in Table appear in multiple publications). Their generation process was originally described at:  \url{https://sites.google.com/site/anhaidgroup/useful-stuff/the-magellan-data-repository/description-of-the-784-data-sets} and repeated in \url{https://github.com/anhaidgroup/deepmatcher/blob/master/Datasets.md}. It involves the following steps for most datasets:

\begin{enumerate}
    \item Given the two tables (tableA.csv and tableB.csv), perform dataset specific blocking to obtain a candidate set $C$.
    \item For each tuple pair in set $C$, if the pair is present in the golden matches file (gold.csv), mark the pair as a match. Else, mark the pair as a non-match.
    \item Randomly split the labeled candidate set C into 3 sets, i.e., train, validation, and test, with ratio 3:1:1.
\end{enumerate}

The procedure is slightly different for $D_{s5}$:
\begin{enumerate}
    \item Crawled HTML pages from the two websites.
    \item Extracted tuples from the HTML pages to create two tables, one per site.
    \item Performed blocking on these tables (to remove obviously non-matched tuple pairs), producing a set of candidate tuple pairs.
    \item Took a random sample of pairs from the above set and labeled the pairs in the sample as ``match'' / ``non-match''.
\end{enumerate}

In both cases, there is no reference to the blocking method that was used and to its parameter configuration (e.g., which attributes were considered in the creation of blocks?). The blocking configuration typically has a significant impact on the resulting performance \cite{DBLP:journals/pvldb/ZeakisPSK23,DBLP:journals/corr/abs-2202-12521}. This means that the dataset generation process is not reproducible and that there is no way of evaluating its outcomes. In general, a loose blocking approach achieves high recall, ensuring that all positive pairs are included in the labeled instances, at the cost of including many negative pairs with low similarity, which can thus be easily discarded by a learning-based matching algorithm, even a linear one. In contrast, a strict blocking approach might sacrifice a small part of the positive pairs, but mostly includes highly similar negative pairs, which are harder to be classified, thus requiring more effective, complex and non-linear learning-based matching algorithms. DeepBlocker, with its high effectiveness, ensures that only negative pairs involving nearest neighbors are included in the labeled instances, without sacrificing a significant portion of the positive 

As a result of this unclear generation pnrocess, the imbalance ratio of the datasets generated by our methodology in Section \ref{sec:methodology}, which leverages DeepBlocker, is much lower than that in the existing benchmark datasets, even though the difference in recall is low. More specifically, we focus on the datasets in Tables \ref{tb:commonDatasets} and \ref{tb:newDatasets} that have the same origin, which are $D_{s1}$-$D_{n3}$, $D_{s2}$-$D_{n8}$, $D_{s4}$-$D_{n7}$, $D_{s6}$-$D_{n2}$ and $D_{t1}$-$D_{n1}$. Note that for $D_{s1}$, $D_{s2}$, $D_{s6}$ and $D_{t1}$, we rely on the original version of the datasets released by \cite{DBLP:journals/pvldb/KopckeTR10}. For $D_{s6}$, we use the statistics of the website that provides this dataset. For the corresponding new benchmarks datasets, we use the clean versions employed in a series of past publications (e.g., \cite{DBLP:journals/pvldb/GagliardelliPSB22,DBLP:journals/pvldb/ZeakisPSK23,DBLP:journals/corr/abs-2202-12521}).

Table \ref{tb:existingVsNewCharacteristics} compares the number of entities and duplicates per dataset. 

\begin{table*}[t]\centering
\renewcommand{\tabcolsep}{5pt}
\caption{{\small Characteristics of the datasets lying at the core of ER benchmarks}}
\vspace{-7pt}
{\small
\begin{tabular}{ | l | r | r | r || r | r | r | c | c |}
\hline
 & \multicolumn{3}{c||}{Original Dataset} & \multicolumn{3}{c|}{Clean Version} & Existing & New\\
 & $D_1$ Entities  & $D_2$ Entities & Duplicates
 & $D_1$ Entities  & $D_2$ Entities & Duplicates &
 Dataset & Dataset \\
 \hline
 \hline
Abt-Buy & 1,081 & 1,092 & 1,097 & 1,076 & 1,076 & 1,076 & $D_{t1}$ & $D_{n1}$ \\
DBLP-ACM & 2,616 & 2,294 & 2,224 & 2,616 & 2,294 & 2,224 & $D_{s1}$ & $D_{n3}$ \\
DBLP-GS & 2,616 & 64,263 & 5,347 & 2,516 & 61,353 & 2,308 & $D_{s2}$ & $D_{n8}$ \\
Walmart-Amazon & 2,554 & 22,074 & 962 & 2,554 & 22,074 & 853 & $D_{s4}$ & $D_{n7}$ \\
Amazon-GP & 1,363 & 3,226 & 1,300 & 1,354 & 3,039 & 1,104 & $D_{s6}$ & $D_{n2}$ \\
\hline
	\end{tabular}
	}
	\vspace{-14pt}
	\label{tb:existingVsNewCharacteristics}
\end{table*}

We observe that the two versions agree only in DBLP-ACM. For Abt-Buy, DBLP-GS and Amazon-GP, the clean version involves both fewer entities and fewer duplicates than the original one, because the latter matches multiple entities from D1 with the same entity in D2 and vice versa. This is evident in Abt-Buy and DBLP-GS by the fact that the number of duplicates is higher than the number of D1 and/or D2 entities. These settings are in conflict with the definition of Record Linkage (aka Clean-Clean ER) \cite{DBLP:series/synthesis/2015Christophides}, which requires that every data source is individually duplicate-free. For this reason, the clean versions of these datasets have removed all entities involved in the contradictory matches. In the case of the Walmart-Amazon dataset, we kept all entities, but removed the contradictory matches from the ground truth.

Based on these generic characteristics, Table \ref{tb:existingVsNew}  compares the ER-specific measures per dataset.

We observe that in Abt-Buy and DBLP-GS, the existing benchmark outperforms the new one both with respect to recall and precision, even though the latter uses a much more effective blocking method than the former \cite{DBLP:journals/pvldb/Thirumuruganathan21} (note also that the configuration of DeepBlocker has been optimized through grid search in order to maximize the precision of the new benchmarks). Therefore, the higher precision of the existing benchmarks is most likely achieved due to the removal of negative pairs. 

For the Walmart-Amazon and Amazon-GP pairs of datasets, the existing benchmarks offer a different balance between recall and precision. In the former pair, the recall of the existing benchmark is lower than the new one by just 6\%, while its precision is higher by a whole order of magnitude. In the latter pair, the recall of the existing benchmark is lower than the new one by just 1.2\%, while its precision is higher by 3.3 times. These tradeoffs are not common in blocking over these two particular datasets and could be caused by removing a large portion of negative pairs.

The opposite is true in the case of DBLP-ACM, where the new benchmark exhibits much higher precision (almost by 7 times) than the existing one, even though their difference in recall is just 1.5\%. Given that a wide range of blocking methods achieves exceptionally high precision in this bibliographic dataset, the low precision of the existing benchmark could be caused by including a large number of easy, negative pairs.

Overall, all five dataset pairs seem to involve an undocumented approach for inserting or removing an arbitrary number of negative pairs.

\end{document}